\documentclass[]{aa}

\usepackage{graphicx}
\usepackage{xcolor}
\makeatletter\@fleqnfalse\makeatother

\usepackage{amsmath}
\usepackage{nicefrac}
\usepackage{xspace}
\usepackage{nccmath,gensymb}
\usepackage{hyperref}

\providecommand{\orcid}[1]{}   
\DeclareRobustCommand{\eqmark}{\footnotemark[1]}   
\makeatletter
\renewcommand*\@fnsymbol[1]{\ensuremath{\ifcase#1\or\dagger\or\star\or{\star}{\star}\or{\star}{\star}{\star}\or\ddagger\or\dagger\dagger\or\ddagger\ddagger\else\@ctrerr\fi}}
\makeatother

\newcommand{\kms}{\ensuremath{\nicefrac{\mathrm{km}}{\mathrm{s}}}}
\newcommand{\prob}[1]{\mathcal{P}\left(#1\right)}

\newcommand{\NvisAll}{375{,}100}
\newcommand{\NstarAll}{218{,}235}
\newcommand{\NvisFit}{236{,}949}
\newcommand{\NstarFit}{142{,}666}
\newcommand{\NvisCat}{2{,}618{,}012}      
\newcommand{\NvisMask}{400{,}469}         
\newcommand{\NvisFracCut}{16{,}443}       
\newcommand{\NstarFracCut}{12{,}799}      
\newcommand{\NvisDistCut}{283{,}324}      
\newcommand{\NstarDistCut}{171{,}874}
\newcommand{\Nsamp}{48}                 
\newcommand{\NsampMock}{24}             
\newcommand{\Rzero}{8.2}                
\newcommand{\rLocal}{1.25}              
\newcommand{\binLin}{4.0}               
\newcommand{\rLin}{247}                 
\newcommand{\binMax}{95}                
\newcommand{\Nrad}{260}                 
\newcommand{\calDens}{1.37}             
\newcommand{\calVel}{1.50}              
\newcommand{\calLo}{1.3}                
\newcommand{\calHi}{1.5}                
\newcommand{\nwMockEW}{1.02}            
\newcommand{\nwMockV}{1.06}             
\newcommand{\nwRef}{1.04}               
\newcommand{\nwRealEW}{0.85}            
\newcommand{\nwRealV}{0.90}             
\newcommand{\nwRealEWscaled}{0.95}      
\newcommand{\recalScale}{0.87}          
\newcommand{\respReal}{0.87}            
\newcommand{\respBelieved}{80\%}        
\newcommand{\respRejected}{8.5\%}       
\newcommand{\respMiddle}{12\%}          
\newcommand{\respNondet}{13\%}          
\newcommand{\respMockBelieved}{74.6\%}  
\newcommand{\respMockTrue}{70.0\%}      
\newcommand{\respDet}{0.985}            
\newcommand{\respNoise}{0.19}           
\newcommand{\priorNoise}{0.43}          
\newcommand{\RdFit}{4.10 \pm 0.15}        
\newcommand{\zhFit}{93.3 \pm 1.8}         
\newcommand{\RflareFit}{11.6 \pm 0.8}     
\newcommand{\RdFull}{3.95 \pm 0.20}       
\newcommand{\zhFull}{110.0 \pm 2.5}
\newcommand{\RflareFull}{11.7 \pm 1.4}
\newcommand{\kReal}{1.12\degr \pm 0.11\degr}    
\newcommand{\kMock}{0.98\degr \pm 0.25\degr}    
\newcommand{\zhMock}{109.3 \pm 4.5}             
\newcommand{\zenodoDOI}{10.5281/zenodo.24136276}        
\newcommand{\zenodoURL}{https://doi.org/10.5281/zenodo.24136276}
\newcommand{\zhMockFlare}{150 \pm 5}            
\newcommand{\zhTruthFlare}{152}                 
\newcommand{\resFloor}{29}              
\newcommand{\sTwoSlope}{0.64 \pm 0.04}  
\newcommand{\sigOnePcPrior}{2.5}        

\newcommand{\rMaserRes}{0.46}
\newcommand{\rClusterRes}{0.28}

\newcommand\apMADGICS{\texttt{apMADGICS.jl\xspace}}
\newcommand{\apDIB}{15{,}273~\AA\ DIB\xspace}

\begin{document}

\title{Kinetic Tomography with Diffuse Interstellar Bands in APOGEE DR17}

\titlerunning{Kinetic tomography with DIBs in APOGEE DR17}
\authorrunning{P. Frank, A. K. Saydjari et al.}

\author{Philipp Frank\inst{1,2,3}\orcid{0000-0001-5610-3779}\fnmsep\thanks{P.F. and A.K.S. contributed equally to this work.}\fnmsep\corrauth{philipp@mpa-garching.mpg.de}
  \and Andrew K. Saydjari\inst{4,5}\orcid{0000-0002-6561-9002}\fnmsep\eqmark\fnmsep\corrauth{aksaydjari@gmail.com}\fnmsep\thanks{Hubble Fellow}
  \and Catherine Zucker\inst{6}\orcid{0000-0002-2250-730X}\email{catherine.zucker@cfa.harvard.edu}
  \and Susan E. Clark\inst{2,3}\orcid{0000-0002-7633-3376}\email{susanclark@stanford.edu}
  \and Douglas Finkbeiner\inst{6}\orcid{0000-0003-2808-275X}\email{dfinkbeiner@cfa.harvard.edu}
  \and Torsten En{\ss}lin\inst{1,7,8,9}\orcid{0000-0001-5246-1624}\email{ensslin@mpa-garching.mpg.de}
  \and Andrew R. Casey\inst{10}\orcid{0000-0003-0174-0564}\email{andrew.casey@monash.edu}
  \and Alexandre Roman-Lopes\inst{11}\orcid{0000-0002-1379-4204}\email{aroman@userena.cl}
  \and Jos\'e G. Fern\'andez-Trincado\inst{12}\orcid{0000-0003-3526-5052}\email{jose.fernandez@ubo.cl}
  \and Dhanesh Krishnarao\inst{13}\orcid{0000-0002-7955-7359}\email{dkrishnarao@coloradocollege.edu}
  \and John C. Wilson\inst{14}\orcid{0000-0001-7828-7257}\email{jcw6z@virginia.edu}}

\institute{Max Planck Institute for Astrophysics, Karl-Schwarzschild-Str. 1, 85748 Garching, Germany
  \and Kavli Institute for Particle Astrophysics \& Cosmology, P.O. Box 2450, Stanford University, Stanford, CA 94305, USA
  \and Department of Physics, Stanford University, Stanford, CA 94305, USA
  \and Department of Astrophysical Sciences, Princeton University, Princeton, NJ 08544, USA
  \and Department of Astronomy, New Mexico State University, Las Cruces, NM 88001, USA
  \and Center for Astrophysics $|$ Harvard \& Smithsonian, 60 Garden Street, Cambridge, MA 02138, USA
  \and Deutsches Zentrum f\"ur Astrophysik, Postplatz 1, 02826 G\"orlitz, Germany
  \and Ludwig-Maximilians-Universit\"at M\"unchen, Geschwister-Scholl-Platz 1, 80539 Munich, Germany
  \and Excellence Cluster ORIGINS, Boltzmannstr. 2, 85748 Garching, Germany
  \and School of Physics \& Astronomy, Monash University, Wellington Road, Clayton, Victoria 3800, Australia
  \and Departamento de Astronom\'ia, Universidad de La Serena, Av. Ra\'ul Bitr\'an 1305, La Serena 1700000, Coquimbo, Chile
  \and Centro de Investigaci\'on en Astronom\'ia, Facultad de Ingenier\'ia, Ciencia y Tecnolog\'ia, Universidad Bernardo O'Higgins, Av. Viel 1497, Santiago, 8370993, Chile
  \and Department of Physics, Colorado College, 14 East Cache la Poudre St., Colorado Springs, CO 80903, USA
  \and Department of Astronomy, University of Virginia, Charlottesville, VA 22904-4325, USA}

\date{Received ---; accepted ---}

\abstract
{Mapping gas flows in the interstellar medium is essential for understanding the physical processes shaping the formation of structure in our Milky Way. Methods that simultaneously reconstruct clouds' distances and line-of-sight velocities are known as kinetic tomography (KT). Diffuse interstellar bands (DIBs) are powerful tracers for such maps because they simultaneously trace the integrated dust density and the dust-weighted velocity. Previous DIB-KT works had small sample sizes with coarse kpc-scale resolution and could not capture vertical disk structure.}
{Make 4D (3D position + 1D radial velocity) maps of the Milky Way from new, high-quality \apDIB detections in APOGEE DR17.}
{We combine recent advances in variational inference and Gaussian processes with a mixture likelihood for low signal-to-noise detections and a full marginalization over distance uncertainties. We calibrate the model on mock fields and spectral injection tests and validate the resulting map against independent tracers of the dust density (photometric 3D dust maps) and velocity substructure (masers and young stellar clusters).}
{We recover a full posterior distribution over 3D maps of the density and velocity structure of the \apDIB in a sphere of radius 6~kpc around the Sun. We measure the large-scale structure of the DIB disk (scale length, scale height and flare scale) and detect vertical distortions consistent with the warp. On smaller scales, the map is able to resolve distance ambiguities in gas longitude--velocity diagrams. From the second-order velocity structure function, we measure a slope ($0.64\pm0.04$) consistent with Kolmogorov turbulence, amplitude consistent with expectations for molecular clouds, and a preliminary ratio of compressive-to-solenoidal modes.}
{With the larger DIB catalogs expected from SDSS-V, this framework should resolve kinematic substructure at the scales necessary to constrain the formation mechanisms and dynamical processes of individual clouds and voids. Our maps are publicly available at \href{\zenodoURL}{doi:\zenodoDOI}.}


\maketitle
\nolinenumbers 
\clearpage

\section{Introduction}

The motion of gas within galaxies is one of the fundamental quantities governing how structure in the interstellar medium (ISM) forms, evolves, and dissipates. The velocity field determines the rate at which diffuse atomic gas is swept up, compressed, and cooled into the dense molecular clouds that form stars, and likewise dictates how those clouds are destroyed by shear and stellar feedback \citep{Ballesteros_2020, Chevance_2023}. On larger scales, the streaming motions of gas across the Galactic disk encode the gravitational potential of the Milky Way and the nature of its spiral structure. For example, whether the arms are long-lived, quasi-stationary density waves or more transient features leaves a distinct kinematic imprint on the gas as it flows into and out of the arms \citep{Dobbs_2014, Shu_2016}. A spatially resolved, 3D view of how the ISM is moving is therefore a prerequisite for connecting the small-scale physics of cloud formation and destruction to the large-scale architecture of the Milky Way.

Historically, the kinematics of the ISM have been traced almost exclusively through CO and \ion{H}{I} spectral-line emission. These surveys provide position--position--velocity (PPV) cubes that map the radial velocity of the gas along the line of sight. They do not, however, indicate \emph{where} along the line of sight that gas is located. Converting an observed line-of-sight radial velocity into a distance requires the assumption of a Galactic rotation curve \citep{Nakanishi_2016, Roman_Duval_2009}, an approach that breaks down where the dynamics are most interesting --- in the presence of non-circular motions \citep{Hunter_2024}. Toward the inner Galaxy, a single radial velocity can also map to two different distances, constituting the well-known near/far kinematic distance ambiguity within the solar circle \citep{Wienen_2015, Kolpak_2003}. As a result, the PPV view of the Milky Way alone makes it difficult to reliably reconstruct the 3D flow of gas. Recent works like \citet{2025A&A...693A.139S} have made progress in this direction by reconstructing the 3D gas density and velocity fields across the disk through forward-modeling gas emission with a Gaussian-process-based velocity field anchored to a rotation curve \textit{and} maser and YSO parallax constraints.

Over the past decade, however, a complementary approach known as \emph{kinetic tomography} has emerged to overcome this limitation. The core idea \citep[cf.][]{Tchernyshyov_2017, Tchernyshyov_2018} is to combine a tracer of distance with a tracer of velocity to recover the full 4D position--position--position--velocity (PPPV) distribution of the ISM. Two broad approaches to kinetic tomography have been developed. 

The first technique couples 3D dust maps to spectral-line emission by assuming an approximately constant gas-to-dust ratio. This approach was introduced by \citet{Tchernyshyov_2017}, who fused distance-resolved 3D dust maps with CO and \ion{H}{I} emission cubes to solve for the line-of-sight velocity field that self-consistently relates the spatially resolved dust to the velocity-resolved gas across much of the Galactic plane. Building on the advent of parsec-scale 3D dust reconstructions enabled by \textit{Gaia} \citep{2020A&A...639A.138L,2024A&A...685A..82E}, \citet{Soler_2023} used the histogram of oriented gradients method to combine the distance information in 3D dust with \ion{H}{I} and CO emission to constrain the dynamics of the Taurus molecular cloud. \citet{Soler_2025} then extended this approach to the Galactic plane within \rLocal{}~kpc of the Sun using the \citet{2024A&A...685A..82E} 3D dust map. Shortly thereafter, \citet{McCallum2026_HI} likewise combined the \citet{2024A&A...685A..82E} 3D dust map with \ion{H}{I} emission from the HI4PI survey within the framework of Information Field Theory \citep{ensslin2009information,ensslin2018information}
to infer the local distance-resolved velocity field at even higher angular and distance resolution. These 3D-dust-based spectral-line approaches are powerful, but they rely on morphologically matching two physically distinct tracers -- dust seen in extinction and gas seen in emission -- across distance and velocity and are particularly limited in the presence of morphological differences \citep[e.g. due to significant variations in the dust-to-gas ratio, as seen locally in recent work;][]{ONeill_2026} or if the 3D dust does not trace the full column of gas due to the sensitivity and depth of current maps. 

The second technique sidesteps this morphological matching challenge by using a single tracer that carries both distance and velocity information at once: diffuse interstellar bands (DIBs). DIBs are broad (Gaussian widths of $\sim 1$--$2$~\AA) absorption features, imprinted on stellar spectra by the intervening ISM, whose carriers remain largely unidentified but reliably trace interstellar material \citep{Zasowski_2015, Saydjari_2023_ApJ}. Because a DIB is measured in absorption against a background star, its strength constrains the amount of intervening medium while its central wavelength yields a density-weighted average of the integrated line-of-sight velocity. Pairing that absorbing star with a stellar distance then localizes that measurement along the line of sight. DIB-based kinetic tomography is thus uniquely direct, mapping the kinematics of the ISM from a single, internally consistent set of absorption measurements. Prior work by \citet{Tchernyshyov_2018} used APOGEE DR14 DIB measurements to reconstruct the line-of-sight velocity field of the northern Galactic plane, and compared the result to simulations of density-wave and dynamic spiral structure, finding the data more consistent with transient, dynamically evolving arms.

\begin{figure*}[t]
    \centering
    \includegraphics[width=\linewidth]{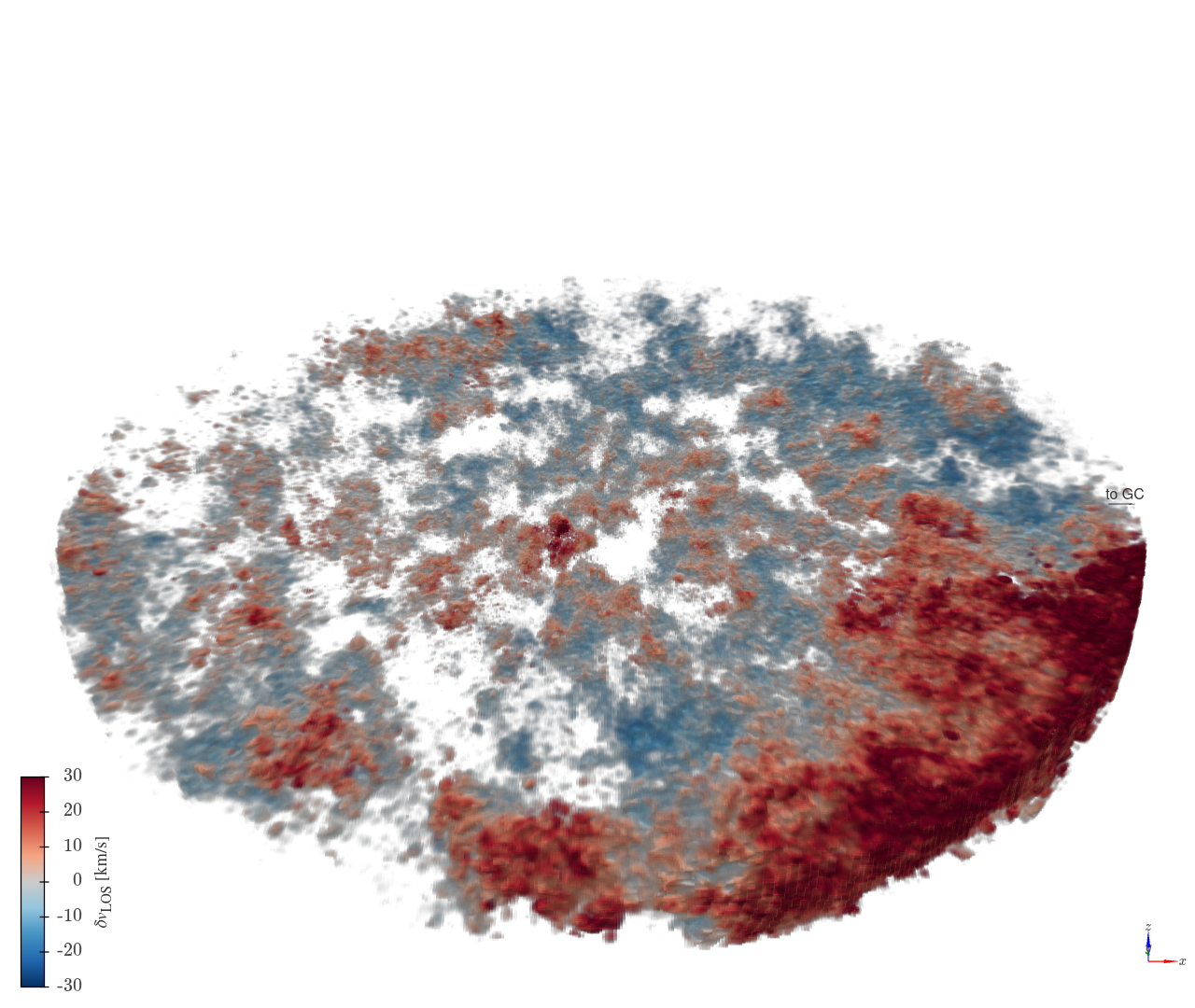}
    \caption{3D rendering of one posterior sample of the differential DIB line-of-sight velocity field ($\delta v_{\rm LOS}$) in the heliocentric frame. The opacity scales with DIB density and is clipped at its $99.5$th percentile. An interactive version of this rendering is available at \url{https://www.ph-frank.de/resources/dib_kt_dr19_2026.html}.}
    \label{fig:money_plot}
\end{figure*}

In this work, we revisit DIB-based kinetic tomography with a dramatically larger and higher-quality sample of DIB measurements. A new reprocessing of APOGEE DR17 spectra using the \texttt{apMADGICS} component-separation pipeline \citep{Saydjari_2025_AJ, Saydjari_APOGEE_DIB_unpub} yields measurements of the 15{,}273~\AA\ DIB --- both its equivalent width and its central velocity --- toward \NstarAll\ unique stars with matched \textit{Gaia} distances \citep{2021AJ....161..147B}. This new sample contains a factor of $\sim\!12$ more stars than the $17{,}546$ midplane stars ($|b| < 1\degr$) used in the earlier DIB-based kinetic-tomography study of \citet{Tchernyshyov_2018}. We combine these DIB measurements with a forward model and a non-parametric Gaussian-process based prior built within the software package Numerical Information Field Theory (NIFTy, \citealt{Selig2013, nifty}) to produce a joint, fully probabilistic reconstruction of both the 3D DIB density and the 3D line-of-sight velocity field of the ISM out to roughly 6 kpc from the Sun. Figure \ref{fig:money_plot} shows a three-dimensional rendering of the resulting velocity field.

This paper is organized as follows. In \S \ref{sec:data} we describe the APOGEE DR17 DIB catalog produced by \texttt{apMADGICS} and the quality cuts that define our working sample. In \S \ref{sec:model} we present our forward model and the inference scheme used to make the reconstruction tractable across the full 12~kpc diameter sphere. In \S \ref{sec:results} we present the resulting 3D maps of DIB density and line-of-sight velocity, validate them against independent tracers and discuss measured properties of our maps. We conclude in \S \ref{sec:conclusions} with a summary of reported properties and validations of our 3D DIB density and velocities.

\section{Data} \label{sec:data}

We use a catalog of measurements of the \apDIB for kinetic tomography, made on \NvisCat\ visit-level APOGEE spectra and produced by a reprocessing of the spectra released as part of DR17 from the Sloan Digital Sky Survey \citep[SDSS, ][]{Abdurrouf_2022_ApJS}. The catalog is available\footnote{\url{https://www.sdss.org/dr19/data_access/value-added-catalogs/?vac_id=10006}} as a value-added catalog (VAC) in SDSS DR19, a data release of SDSS-V \citep{Kollmeier_2026_SDSSV}.\footnote{However, the catalog only contains spectra through DR17.}

The APOGEE spectrographs \citep{Wilson_2019_APOGEE} measure near-infrared spectra from $\sim15{,}000$--$17{,}000$~\AA\ with resolution $R = \lambda/\Delta\lambda \approx 22{,}500$, using three mercury cadmium telluride (HgCdTe) 2048 $\times$ 2048 pixel detectors, which leave two $\sim50$ \AA\ gaps. Each spectrograph is fed with 300 fibers, so it can observe up to 300 targets simultaneously. Two versions of the instrument provide full-sky coverage, APOGEE-N at Apache Point Observatory (APO) on the Sloan Foundation 2.5 m telescope \citep{Gunn_2006_SDSS} and APOGEE-S at Las Campanas Observatory (LCO) on the Irénée du Pont 2.5 m telescope \citep{Bowen_1973_duPont}. APOGEE DR17 contains observations spanning almost a decade, between 09-03-2011 and 01-21-2021. 

The \apDIB catalog was produced with a Bayesian component separation pipeline (\apMADGICS\footnote{\url{https://github.com/andrew-saydjari/apMADGICS.jl}}) that uses a method called ``Marginalized Analytic Data-space Gaussian Inference for Component Separation'' (MADGICS). In addition to producing an APOGEE DIB catalog, these component separations for APOGEE have improved the stellar radial velocity precision from 100~m\,s$^{-1}$ to 30~m\,s$^{-1}$ \citep{Saydjari_2025_AJ}. More broadly, the MADGICS method has also been applied to \textit{Gaia} RVS spectra for DIBs \citep{Saydjari_2023_ApJ} and DESI spectra for galaxies with strong Lyman-alpha emission \citep{Uzsoy_2025_arXiv}. 

The basic idea of MADGICS is to model a spectrum as a linear combination of components $x_k$, where we put a prior on each of these components expressed as a pixel-pixel covariance matrix $C_k$. That is, the components and their priors live in the ``data space'' of spectral wavelength bins. Then, subject to the constraint that the components sum exactly to the data, we obtain a posterior for the components.

\begin{ceqn}
\begin{align} \label{eq:post_mean}
    \hat{x}_{k} &= C_k C_{\rm{tot}}^{-1}\left(x_{\rm{tot}} -\mu_{\rm{tot}}\right) + \mu_k \\
    \label{eq:post_kk}
    \hat{C}_{kk} &= (I - C_k C_{\rm{tot}}^{-1})C_k \\
    \label{eq:post_km}
    \hat{C}_{km} &= -C_k C_{\rm{tot}}^{-1} C_m
\end{align}
\end{ceqn}
where $I$ is the identity matrix, $\hat{x}_k$ is the predicted posterior mean component, $\hat{C}_{kk}$ is the predicted pixel-pixel covariance of pixels in component $k$, $\hat{C}_{km}$ is the predicted covariance of a pixel in component $k$ with a pixel in component $m$, $C_{\rm{tot}}$ is the sum of the covariance priors for all components in the model, and $x_{\rm{tot}}$ is the sum of all components, which is by definition equal to the observed data. For more details see \citet{Saydjari_2023_ApJ} and \citet{Saydjari_MADGICS_unpub}.

For APOGEE, the MADGICS model decomposed the spectra into ``sky continuum'' + ``sky emission lines'' + ``star continuum'' + ``star lines'' + ``DIB'' + ``noise'' components. The DIB component was modeled with a Gaussian lineshape, explicitly sampling over a variable central wavelength and width (first and second moments). \citet{Saydjari_APOGEE_DIB_unpub} only model a single DIB at a time, exploring the two strongest DIBs in the APOGEE wavelength range, though we only consider the strongest here. From these decompositions, we use the equivalent width and the central velocity of the \apDIB\ together with their measurement uncertainties, all obtained from the posterior over the line's central wavelength and width. For high quality DIB catalogs, it is extremely important that the prior for the stellar and sky components be data-driven, so we can fully marginalize over hard-to-model lines \citep[see Figure 1, ][]{Saydjari_2023_ApJ}.

We exclude from the APOGEE DR17 MADGICS catalog DIB measurements likely to contain multiple velocity components (we keep $\sigma_{\text{DIB}} < 3.5$~\AA), in addition to the mandatory quality cuts on $\chi^2$ and grid-edges from \citet{Saydjari_APOGEE_DIB_unpub}. The average $\sigma_{\text{DIB}}$ of the resulting sample is $1.7$~\AA. We deliberately do not remove fits with a zero or negative measured equivalent width. For a weak or absent band the measured equivalent width scatters symmetrically around zero, and discarding one side of that scatter would truncate the noise distribution and bias the noise model of \S\ref{sec:lowsnr}, which has to describe exactly these measurements as non-detections. Such fits account for 13\% of the final catalog and 10\% of the visits included in the reconstruction. We employ an additional weak quality cut at a signal-to-noise ratio ${\rm SNR} > 2$ to remove a few extreme failures at low SNR.

Of the \NvisCat\ visit spectra in the catalog, \NvisMask\ pass these cuts. We further require a distance for each star that has at least one such visit, which we obtain from the catalog of \citet{2021AJ....161..147B} (hereafter CBJ), matched through the \textit{Gaia} EDR3 source identifier. We use the CBJ geometric distances, which combine the EDR3 parallax with a direction-dependent prior on the distance distribution of stars derived from a model of the Galaxy, rather than the photogeometric ones, which also depend on a model of interstellar extinction that the DIBs trace. For every star we take the median and the $16$th and $84$th percentiles of its geometric distance posterior (\texttt{r\_med\_geo}, \texttt{r\_lo\_geo}, \texttt{r\_hi\_geo}). We use the $84$th distance percentiles to obtain a distance selection of stars to be confidently within our reconstruction volume of radius 6 kpc (see \S\ref{sec:distances} for details).

\begin{figure*}
    \centering
    \includegraphics[width=\linewidth]{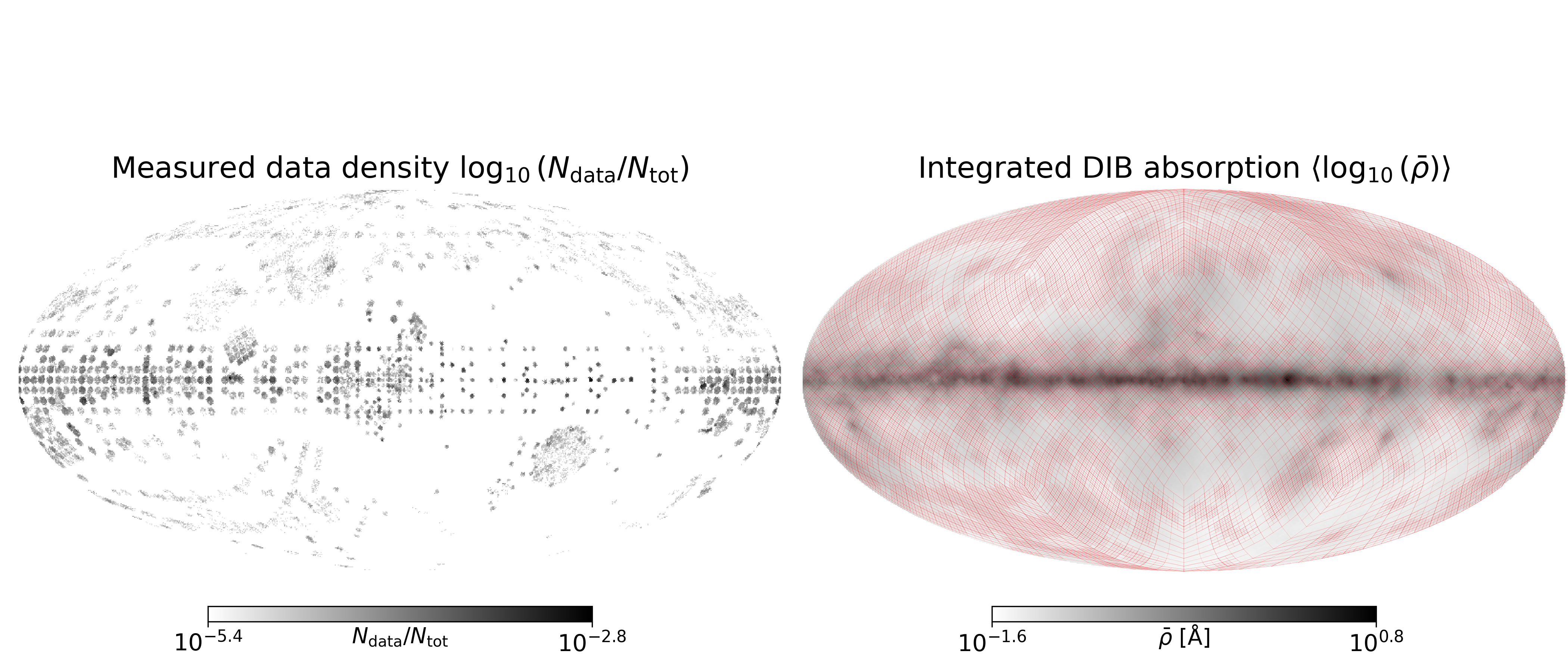}
    \caption{The APOGEE DR17 \apDIB\ footprint and the reconstructed DIB density. \emph{Left:} the fraction of retained DIB measurements per HEALPix pixel ($N_{\rm data}/N_{\rm tot}$). \emph{Right:} the line-of-sight integrated posterior mean DIB density on logarithmic scale, with the sparse Iterative Charted Refinement grid of \S\ref{sec:icr} overlaid in red.}
    \label{fig:allsky_data_plot}
\end{figure*}

Most stars are observed several times, and the interstellar absorption toward a given star is the same in every visit, so a genuine DIB should be recovered in a consistent fraction of a star's visits. For each star we therefore compute the fraction of its visits in which \apMADGICS\ returns a DIB fit that passes the quality cuts, and discard all visits of stars for which this fraction is below $0.1$. This cut removes \NvisFracCut\ visits (4.2\%) toward \NstarFracCut\ stars; these stars have a median of 20 visits and a median recovery fraction of 0.07, against a median of three visits and 0.4 for the stars we keep. The removed measurements also have a median DIB signal-to-noise ratio of 2.0 versus 4.5 for the ones we retained. This yields a final catalog of \NvisFit\ visits toward \NstarFit\ stars that enter the reconstruction. After applying these quality cuts, the survey footprint across the sky can be seen in Figure \ref{fig:allsky_data_plot}.


\section{Model}
\label{sec:model}
To construct the joint likelihood across all stars within a given angular window, we first model the likelihood of each star individually. We combine all of the visits of a given star by treating them in the joint likelihood as independent measurements of DIB absorption towards each star. We treat the plane-of-sky (POS) positions of the stars as exact and account only for the uncertainties of their line-of-sight (LOS) distances (\S\ref{sec:distances}).

For a star at distance $d^*$, we construct a probabilistic forward model for the total amount of DIB absorption ($\bar{\rho}_{d^*}$ as measured in EW) and the DIB's central velocity ($\bar{v}_{d^*}$ in \kms) as measured relative to the local standard of rest \citep{Reid2019}.
For a single measurement, we can connect these quantities to their respective line-of-sight distributions $\rho\left(r\right)$ and $v\left(r\right)$ via
\begin{align}\label{eq:ew_v}
    \bar{\rho}(d^*) &\equiv \int_0^{d^*} \rho\left(r\right) \ \mathrm{d}r \ , \\
    \bar{v}(d^*) &\equiv \frac{1}{\bar{\rho}(d^*)} \int_0^{d^*} v\left(r\right) \rho\left(r\right) \ \mathrm{d}r \ .
\end{align}
Analogously, we can also think of $\bar{\rho}(d^*)$ and $\bar{v}(d^*)$ as the 0th and 1st moment obtained when integrating absorption of local strength $\rho$ at local central velocity $v$ up to a general distance $d^*$. Throughout this paper the \emph{DIB density} $\rho$ denotes this local absorption strength, the differential equivalent width of the \apDIB\ per unit path length, in units of $\nicefrac{\text{\AA}}{\mathrm{pc}}$.
Given a model for the line-of-sight distributions $\rho(r)$ and $v(r)$, we can then connect those with a likelihood of observing $\bar{\rho}_{d^*}$ and $\bar{v}_{d^*}$ for a given star at a given distance $d^*$. This constitutes the core relation on which we build our full forward model, which includes multiple measurements for many stars, distance uncertainties to them, a measurement likelihood respecting the properties of the MADGICS reduction pipeline, and a non-parametric generative 3D prior model for both $\rho$ and $v$ that models all line-of-sight distributions jointly and incorporates correlations in 3D space to connect and interpolate the constraints obtained from the data across the plane of sky and along the line of sight.
\subsection{Data model}
For a single visit of a single star the \apMADGICS\ DIB catalog provides data $d = \left(d_{\rho}, d_v\right)$ and corresponding measurement errors $\sigma = \left(\sigma_\rho, \sigma_v\right)$ as given by the maximum-likelihood (ML) fits and their respective local curvature of the likelihood surface \citep{Saydjari_APOGEE_DIB_unpub}. For a confidently detected DIB they provide a likelihood for $\bar{x}_{d^*} \equiv \left(\bar{\rho}(d^*), \bar{v}(d^*)\right)$ that is close to, but not exactly, Gaussian in the noise-weighted residual $r$. We write that residual, affinely recalibrated, as
\begin{align}\label{eq:nwr}
    r &\equiv \left(\frac{d_\rho - \bar{\rho}_{d^*}}{\sigma_\rho},\ \frac{d_v - \bar{v}_{d^*}}{\sigma_v}\right) \ , \\
    \tilde{r} &\equiv L\left(r - b_{\rm n} - M\,\bar{x}_{d^*}\right) \ , \nonumber
\end{align}
with a constant offset $b_{\rm n}$, a term $M\,\bar{x}_{d^*}$ that absorbs a bias proportional to the true absorption, and a lower-triangular $L$ that rescales the equivalent-width residual and lets the velocity residual depend on it, and evaluate it under a Student-$t$ distribution with $\nu$ degrees of freedom in each component $\tilde{r} = \left(\tilde{r}_\rho, \tilde{r}_v\right)$,
\begin{align}\label{ew:pdetect}
    \prob{d|\bar{x}_{d^*}, \mathrm{detect}} &= \frac{\left|\det L\right|}{\sigma_\rho \sigma_v} \ t_\nu\!\left(\tilde{r}_\rho\right) \ t_\nu\!\left(\tilde{r}_v\right) \ , \\
    t_\nu(x) &\propto \left(1 + \frac{x^2}{\nu}\right)^{-\frac{\nu+1}{2}} \ , \nonumber
\end{align}
so that the small fraction of catastrophic outliers surviving the quality cuts cannot dominate the posterior; we use $\nu = 10$. For an unbiased, correctly scaled measurement (i.e. $L = \mathbf{I}$, $b_{\rm n} = 0$, $M = 0$ and $\nu \to \infty$), this reduces to the product of Gaussians $\mathcal{N}(d_\rho; \bar{\rho}_{d^*}, \sigma_\rho^2)\,\mathcal{N}(d_v; \bar{v}_{d^*}, \sigma_v^2)$. The recalibration parameters are determined from the \apMADGICS\ injection tests in Appendix \ref{ap:lowsnr}. Overall their impact is small. The velocity likelihood is already calibrated and hence unchanged and the equivalent width likelihood obtains a small density dependent bias correction as well as an overall inflation of errors by about 15\%.

\subsubsection{Including low-SNR data}
\label{sec:lowsnr}
For lower SNR, the likelihood above alone loses its validity as a model for the true underlying measurement's likelihood. In essence, as signal decreases (and eventually approaches zero) it becomes more likely that random noise peaks are identified and mapped out as DIB features instead of the true underlying signal. MADGICS essentially performs its ML fit by searching for the global optimum among the locally optimal fits for a window of wavelength bins around the \apDIB central rest wavelength. The best fit and its respective curvature found this way are then reported for the visit. While the reported values remain a valid likelihood, the detection probability of observing the true DIB feature instead of a noise peak decreases with SNR and approaches zero for sufficiently small signals. The zero signal likelihood hence becomes independent of the true signal and follows the distribution of non-detections obtained by measuring values from noise peaks with MADGICS.

Fortunately, MADGICS provides us with two important additional ingredients beyond data and measurement errors: A proxy for the SNR of a given visit, and a suite of injection tests that use fiducial true DIB signals injected into real, DIB-signal-free noise traces from stars with varying levels of residual noise and negligible absorption, together with their corresponding recovered data. They allow us to model the full likelihood in a principled way by mapping out the distribution of non-detections using the MADGICS results on noise traces alone and by modeling the probability of detection using the model signal strength and the SNR proxy. The transition between the high- and low-SNR likelihoods is smooth and the full likelihood is formally defined in terms of a mixture by marginalizing over the probability of detection
\begin{align}\label{ew:visit}
    \prob{d|\bar{x}_{d^*}} &= \prob{d|\bar{x}_{d^*}, \mathrm{detect}} \ \prob{\mathrm{detect}|\bar{x}_{d^*}} 
    \\
    &+ \prob{d|\overline{\mathrm{detect}}} \ \left(1 -\prob{\mathrm{detect}|\bar{x}_{d^*}}\right) \nonumber \ ,
\end{align}
where $\prob{d|\bar{x}_{d^*}, \mathrm{detect}}$ is the likelihood in case of detecting the true DIB feature from \eqref{ew:pdetect}, $\prob{\mathrm{detect}|\bar{x}_{d^*}}$ denotes the detection probability as a function of predicted model parameters, and $\prob{d|\overline{\mathrm{detect}}}$ is the distribution of non-detections. The detection probability is a raised logistic in a model-dependent signal-to-noise ratio,
\begin{align}\label{eq:pdetect_param}
    \prob{\mathrm{detect}|\bar{x}_{d^*}} &= \left(\frac{1}{1 + e^{-\left(a\,\mathrm{SNR} + b\right)}}\right)^{10} \ , \\
    \mathrm{SNR} &\equiv \mathrm{SNR}_{\rm proxy} \left|\frac{\bar{\rho}_{d^*}}{d_\rho}\right| \ , \nonumber
\end{align}
where $\mathrm{SNR}_{\rm proxy}$ is the per-visit signal-to-noise proxy supplied by \apMADGICS. Evaluating it at the \emph{model-predicted} absorption rather than the measured one is what makes this a likelihood rather than a data cut: a visit whose measured equivalent width looks like a noise excursion given the density the model predicts along its sightline is smoothly demoted, and the same visit is promoted if the model predicts real absorption there. Under the non-detection hypothesis the measurement is uninformative about the velocity, so $\prob{d_v|\overline{\mathrm{detect}}} = p_v$ is flat over the velocity range considered, while the reported equivalent width is the largest of many noise excursions and follows a Gumbel-type extreme-value distribution \citep{Gumbel1958}, symmetrized in $|d_\rho|$ because a noise peak is equally likely to be found in absorption or emission:
\begin{align}\label{eq:gumbel}
    \prob{d|\overline{\mathrm{detect}}}
    &= \frac{p_v}{2\sigma_{\rm n}\left(1 - e^{-e^{\mu_{\rm n}/\sigma_{\rm n}}}\right)}
      \exp\left(-t - e^{-t}\right) \ , \\
    t &\equiv \frac{|d_\rho| - \mu_{\rm n}}{\sigma_{\rm n}} \ . \nonumber
\end{align}
The likelihood recalibration parameters of Equation \eqref{eq:nwr} ($L$, $b_{\rm n}$, $M$) together with the non-detection parameters of Equations \eqref{eq:pdetect_param} and \eqref{eq:gumbel} ($a$, $b$, $\mu_{\rm n}$, $\sigma_{\rm n}$) are fit jointly to the \apMADGICS\ injection tests. The fit, its values and the tests of the calibration are in Appendix \ref{ap:lowsnr}. The calibration provides a substantial correction to the sample: 47\% of the modeled visits have $\prob{\mathrm{detect}|\bar{x}_{d^*}} < 0.9$ and 11.5\% fall below $0.5$ when evaluating the detection probability at the measured equivalent width. The turnover of the detection probability lies at ${\rm SNR} = 2.43$ but its width is wide and only reaches $\prob{\mathrm{detect}|\bar{x}_{d^*}} > 0.9$ for ${\rm SNR} > 6$.

Extending the likelihood beyond the well-defined Gaussian regime into the low-SNR regime, where the likelihood of the DIB moments becomes a mixture of detections and non-detections, significantly increases the model complexity and we would like to justify their inclusion. For one, a substantial fraction of the stars are observed more than once: 38.7\% of the stars in our sample have two or more visits and 15.0\% have three or more, up to a maximum of 31 visits. While the per-visit SNR might be too low to significantly detect a DIB signal, the combined information still produces significant signals. Second, for tomographic mapping, being able to use lines of sight that have DIB measurements consistent with zero density gives important lower bounds to the amount of density that can be observed along the LOS, and thus significantly increases the density contrast that can be achieved within the map. Finally, relative to the confidently detected part of our own catalog (${\rm SNR} > 6$), admitting the low-SNR visits increases the total number of visits by a factor of $\sim2.4$ and the number of unique stars by a factor of $\sim2.7$.
The radial distribution of low-SNR visits also differs from the other visits: throughout the volume they add about a factor of two more visits distributed in the same way as the high-SNR ones, while they add up to an order of magnitude more visits within the first $\sim$1 kpc (see Figure \ref{fig:ap_distance} for details). Most of those are measurements consistent with zero and hence significantly help to clean up nearby regions empty of DIB absorption.

\subsubsection{Per-star likelihood}
Equation \eqref{ew:visit} defines the per-visit likelihood of observing DIB absorption and central velocity data $d$ towards a star at distance $d^*$. Hence for each star $i$ having $V$ visits $j \in \left\lbrace 1, ..., V  \right\rbrace$ with data $d_{i,j}$, we can define the per-star dataset $D_i = \left\lbrace d_{i,1}, ..., d_{i,V}\right\rbrace$. Assuming each visit to be an independent measurement of the same underlying density and velocity field we can use the per-visit likelihood to define the likelihood of $D_i$ as
\begin{equation}\label{eq:stardist}
    \prob{D_i | \bar{x}_{d^*}} = \prod_{j = 1}^V \prob{d_{i,j}|\bar{x}_{d^*}} \ .
\end{equation}

\subsubsection{Distance uncertainties}
\label{sec:distances}
Finally, while the plane of sky (POS) positions of stars we consider in the APOGEE footprint are known to a precision far beyond the resolution targeted for the DIB map and hence virtually noiseless, their line-of-sight (LOS) distances obtained from \textit{Gaia} EDR3 carry significant and heterogeneous distance uncertainties that range from a few pc for nearby stars up to a kpc at the edge of our volume and hence have to be fully taken into account as part of the likelihood. To do so, we make use of the CBJ geometric distance posteriors \citep{2021AJ....161..147B, 2021yCat.1352....0B} and use them to construct a functional form for the distribution from their reported quantiles. Specifically we use the $16$th, $50$th, and $84$th posterior percentiles to construct a distribution $\prob{d_i^*|\mathrm{CBJ}}$ as an asymmetric Normal distribution on log-distances (see Appendix \ref{ap:cbj_distances} for further details on constructing the distance posterior and respective model choices). We keep the visits whose $16$th and $84$th distance percentiles both fall inside the reconstructed volume ($2$~pc to $6$~kpc). The median distance of stars in our sample is 2.3~kpc and the median fractional distance uncertainty over the whole sample is around 5\%. The median fractional uncertainty changes rapidly with distance with 2\% at $1$~kpc, 7\% at $3$~kpc and 11\% at $4$--$5$~kpc and is strongly skewed at large distance (the $84$th percentile reaches 25\% at $4$~kpc; Figure \ref{fig:ap_distance}). Note that toward the edge of the volume only stars with relatively precise distances survive the selection.

The full per star likelihood, after marginalization of distance uncertainties, reads
\begin{align}\label{eq:perstar}
    \prob{D_i | \rho, v} &= \int_0^\infty  \prob{D_i | \bar{x}_{d^*}} \ \prob{d_i^*|\mathrm{CBJ}} \ \mathrm{d}d_i^* 
    \\
    &= \int_0^\infty  \left(\prod_{j = 1}^V \prob{d_{i,j}|\bar{x}_{d^*}}\right) \prob{d_i^*|\mathrm{CBJ}} \, \mathrm{d}d_i^* \, , \nonumber
\end{align}
where $\bar{x}_{d^*} \equiv \left(\bar{\rho}(d^*), \bar{v}(d^*)\right)$ is the pair of moments of Equation \eqref{eq:ew_v} evaluated at the distance $d^*$ that is integrated over, obtained from the two 3D distributions of DIB density $\rho$ and velocity $v$.
The joint likelihood of all stars $i \in \left\lbrace1, ..., N^*\right\rbrace$ is obtained by the product of all per star measurements as
\begin{equation}\label{eq:full_lh}
    \prob{D | \rho, v} = \prod_{i = 1}^{N^*} \prob{D_i | \rho, v} \ .
\end{equation}
We note that from a DIB perspective this likelihood can be fully expressed in terms of $\rho$ and $v$. Hence we can think of it as a joint likelihood that takes in the 3D (POS + LOS) distributions, reads out the moment 0 and 1 distributions along the line of sight for each star $i$ (no POS uncertainty), and then proceeds to evaluate and integrate the per star likelihood, taking into account all visits, at every distance weighted by the respective star distance posterior. Once the 3D volume is discretized and the integrals become sums, this fully specifies the likelihood implemented for the analysis in this work (see Appendix \ref{ap:grid} for discretization details).

As is the case for low-SNR DIB data, a full treatment of distance uncertainties beyond the linear regime adds a level of complexity to the model we want to justify and comment on. Compared with previous \textit{Gaia}-based 3D dust-mapping studies \citep{2024A&A...685A..82E, 2020A&A...639A.138L}, which used a linearized likelihood treating distance uncertainties as first and second order corrections in a Gaussian (or Student-$t$) likelihood, the APOGEE DIB footprint is significantly sparser in both star density as well as sky coverage, while we simultaneously probe a much larger volume spanning a sphere with radius $6$ kpc compared to, e.g., the $\rLocal{}$ kpc radius used in \citet{2024A&A...685A..82E}.
In the larger volume of this work, a significant part of the stars have distance uncertainties comparable to or even larger than the typical correlation scale of the measured density, which rules out a perturbative treatment of the distance uncertainties. 
While we do not know the 3D spatial correlation of DIBs a priori and will also infer it from data (see \S\ref{sec:cf} for details), for the sake of a rough scaling we may use the correlation scale of 3D dust as an estimate. Following \citet{2020A&A...639A.138L} we find that local dust extinction, on average, is spatially correlated at most up to $\sim 100$ pc scales, whereas \citet{2021yCat.1352....0B} gives comparable $\sim 100$ pc distance uncertainties already at a stellar distance of $2$ kpc and hence puts most measurements in our volume in the ``uncertainty beyond correlation'' regime. Many stars probing the same correlated region of the DIB density field partially mitigate this effect, which is one of the strengths of full 3D tomography, but at the given sampling density they do not change the picture significantly.

Overall this provides us with a joint likelihood across the full DIB catalog produced by the \apMADGICS\ pipeline, while taking into account multiple visits, calibrated measurement uncertainties, a model for non-detections, and the uncertainty of the \textit{Gaia}-based stellar distances.

\subsection{Density and velocity prior models}
Complementing the likelihood, we incorporate prior knowledge about the fields of interest in order to regularize the inference of the three-dimensional maps in regions of sparse or absent data. We make use of Gaussian Processes (GPs) \citep{rasmussen2006gaussian} as non-parametric and flexible universal function approximators to model the large-scale distribution of the 3D DIB density and velocity fields. Specifically, we adopt and extend the Information Field Theory  framework \citep{ensslin2009information,ensslin2018information}, which has been successfully applied to a range of tomographic reconstructions of the local interstellar medium \citep[e.g.][]{2020A&A...639A.138L, 2024A&A...685A..82E, 2025A&A...693A.139S, McCallum2026_HI, McCallum2026_Bfield, Zandinejad2026_CR} over the last years.

Within this framework, we model the logarithmic DIB density $\tau(\mathbf{x}) \equiv \log \rho(\mathbf{x})$
and the line-of-sight velocity field \(v(\mathbf{x})\) (relative to the local standard of rest) as Gaussian random fields with prior distributions
\begin{equation}
    \prob{\tau} = \mathcal{N}\!\left(\bar{\tau},\, C_\tau\right), 
    \qquad 
    \prob{v} = \mathcal{N}\!\left(\bar{v},\, C_v\right),
\end{equation}
where \(\bar{\tau}\) and \(\bar{v}\) are the prior means and \(C_\tau\) and \(C_v\) the prior covariance operators. Both the mean fields and the covariance structures are treated as unknowns and are inferred jointly with the signal fields in a hierarchical Bayesian fashion.

For the logarithmic density field \(\tau\), we adopt a spatially constant scalar prior mean \(\bar{\tau}\), which represents the overall average DIB density within the modeled volume. This reflects a conservative choice and ensures that spatial structure in the reconstructed density field is entirely data driven. We treat \(\bar{\tau}\) as an additional scalar variable which we jointly infer with all other parameters. See Table \ref{table:params} for a full list of parameters and their respective (hyper-)priors.
Deviations from the mean are governed by the covariance operator \(C_\tau\), which we assume to be statistically homogeneous and isotropic. Under this assumption, the covariance depends only on spatial Euclidean distance,
\begin{equation}
    C_\tau(\mathbf{x}, \mathbf{x}') = C_\tau(|\mathbf{x}-\mathbf{x}'|),
\end{equation}
and is diagonal in Fourier space and fully described by a respective power spectrum $P_\tau(|\mathbf{k}|)$ for Fourier space coordinates $\mathbf{k}$. As for the mean, we treat this power spectrum as unknown and infer it from the data together with the density field (see \S\ref{sec:cf} for details on the spectral prior model).

The prior mean \(\bar{v}(\mathbf{x})\) of the line-of-sight velocity field is chosen to encode the dominant large-scale Galactic kinematics projected along the line of sight. Specifically, we adopt a flat Galactic rotation curve,
\begin{equation}
    \bar{v}(\mathbf{x}) = v_0\,\left[\hat{\boldsymbol{\phi}}(\mathbf{x}) - \hat{\boldsymbol{\phi}}_\odot\right] \cdot \hat{\boldsymbol{n}}(\mathbf{x}),
\end{equation}
where \(v_0\) denotes the circular velocity, \(\hat{\boldsymbol{\phi}}\) denotes the azimuthal unit vector in Galactocentric cylindrical coordinates, \(\hat{\boldsymbol{\phi}}_\odot\) its value at the position of the Sun, and \(\hat{\boldsymbol{n}}(\mathbf{x})\) is the line-of-sight direction from the observer to position \(\mathbf{x}\). Subtracting the circular motion at the Sun's position refers the rotation curve to the local standard of rest, the frame of the data. As for \(\bar{\tau}\), we treat \(v_0\) as an unknown parameter and jointly infer it with the rest.
This choice captures the leading-order contribution of Galactic rotation to the observed absorption central velocities and allows the GP prior to focus on spatially correlated deviations from circular rotation.
The covariance \(C_v\) describes velocity fluctuations around this mean flow. As for the density field, we assume statistical homogeneity and isotropy, and model it in terms of the power spectrum in Fourier space. No explicit coupling between the density and velocity fields is imposed at the prior level.

Both power spectra, $P_\tau$ and $P_v$, are inferred from the data together with the fields (\S\ref{sec:cf}), so the smoothness and fluctuation amplitude of the reconstruction are learned rather than imposed.

\subsubsection{(Sparse) iterative charted refinement}
\label{sec:icr}
To utilize the prior models in practice, we require a computationally scalable way of generating and inferring the values of GPs that cover the $12$~kpc diameter sphere while maintaining high and very irregular resolution across the volume due to survey selection and radially increasing distance uncertainties.
Following and extending upon \citet{2024A&A...685A..82E}, we implement a variant of Iterative Charted Refinement (ICR), a flexible and scalable way of defining and generating GPs on arbitrary grids. It utilizes a multi-resolution pixelation of space and proceeds to approximate a GP by first generating values at coarse resolution and then generating values at higher resolution conditional on those coarse values, restricting conditioning to a local neighborhood at the previous level. By repeating the refinement procedure over multiple levels, we obtain an exponential increase in resolution per level, taking into account both short- and long-range correlations while restricting long-range correlations to large spatial scales, with the latter being the main axis of approximation of ICR. 
ICR has already been applied in the context of 3D tomography of the Milky Way, specifically for 3D dust mapping \citep{2024A&A...685A..82E}, as well as kinetic tomography using line-emission data \citep{2025A&A...693A.139S}. Following these prior works, we choose to pixelate 3D space using a spherical coordinate system and use the HEALPix \citep{Gorski2005} pixelation at an $N_{\rm side}$ of $128$ for angular space together with a hybrid linear-logarithmic distance binning for the radial axis to balance resolution in the nearby volume with computational tractability at large distances (see \citet{2025A&A...693A.139S} for details on the radial binning). We use linear distance binning with $\binLin{}\,\mathrm{pc}$ radial bins out to $\rLin{}\,\mathrm{pc}$. Beyond this distance, the binning transitions to logarithmic spacing, with bin widths increasing exponentially such that the outermost radial bins reach a maximum size of $\binMax{}\,\mathrm{pc}$ at the $6$~kpc edge (Appendix \ref{ap:grid}). This choice reflects the rapidly increasing distance uncertainties and decreasing effective spatial resolution at large heliocentric distances, while maintaining a manageable number of degrees of freedom for GP inference across the full volume.

In this work, we extend the ICR framework in two key directions. First, we introduce a sparse variant of ICR, in which refinement is performed only in regions of the angular–radial grid that are supported by data. Rather than refining the full HEALPix grid uniformly at each level, we selectively refine only those pixels that intersect the survey footprint, resulting in a sparse, data-adaptive multi-resolution representation of the GP. Figure \ref{fig:allsky_data_plot} displays the grid implied by the footprint. This substantially reduces the number of active degrees of freedom at the highest resolutions, while preserving the ability of ICR to propagate correlations across the full volume. Note that due to the hierarchical nature of ICR, this allows up-sampling to a full, dense grid after the reconstruction is fully converged, while being mathematically equivalent to a fit performed on a dense grid from the start, which leads to an overall better performance in both memory and runtime of the reconstruction algorithm.

Second, we enable kernel inference within the ICR hierarchy. In contrast to prior applications of ICR, in which the GP covariance kernel was fixed and not jointly inferred from the data, we allow the prior to be described by an arbitrary stationary kernel whose power spectrum is learned directly from the observations. This removes the need to commit to a specific correlation structure in advance and allows the data to inform both the overall shape and characteristic scales of the prior covariance in a non-parametric manner. The inferred kernel remains fully compatible with the hierarchical conditional structure of ICR, allowing kernel learning and multi-resolution refinement to be performed jointly in a scalable fashion.

\subsubsection{The correlated field model}
\label{sec:cf}

To model the unknown spatial correlation structure of the DIB density and velocity fields in a flexible but well-regularized manner, we adopt the \emph{correlated field} framework in the form initially developed in \citet{eht}, which is the version used in this work and the reference for all details of the model. It models power spectra, specifically $P_\tau(k)$ and $P_v(k)$ in a non-parametric fashion. We restrict the discussion to $P_\tau(k)$ in the following, an analogous derivation holds for $P_v(k)$.

The correlated field model places a hierarchical, non-parametric prior directly on the logarithm of $P_\tau(k)$ as smooth functions of $\log k$. Specifically, each spectrum is parameterized as a power law plus smooth deviations,
\begin{equation}
    \log P_\tau(k) = \alpha_\tau + \beta_\tau \log k + \sigma_\tau\, \psi_\tau(\log k) \ .
\end{equation}
Here $\alpha_\tau$ controls the overall amplitude, $\beta_\tau$ the spectral slope, and $\psi_\tau(\log k)$ is a zero-mean latent Gaussian field capturing deviations from pure power-law behavior. Smoothness of $\psi_\tau$ is enforced by an integrated Wiener process prior in $\log k$: $\psi_\tau$ is the twice-integrated white-noise process, whose covariance operator $C_{\psi_\tau}$ penalizes rapid variations in $\log k$ and whose amplitude is the \emph{flexibility} hyperparameter of Table \ref{table:params},
\begin{equation}
    \mathcal{P}(\psi_\tau)
    =
    \mathcal{N}\!\left(0,\, C_{\psi_\tau}\right) \ .
\end{equation}

All spectral degrees of freedom, including the latent deviation fields and their amplitudes, are inferred jointly with the signal fields within one global posterior. It enables the reconstruction to learn characteristic correlation scales and relative power across spatial scales directly from the data, and use those as weakly informative and robust regularization in regions of sparse observational coverage across scales.

The combination of a non-linear line-of-sight likelihood, hierarchical correlated-field priors, and joint inference of two fields results in a highly non-Gaussian posterior over millions of coupled degrees of freedom. In particular, line-of-sight integration, heterogeneous distance uncertainties, and kernel inference induce strong correlations between spatial locations and model parameters that preclude simple analytic treatments. To perform scalable Bayesian inference in this setting while retaining calibrated uncertainty estimates, we adopt the Geometric Variational Inference (geoVI; \citealt{geovi}) framework. It provides a scalable non-Gaussian approximation to Bayesian inference within an expressive variational family that allows for well-calibrated uncertainty estimates.

\subsection{Geometric Variational Inference}
\label{sec:gvi}
In variational inference, the true posterior $\mathcal{P}(\xi|D)$ over all latent variables $\xi$ is approximated by a tractable distribution $\mathcal{Q}(\xi)$ by minimizing the Kullback--Leibler (KL) divergence
\begin{equation}
    \mathrm{KL}\!\left[\mathcal{Q}(\xi)\,\|\,\mathcal{P}(\xi|D)\right]
    =
    \int \mathcal{Q}(\xi)\,\log\frac{\mathcal{Q}(\xi)}{\mathcal{P}(\xi|D)}\,\mathrm{d}\xi ,
\end{equation}
which is equivalent to maximizing the evidence lower bound (ELBO),
\begin{equation}
    \mathcal{L}_{\mathrm{ELBO}}
    =
    \big\langle \log \mathcal{P}(D|\xi) \big\rangle_{q}
    -
    \mathrm{KL}\!\left[\mathcal{Q}(\xi)\,\|\,\mathcal{P}(\xi)\right] \ .
\end{equation}
Here $\xi$ denotes all variables in the global inference problem re-parametrized such that they follow a standard normal distribution $\mathcal{P}(\xi) = \mathcal{N}\left(\xi; 0, \mathbf{I}\right)$ a priori.

The geoVI method constructs a variational family for $\xi$ by exploiting the Riemannian geometry induced by a positive definite posterior metric of the form
\begin{equation}\label{eq:metric}
    M(\xi)\;\equiv\; M_{\rm lh}(\xi) + \mathbf{I},
\end{equation}
where $M_{\rm lh}$ is the Fisher information metric of the likelihood and $\mathbf{I}$ is the curvature of the prior. This choice yields a symmetric positive-definite metric for all $\xi$, and thereby equips the parameter space with a Riemannian manifold structure.
The method then constructs (locally) a set of Riemann coordinates $y$ defined via the invertible transformation $y=g(\xi; \bar{\xi})$. In these coordinates the metric becomes approximately Euclidean in a neighborhood of an expansion point $\bar{\xi}$. Explicitly \citep{geovi},
\begin{align}\label{eq:geovi_g}
    g(\xi; \bar{\xi}) &= \bar{M}^{-1/2}\left[\left(\xi - \bar{\xi}\right) + \bar{J}^{\rm T}\left(x(\xi) - x(\bar{\xi})\right)\right] , \\
    \bar{M} &\equiv M(\bar{\xi}) , \qquad \bar{J} \equiv \left.\frac{\partial x}{\partial \xi}\right|_{\bar{\xi}} , \nonumber
\end{align}
where $x(\xi)$ is a standardized likelihood variable: a function of the latent parameters for which the Fisher metric of the likelihood is the pullback of the Euclidean metric, $M_{\rm lh} = (\partial x/\partial\xi)^{\rm T}(\partial x/\partial\xi)$. For example for a Gaussian likelihood $x$ would be the noise-weighted residual. For the mixture likelihood of \S\ref{sec:lowsnr} we use an approximation to its Fisher metric: the recalibrated Student-$t$ metric, with its Fisher factor $(\nu+1)/(\nu+3)$, weighted by the detection probability evaluated at the measured rather than the model equivalent width, with the line-of-sight response combined over the distance-bin weights of each star. The transformed distance $g(\xi;\bar{\xi}) - g(\bar{\xi};\bar{\xi})$ thus combines the distance in the prior measure, $\xi - \bar{\xi}$, with the distance in the likelihood measure, $x(\xi) - x(\bar{\xi})$, back-projected into the latent space by the Jacobian at the expansion point and scaled by the local curvature $\bar{M}$; by construction its Jacobian reproduces the metric there, $(\partial g/\partial\xi)^{\rm T}(\partial g/\partial\xi)|_{\bar{\xi}} = \bar{M}$. In these $y$-coordinates, the transformed posterior becomes geometrically simple, and geoVI approximates it by a unit Gaussian,
\begin{equation}
    \mathcal{Q}_y(y) = \mathcal{N}(y;0,\mathbf{I}).
\end{equation}
The induced approximation in the original coordinates is therefore the pushforward
\begin{equation}
    \mathcal{Q} = g^{-1} \bullet \mathcal{Q}_y,
\end{equation}
where $g^{-1} \bullet \mathcal{Q}_y$ denotes the distribution of $g^{-1}(y)$ for $y$ drawn from $\mathcal{Q}_y$. It is generally \emph{non-Gaussian} in $\xi$ because $g^{-1}$ is non-linear.

Optimization of the KL divergence under this family then reduces to finding the optimal expansion point $\bar{\xi}$ which we do using second order natural gradient descent with the metric \eqref{eq:metric} serving as the curvature proxy. Numerical optimization with geoVI is performed using the NIFTy software package, in its JAX-based re-implementation NIFTy.re \citep{nifty}.\footnote{NIFTy.re version 9.1.0 (development commit \texttt{1db48360} of 2025 December 8, 132 commits past the 9.1.0 release) on JAX and jaxlib 0.8.1.}

\section{Results and discussion}
\label{sec:results}
\begin{figure*}[t!]
    \centering
    \includegraphics[width=0.49\linewidth]{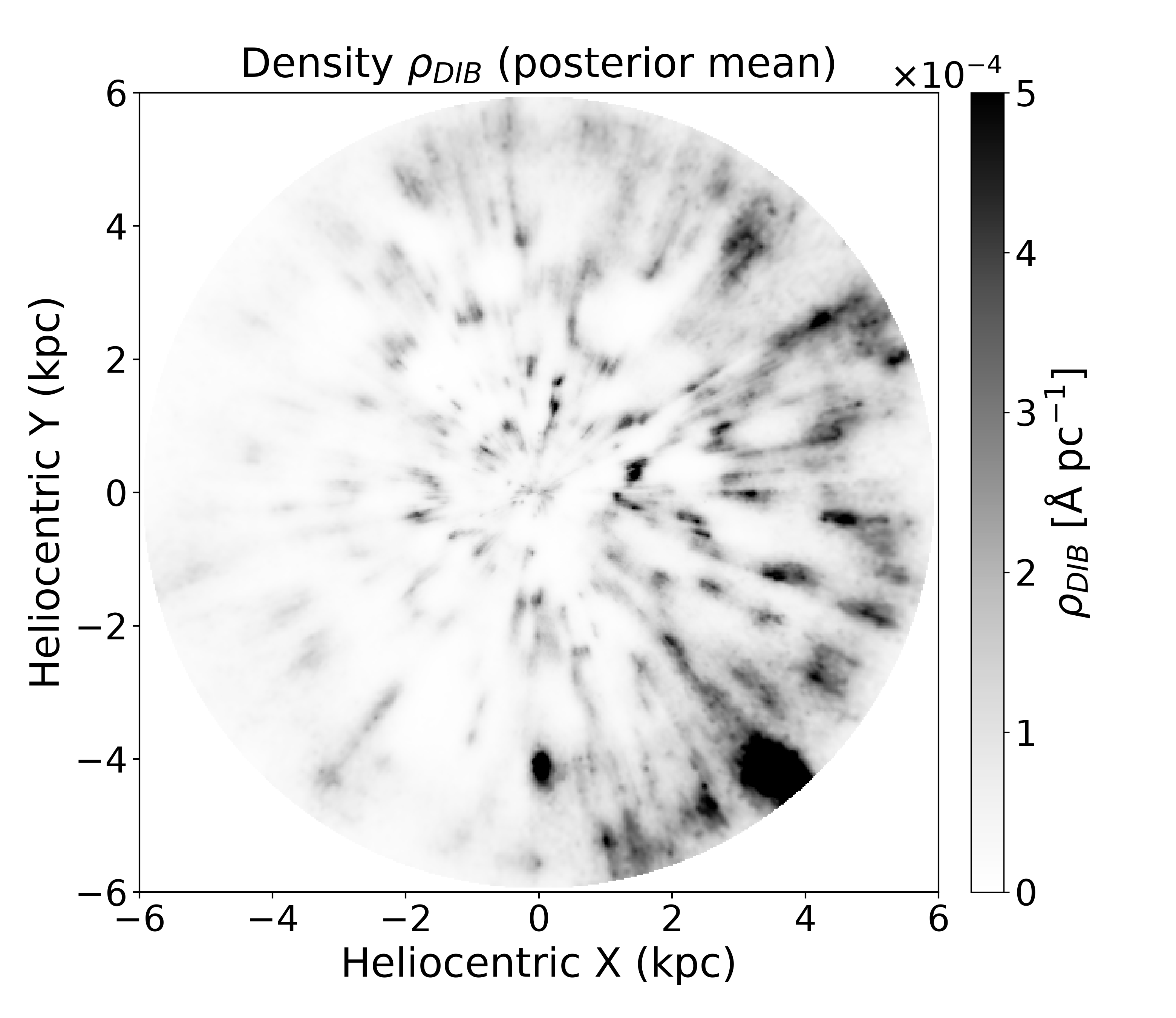}
    \includegraphics[width=0.49\linewidth]{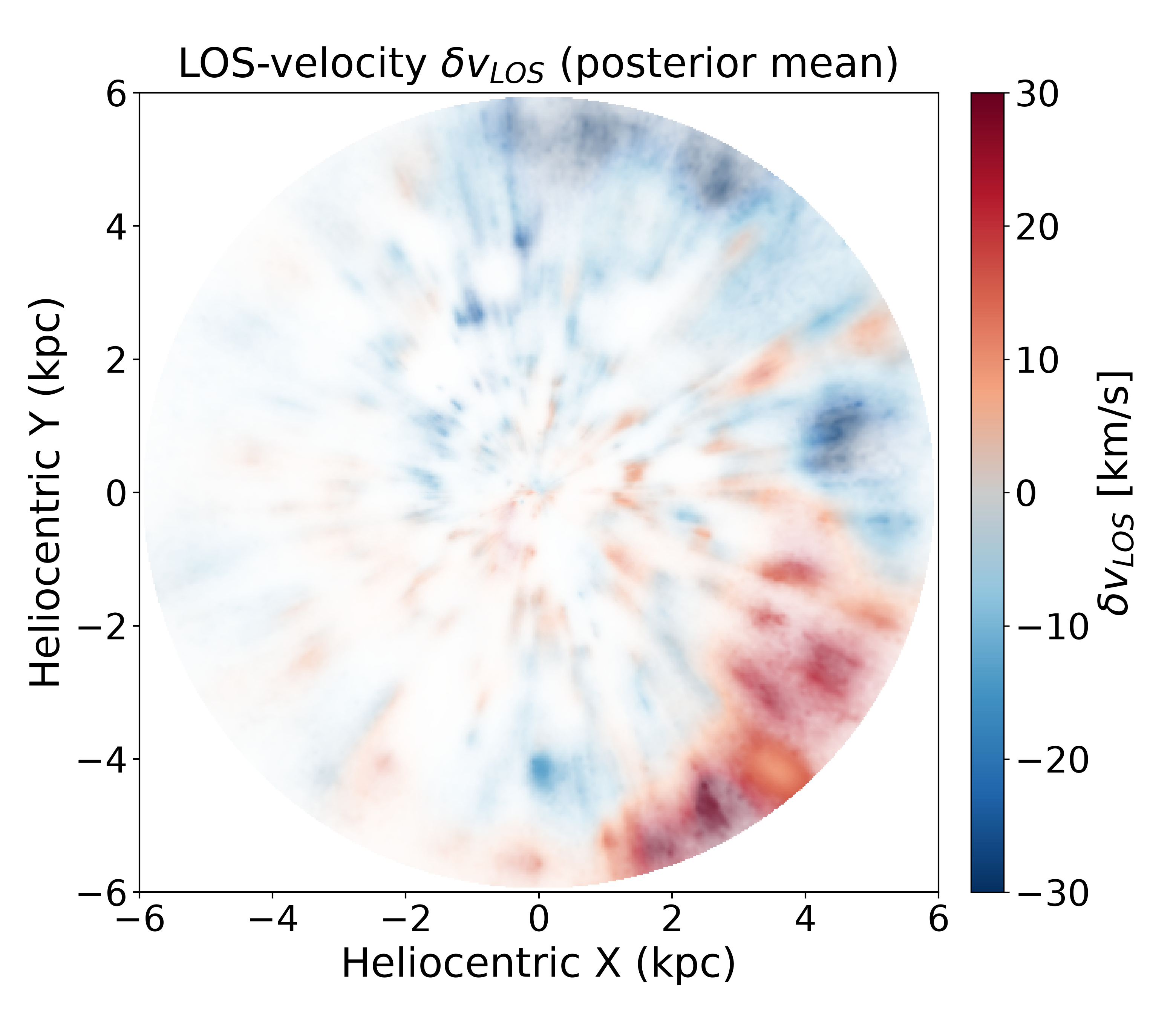}\\
    \includegraphics[width=0.49\linewidth]{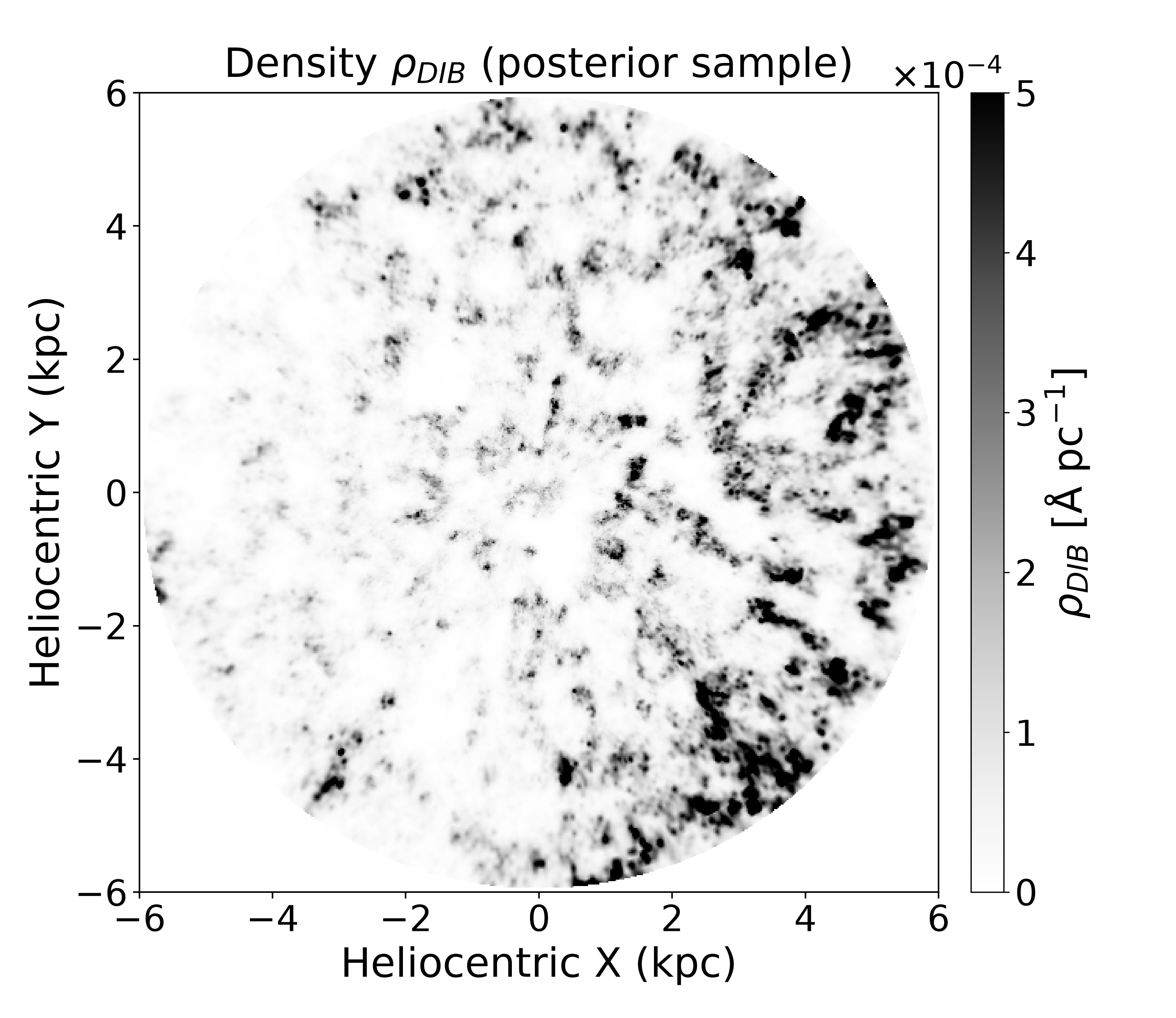}
    \includegraphics[width=0.49\linewidth]{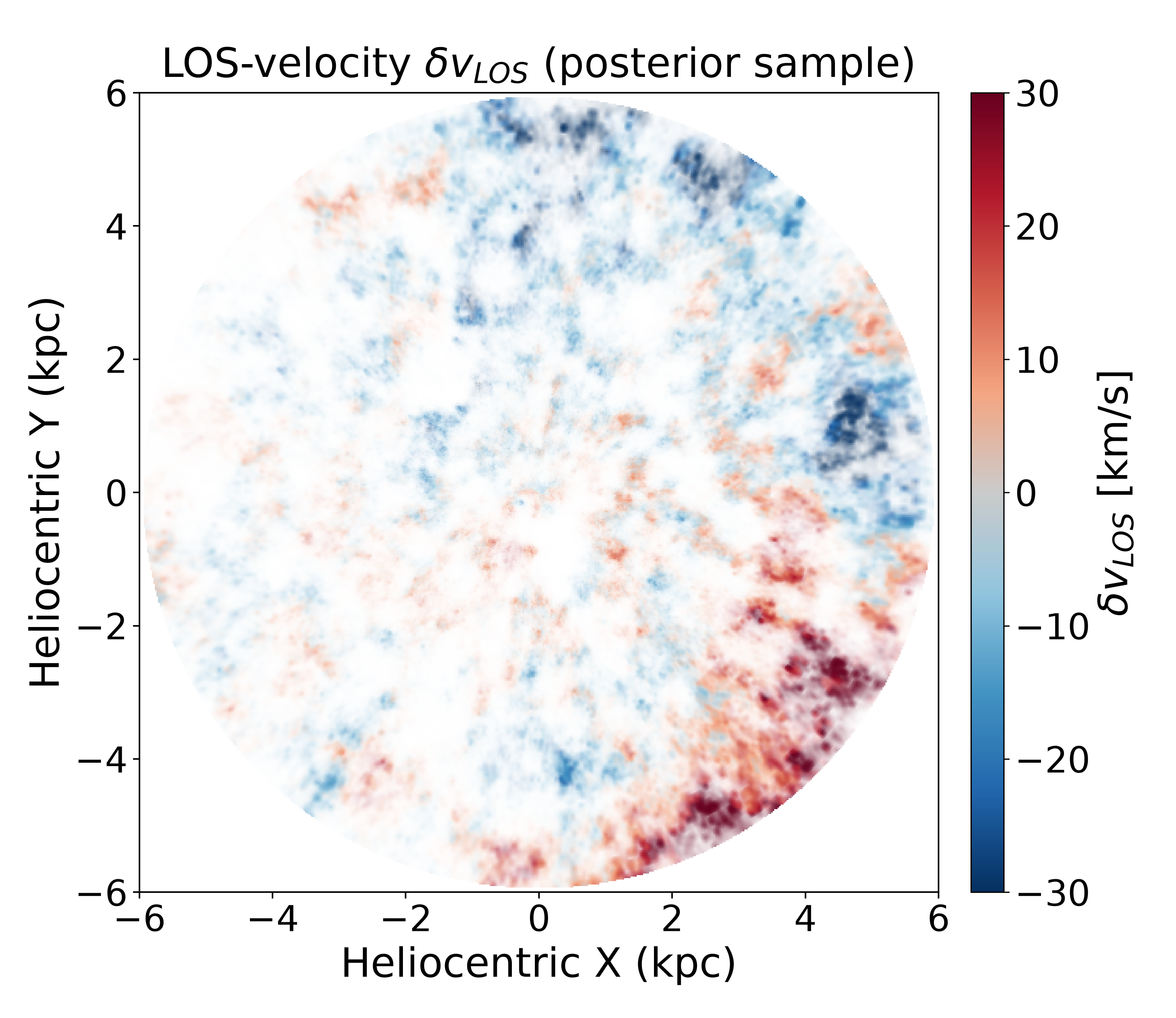}
    \caption{The two joint products of the reconstruction, shown as a $260$~pc thick slice through the Galactic plane centered on the Sun. \emph{Left column:} the DIB density $\rho_{\rm DIB}$. \emph{Right column:} the differential line-of-sight velocity field $\delta v_{\rm LOS}$ in the \citet{Reid2019} solar frame, weighted by DIB density. \emph{Top row:} the posterior mean of both fields. \emph{Bottom row:} a single sample of the joint posterior on DIB density and differential line-of-sight velocity. An animated version of this figure showing all posterior samples is available at \url{https://www.ph-frank.de/resources/dib_release_18_11_26/full_density_velocity_samples_white.html}.
}
    \label{fig:density_velocity}
\end{figure*}

The reconstruction returns two three-dimensional fields on the same grid: the DIB density $\rho \equiv \rho_{\rm DIB}$ and the line-of-sight velocity $v \equiv v_{\rm LOS}$ of the DIB-bearing medium, with their joint posterior distribution represented as a set of \Nsamp{} geoVI approximate posterior samples. Figure \ref{fig:density_velocity} shows both in the Galactic plane, as the posterior mean and as a single sample. The difference between the two is a first impression of the posterior uncertainty. The velocity is shown in terms of residual velocity $\delta v_{\rm LOS}$ after subtraction of the flat rotation curve of \citet{Reid2019} ($\Theta_0 = 236$~\kms, $R_0 = \Rzero{}$~kpc) and is transparency weighted by DIB density. These fields, their posterior summaries and the samples are released as the data products of this paper (see the Data availability section).

The rest of this section presents the two fields in turn. In \S\ref{sec:velocity} we compare the velocity field with independent kinematic tracers: masers and young clusters, the $\ell$--$v$ diagrams of CO and \ion{H}{I}, and other published kinetic tomography maps. Then we use it to quantify the kinematic decoupling of young clusters from their natal medium. In \S\ref{sec:density} we do the same for the density, comparing to three-dimensional photometric dust maps, and then measure the recovered large-scale Galactic structure of the disk. In \S\ref{sec:spectra} we turn to the small scales: the power spectra of both fields against their priors and against either dust maps or theoretical velocity structure functions. Finally, \S\ref{sec:mock} summarizes what reconstruction of a mock catalog says about the fidelity of our results, the details of which are described in Appendix \ref{ap:mock}.

\subsection{Velocity}
\label{sec:velocity}

\begin{figure*}
    \centering
    \includegraphics[width=0.65\linewidth]{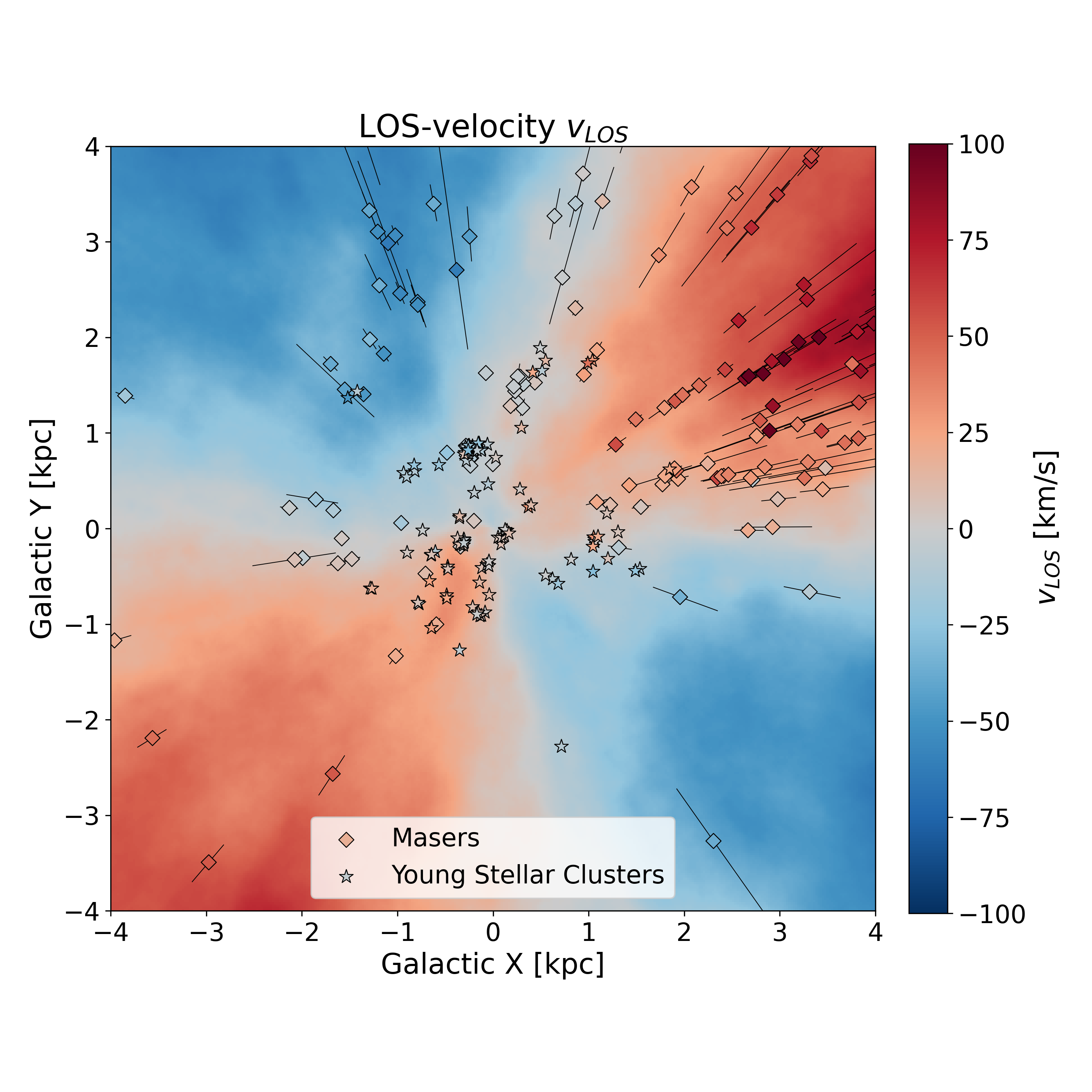}
    \includegraphics[width=0.3\linewidth]{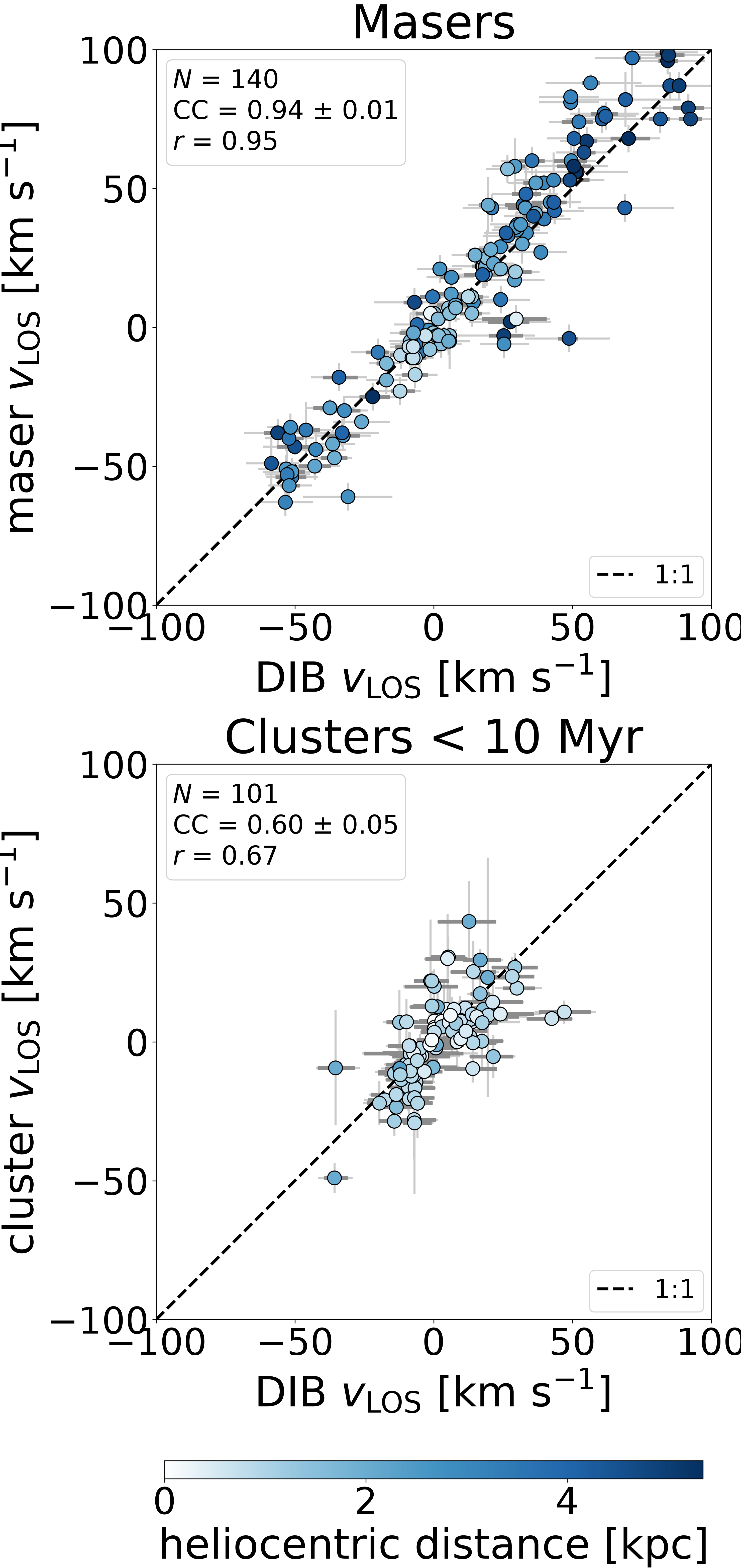}
    \caption{Comparison of the reconstructed DIB line-of-sight velocity field with independent kinematic tracers. \emph{Left:} the posterior mean $v_{\rm LOS}$ field in a $260$~pc thick slice centered on the Sun, with maser positions \citep[diamonds;][]{Reid2019} and young ($<10$~Myr) stellar clusters \citep[stars;][]{Hunt} overplotted and color-coded by their own line-of-sight velocity. \emph{Right:}  DIB velocity at the tracer position for the masers (top) and young clusters (bottom), colored by heliocentric distance. Each panel summarizes the number of tracers $N$ and the raw Pearson correlation coefficient calculated from the samples, $\mathrm{CC}$, as well as the plain coefficient of the sample means, $r$.
}\label{fig:v_clusters}
\end{figure*}

\begin{figure*}
    \centering
    \includegraphics[width=0.65\linewidth]{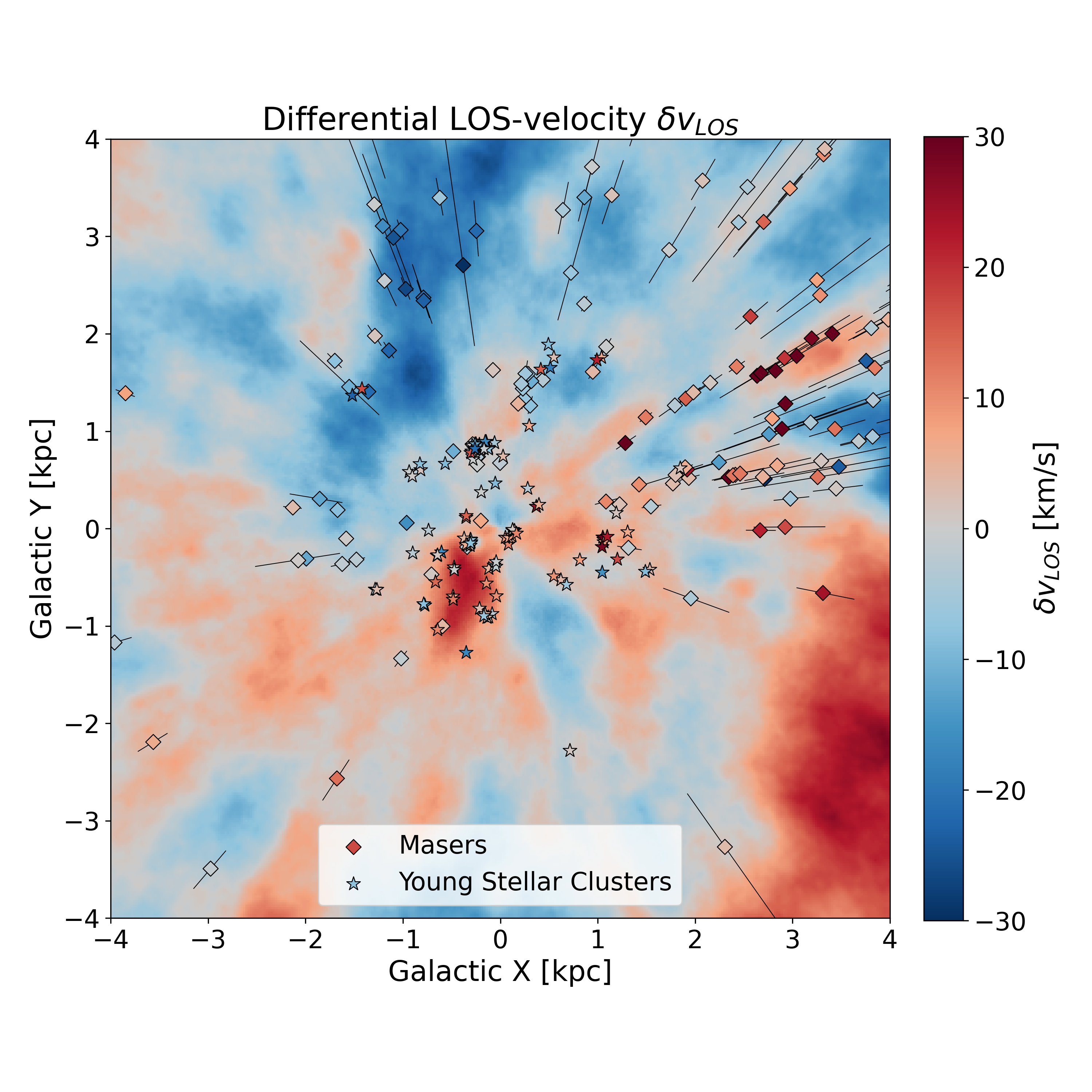}
    \includegraphics[width=0.3\linewidth]{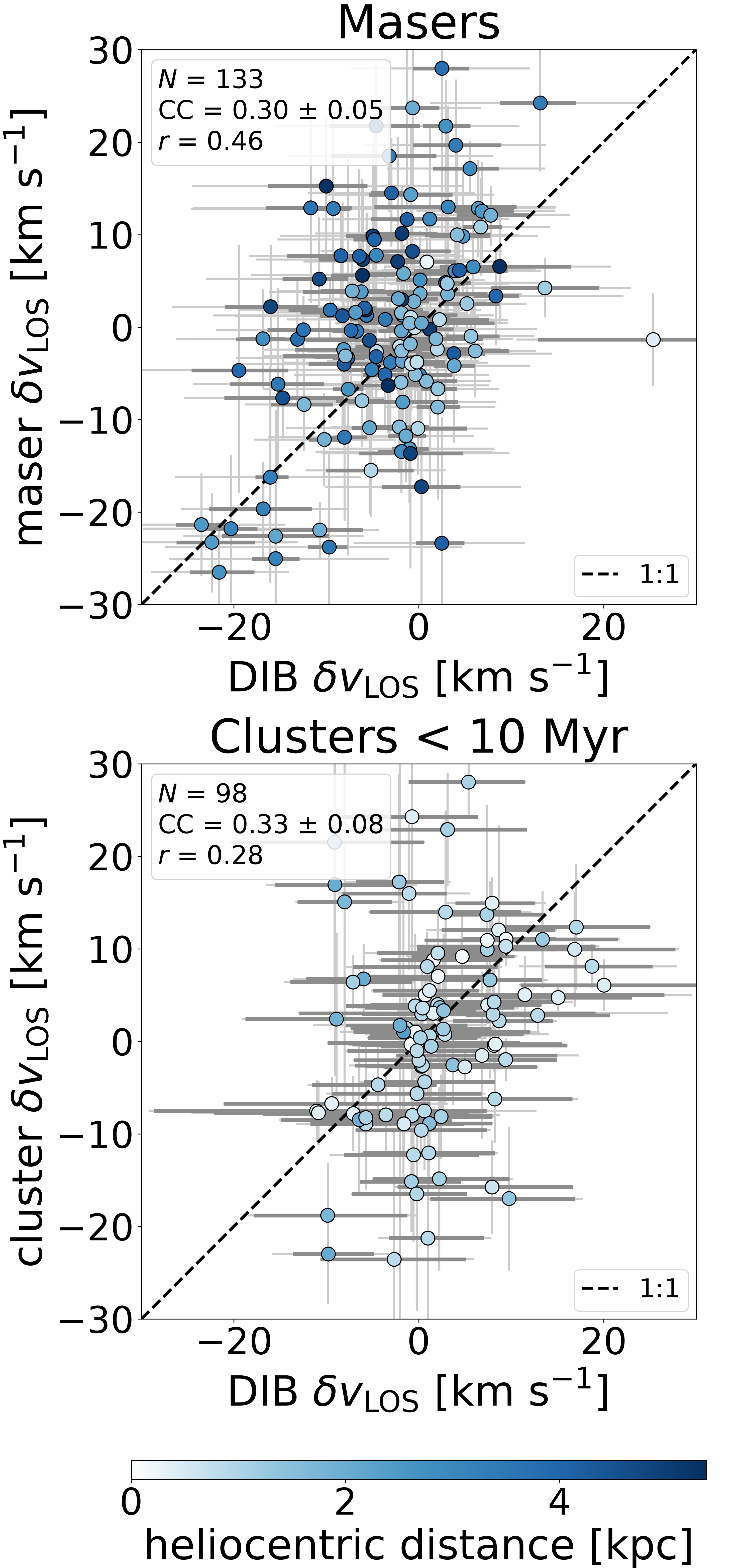}
    \caption{Same as in Figure \ref{fig:v_clusters}, but for the differential velocity $\delta v_{\rm LOS}$ in the \citet{Reid2019} frame.
}
\label{fig:v_residual_clusters}
\end{figure*}

\subsubsection{Maser and young stellar cluster comparisons}
\label{sec:maser_yso}
 
Independent probes of the line-of-sight velocity field offer the most direct external validation of our reconstruction. We compare against two probes with both kinematic and distance information: high-mass star-forming regions (masers) with radial velocities and VLBI parallaxes compiled by \citet{Reid2019}, and the young ($<10$~Myr) stellar clusters with \textit{Gaia}-based 6D phase-space measurements from \citet{Hunt}. Because of their youth, both populations should still share the kinematics of the interstellar medium from which they formed. Thus, these are ideal kinematic tracers to compare with our DIB velocity fields, which we expect to trace the diffuse ISM.
 
Figure \ref{fig:v_clusters} shows the comparison with these probes for the total line-of-sight velocity $v_{\rm LOS}$. The DIB velocity evaluated with respect to a given tracer (the maser or young stellar cluster) is the posterior field averaged along the tracer's distance uncertainty. For each posterior sample we compute the Pearson correlation between the tracers' measured velocities and that sample's DIB velocities. We report the  average of this correlation over the samples and use the spread between samples as our reported uncertainty (Appendix \ref{ap:correlation}). For simplicity, we make no attempt here to correct this measured correlation for the noise in either the DIB reconstruction or the tracer catalogs. Our measured correlation therefore underestimates the true correlation of the underlying fields, since the radial velocity errors and the posterior scatter of the reconstruction impact the correlation measurement.

When comparing velocities that include Galactic rotation, the masers track the DIB field exceptionally well, with $\mathrm{CC} = 0.94 \pm 0.01$ and a tight, unbiased scatter about the one-to-one line across the full $\pm 100$~\kms\ dynamic range. The young stellar clusters, whose velocities are concentrated within $\sim\!25$~\kms\ of the local standard of rest, correlate more weakly, at $0.60 \pm 0.05$. The contrast is mainly due to a dynamic range effect, because the maser sample spans a wide range of Galactocentric azimuth and distance and therefore samples a large-amplitude signal from bulk Galactic rotation. In both cases, the agreement confirms that the bulk of the recovered DIB velocity signal reflects real kinematics of the intervening medium rather than artifacts of the inference.
 
Because raw kinematics are dominated by the large-scale signature of Galactic rotation, the raw correlation in Figure \ref{fig:v_clusters} alone is not a good test of our ability to detect the spatially resolved non-circular velocity structure. To isolate the non-circular motions we subtract a flat rotation curve from both the DIB field and the tracers, using the \citet{Reid2019} rotation curve ($\Theta_0 = 236$~\kms, $R_0 = \Rzero{}$~kpc), and compare the residual differential velocities $\delta v_{\rm LOS}$ in Figure \ref{fig:v_residual_clusters}. As expected, removing rotation substantially reduces the correlation for both the masers and young clusters to $0.30 \pm 0.05$ and $0.33 \pm 0.08$ respectively.
That the two populations have similar correlation amplitudes once rotation is removed suggests that the DIB field recovers genuine, spatially coherent departures from circular rotation at the $\sim\!10$~\kms\ level. De-correlation occurs partially due to uncertainties in the tracers and the $\delta v$ field, but also likely due to astrophysical differences in the $\delta v$ fields probed by these three tracers. Further disentangling these contributions and modeling the true underlying $\delta v$ correlation is beyond the scope of this work. Note that correlating the tracers with the posterior-mean DIB velocity instead of with the individual samples gives $r = \rMaserRes{}$ for the masers and $\rClusterRes{}$ for the clusters (a centered, unweighted coefficient; Appendix \ref{ap:correlation}).
 
\subsubsection{Longitude--velocity diagram comparison}
\label{sec:gas_lv}

\begin{figure*}
    \centering
    \includegraphics[width=\linewidth]{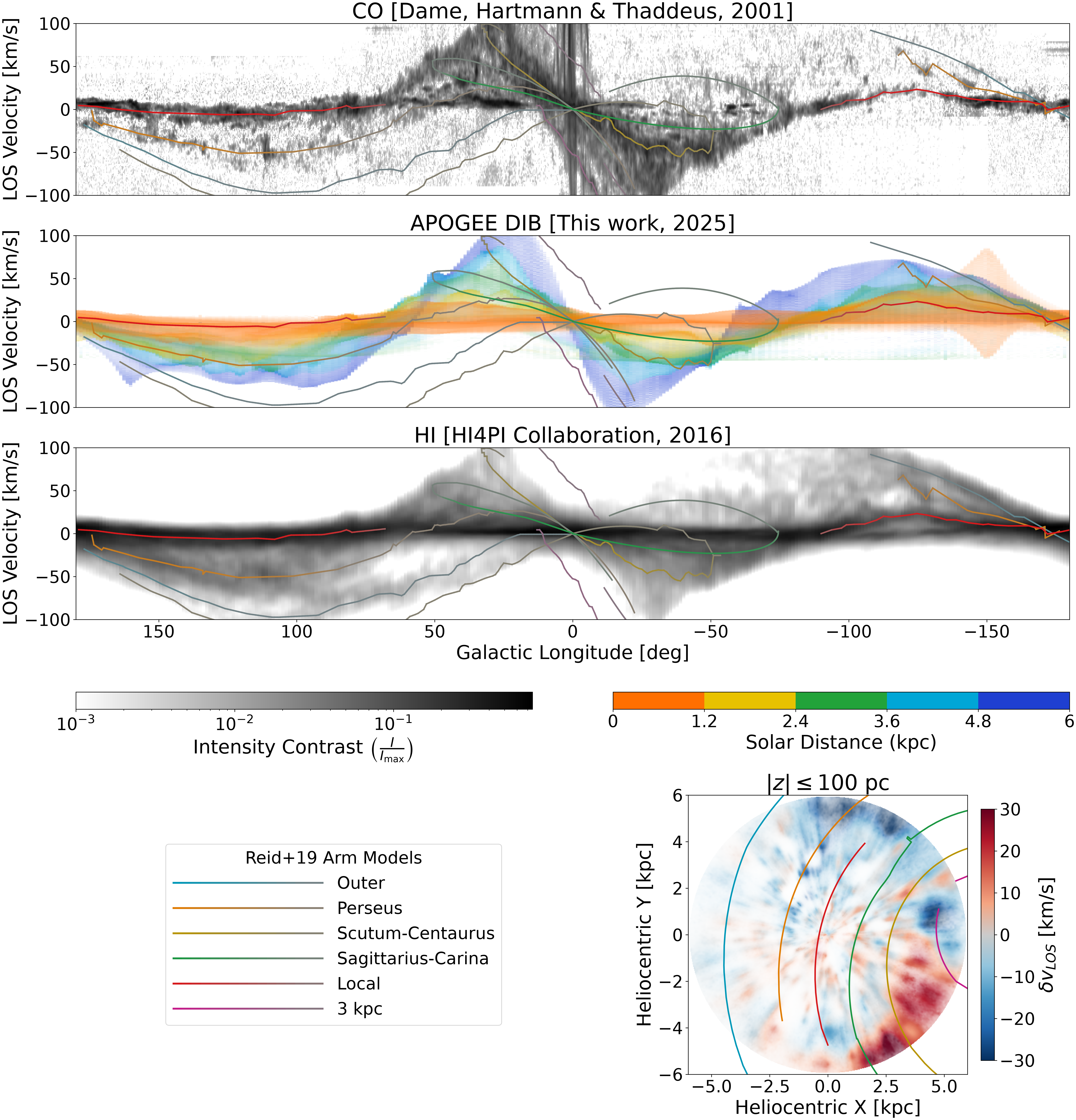}
    \caption{Longitude--velocity ($\ell$--$v$) comparison. \emph{Top three panels:} CO emission \citep{Dame2001}, the DIB absorption (colored by distance), and \ion{H}{I} emission \citep{HI4PI2016}, with the spiral arm models of \citet{Reid2019} overlaid. The colors of the arm models fade to gray once they leave the reconstructed DIB volume to aid visual comparison.
    \emph{Bottom right:} a top down view of the posterior mean density-weighted differential line-of-sight velocity $\delta v_{\rm LOS}$ in the Galactic plane ($|z| \leq 100$~pc), with the same arm models overlaid in X-Y space.}
\label{fig:vl}
\end{figure*}

While gas emission surveys lack distance information, the longitude--velocity ($\ell$--$v$) diagram encodes the integrated kinematics of the gas along each line of sight. We can take our 4D map, which does contain information about distance, and create a corresponding density-weighted $\ell$--$v$ diagram. In Figure \ref{fig:vl}, we compare the $\ell$--$v$ from our DIB kinetic tomography map (middle), the CO survey of \citet[top,][]{Dame2001}, and the \ion{H}{I} survey of \citet[bottom,][]{HI4PI2016}. This provides an independent kinematic check of the DIB velocity, but also an opportunity to disentangle the features coming from different distances within the $\ell$--$v$ diagram, for those that lie within our reconstruction volume.
 
In the middle panel, we color-coded the DIB absorption by the heliocentric distance of the average DIB density contributing to a given pixel. The sub-panel at bottom right shows our reconstructed DIB density in the Galactic plane relative to the spiral arm models of \citet{Reid2019}. The DIB $\ell$--$v$ track reproduces the dominant large-scale sinusoidal signature of Galactic rotation seen in both CO and \ion{H}{I}. However, because the DIB track carries distance information through the color coding, it breaks the velocity--distance degeneracy intrinsic to traditional $\ell$--$v$ diagrams. Features that overlap in velocity in the CO and \ion{H}{I} panels are cleanly separated in distance in the DIB panel, directly demonstrating the added value of kinetic tomography over PPV surveys alone.
 
For visual comparison, we overlay the spiral arm models of \citet{Reid2019} on every panel of Figure \ref{fig:vl}. The arm traces align with some regions of enhanced DIB density with the Local, Sagittarius--Carina, and Perseus arms identifiable as coherent distance-resolved features in the DIB $\ell$--$v$ diagram. More interesting are regions that deviate from the spiral arm models in multiple tracers (CO, \ion{H}{I}, DIB). For example, the wiggle away from the Sagittarius--Carina arm near $\ell \approx-40\degr$ is seen in both the DIBs and CO. With the DIB map, we can localize the source of this feature to be $\sim2$~kpc away. Another interesting feature of Figure \ref{fig:vl} are the substructures we see in the DIB map that overlap with or have previously been assigned to spiral structures at distances beyond the edge of our map (grayed out portions of the arm models). For example, we place the over-density at $\ell \approx 30\degr$ and $v = 75$~\kms\ at $\sim5$~kpc and not at the distance of the \citet{Reid2019} outer arm model. In fact, while we in general agree with the kinematic substructure previously assigned to spiral arms in the $\ell$--$v$ diagram, the bottom right panel does not show coherent spiral arm-like over-densities. This is observational evidence that while grand-design spiral arms are one possible 4D geometry that can explain the observed $\ell$--$v$ diagram, there are other 4D geometries that explain this projection and we likely live in one of those, such as the reconstruction of this work.

\subsubsection{Comparison with existing kinetic tomography maps}
\label{sec:kt_comparison}

\begin{figure*}
    \centering
    \includegraphics[width=\linewidth]{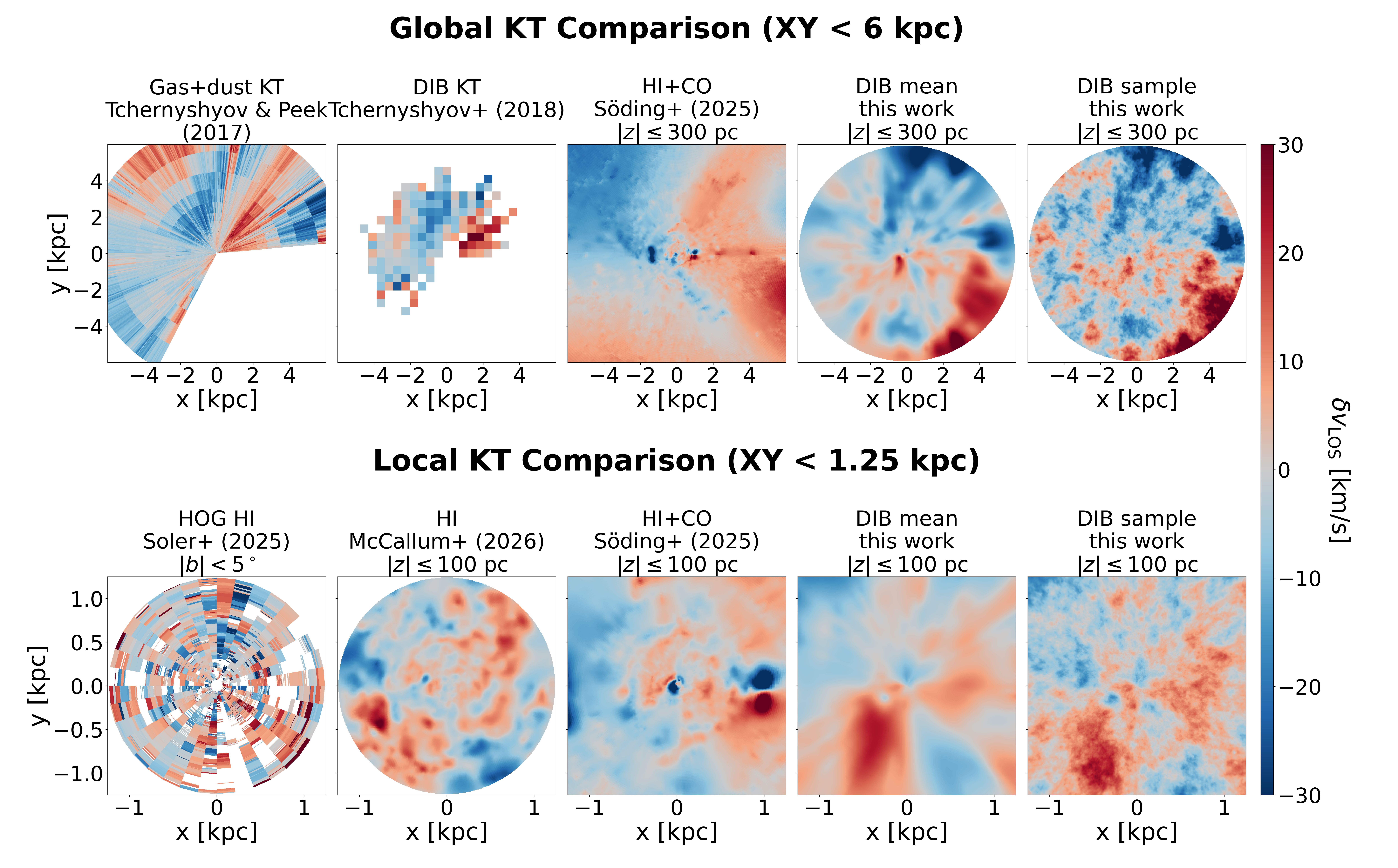}
    \caption{Comparison with published kinetic tomography maps in the literature. \emph{Top row, $\pm 6$~kpc:} the gas${+}$dust reconstruction of \citet{Tchernyshyov_2017}, the DIB reconstruction of \citet{Tchernyshyov_2018}, the \ion{H}{I}${+}$CO reconstruction of \citet{2025A&A...693A.139S}, and the posterior mean and a single posterior sample of this work. \emph{Bottom row, local $\pm \rLocal{}$~kpc:} the HOG \ion{H}{I} map of \citet{Soler_2025}, the \ion{H}{I} map of \citet{McCallum2026_HI}, \citet{2025A&A...693A.139S}, and again the mean and a sample of this work. The top row averages over $|z| \leq 300$~pc and the bottom row $|z| \leq 100$~pc. An animated version of this figure showing all posterior samples is available at \url{https://www.ph-frank.de/resources/dib_release_18_11_26/kt_comparison_samples.html}.}
\label{fig:kt_comparison}
\end{figure*}

We compare our reconstruction with previous kinetic tomography maps of the Galactic plane. Because each published map differs in tracer, resolution, volume, and reference frame, we transform every map into the same \citet{Reid2019} solar frame before comparing (see the details on individual maps in Appendix \ref{ap:external_maps}). Figure \ref{fig:kt_comparison} shows these comparisons on two scales. In the top row, we show the large-volume maps within $\pm 6$~kpc. In the bottom row, we show comparisons at the $\pm \rLocal{}$~kpc scale. The top-down 2D views in Figure \ref{fig:kt_comparison} are obtained by symmetrical, density-weighted, $z$-plane slab averages, where we use the DIB density to weight our maps, and external ones by their own density where they are not already 2D -- the gas density proxy of \citet{2025A&A...693A.139S} and the \ion{H}{I} density of \citet{McCallum2026_HI}.

On the $6$~kpc scale, our map is the only one that over a large area contiguously resolves spatial sub-structure. The earlier DIB reconstruction of \citet{Tchernyshyov_2018} covers structures out to $5$~kpc in the northern sky, but without resolving a third dimension along the $z$-axis and with a coarse spatial pixelation. In the gas${+}$dust reconstruction of \citet{Tchernyshyov_2017} each line of sight is fit largely independently and hence displays radial smearing. Our reconstruction resolves spatially coherent velocity structure over a larger volume and the full sky, which is a direct consequence of the joint prior coupling neighboring sightlines and our input data also covering the southern sky.

Across different kinetic tomography maps, we find several kiloparsec-scale features agree well, despite using different tracers or methods. For example, the most conspicuous shared feature is the large coherent region of negative (blue) residual velocity toward the outer Galaxy just past $\ell = 90\degr$. Visually, \citet{Tchernyshyov_2017}, \citet{Tchernyshyov_2018}, and our map share the least common features with the map produced by \citet{2025A&A...693A.139S}, which we attribute to the fact that this map obtains its 3D spatial structure purely from gas emission and information from a few local velocity tracers via masers and YSOs. 

\begin{figure}
    \centering
    \includegraphics[width=\linewidth]{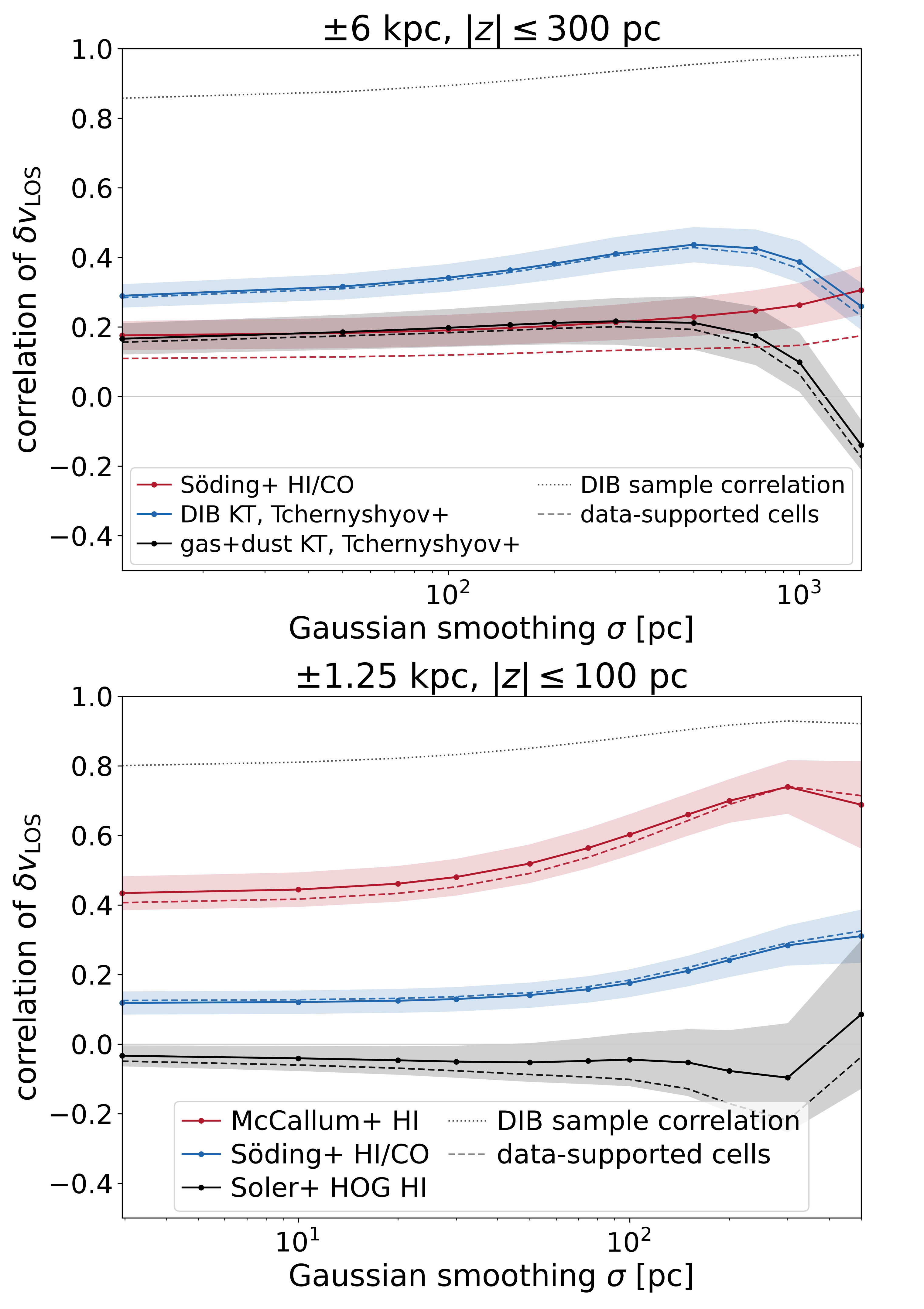}
    \caption{Correlation between our DIB differential velocity and published kinetic tomography maps, computed over $\pm 6$~kpc (top) and $\pm 1.25$~kpc (bottom). Each curve shows the Pearson correlation between our map and one external map (see legend) after smoothing both with a Gaussian kernel of width $\sigma$. Shaded bands show the spread over our posterior samples. Solid lines use all cells covered by both maps, while dashed lines use only cells with a catalog star behind them. The dotted line marks the highest correlation achievable given our own posterior uncertainty. }
\label{fig:kt_agreement}
\end{figure}

We quantify the agreement between the kinetic tomography maps by measuring their correlation with our map as a function of scale in Figure \ref{fig:kt_agreement}. Specifically, we apply a Gaussian smoothing kernel of increasing size to the two-dimensional slab averaged maps from Figure~\ref{fig:kt_comparison}. With the earlier DIB reconstruction of \citet{Tchernyshyov_2018} the correlation rises from 0.29 at native resolution to 0.44 at $\sigma = 500$~pc and then falls at larger scales. The gas${+}$dust map of \citet{Tchernyshyov_2017} has a much lower peak at 0.22 near $300$~pc. For the \citet{2025A&A...693A.139S} map, the correlation rises monotonically from 0.18 to 0.31 at the largest smoothing scale of $1.5$~kpc.

In the local volume, the \ion{H}{I} atlas of \citet{McCallum2026_HI} shows the strongest agreement with our reconstruction, climbing from 0.43 to 0.74 at $\sigma = 300$~pc. For the local part of the \citet{2025A&A...693A.139S} map, the correlation rises from 0.12 to 0.31. The HOG map does not show any correlation with our map at any scale (between $-0.03$ and $-0.10$, with only the largest smoothing scale reaching $+0.09 \pm 0.21$). This may be because the HOG method assigns each $10\degr$ tile a \textit{single} velocity based on morphological matching of \ion{H}{I} channel maps with dust, which can be subject to variability arising from choices in defining morphological matching. For reference, the best-possible correlation between maps given DIB posterior uncertainties would be $0.8 - 0.9$ at native resolution and approaches unity with smoothing (Figure~\ref{fig:kt_agreement}, dotted line).

\subsubsection{How stars leave home}
\label{sec:stars_leave_home}

\begin{figure*}
    \centering
    \includegraphics[width=\linewidth]{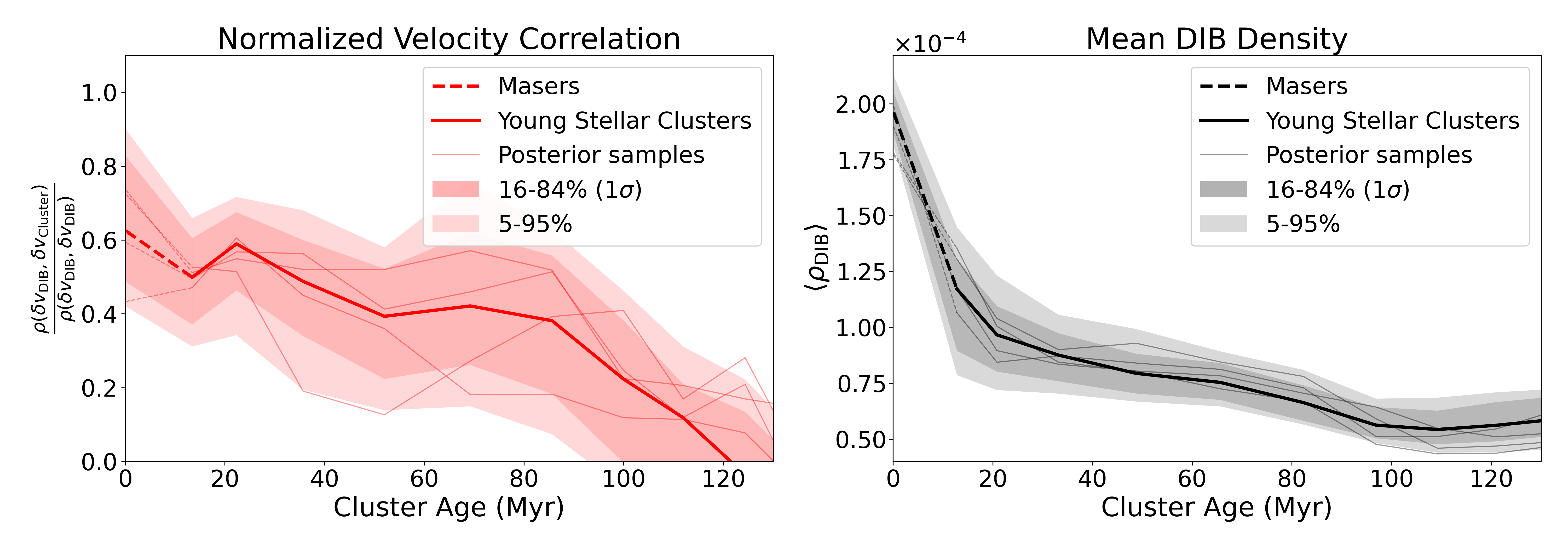}
    \caption{Decoupling of young stellar populations from their birth clouds as a function of age. \textit{Left}: correlation between the residual velocities of young clusters \citep{Hunt} and the DIB velocity field at their positions, normalized so that a tracer moving exactly with the DIB field would sit at unity (Appendix~\ref{ap:correlation}). \textit{Right}: mean DIB density at the cluster positions. Solid lines show the posterior median for the clusters, computed in overlapping rolling windows of 200 clusters. Dashed lines show the \citet{Reid2019} masers, plotted at zero age. Shaded bands are the 16--84\% and 5--95\% posterior intervals, and thin lines are individual posterior samples.}
\label{fig:v_residual_clusters_age} 
\end{figure*}
 
The residual velocity comparison of \S\ref{sec:maser_yso} established that young tracers of star formation on average share the non-circular motions traced by DIBs. A natural interpretation of this correlation is that stars and clusters are kinematically coupled to their natal gas at birth, but that progressive decoupling from that gas as they age might drive an age-dependent decrease in kinematic correlation. We might expect  stellar feedback to disperse the parent cloud or the cluster to drift under the Galactic potential \citep[see e.g. discussion in][]{Lada2003, Krumholz2019}.
 
Figure \ref{fig:v_residual_clusters_age} tests this prediction by measuring how the correlation with tracers evolves as a function of the tracer age. The left panel measures the differential-velocity correlation $\rho(\delta v_{\rm DIB}, \delta v_{\rm Cluster})$ as a function of cluster age, with the raw estimator of \S\ref{sec:maser_yso}. Here, the correlation in each age and tracer window is normalized by what would constitute the correlation coefficient of a perfectly coupled tracer drawn from the DIB velocity field in that window. This includes taking into account the velocity scatter in the DIB posterior samples as well as uncertainties on the tracer distance and velocity measurements (see Appendix \ref{ap:correlation}). A more rigorous forward model treatment of these effects is beyond the scope of this work.

We clearly see the expected age-dependent decline in correlation. The error bands and individual posterior samples of the DIB velocity field confirm that the decline is robust to the posterior uncertainty in the reconstruction and is not an artifact of a single point estimate. The masers, associated with the youngest and most deeply embedded phase of high-mass star formation, reach a normalized correlation of $0.63^{+0.20}_{-0.14}$. The correlation reaches zero ($0.00^{+0.13}_{-0.15}$) at $125$~Myr, at which point the clusters are effectively kinematically decoupled from the present-day diffuse ISM traced by DIBs.

The right panel of Figure \ref{fig:v_residual_clusters_age} instead asks how the correlation of the tracers with ISM density evolves over time. To do this, we evaluate the DIB density at the tracer positions. The density  decline is steep over the first tens of Myr, where the masers and clusters are still embedded in material and flattens beyond $\sim\!100$~Myr at a floor of $0.55 \times 10^{-4} \nicefrac{\text{\AA}}{\mathrm{pc}}$, consistent with the average density of the disk at those positions. This density decline with age is unlikely to be due to a survey selection effect. While \textit{Gaia} is less sensitive to clusters behind large amounts of extinction (e.g. towards dense clouds) that insensitivity applies to clusters of every age and would bias our correlation against the youngest, most deeply embedded clusters. Further, dust obscuration should not significantly impact masers, which are measured in the radio and anchor the curve at the youngest ages and correspond to the highest DIB densities. The decline is therefore most naturally explained by the progressive dispersal of the natal cloud, in conjunction with clusters drifting away from their birth sites.
 
The decorrelation timescales for both the density and the velocity are physically reasonable. We find that the density correlation declines steeply within $\approx 10$--$20$ Myr of a cluster's birth, which is consistent with the dispersal timescale of molecular clouds by stellar feedback measured in simulations \citep[cf. e.g.][]{Kruijssen2019}. Compared to the density correlation, the velocity correlation (left panel of Figure \ref{fig:v_residual_clusters_age}) has a shallower decline over $\approx 60- 80$ Myr from a cluster’s birth, which is broadly consistent with the timescale on which stellar orbits in a Galactic disk are perturbed away from their birth orbits \citep{Arunima2025}. In the simulation of \citet{Arunima2026}, the orbital actions of individual stars undergo significant evolution within $\lesssim 100$ Myr of birth, driven by gravitational encounters with molecular clouds and spiral structure. The kinematic imprint of a star’s natal gas is therefore expected to be erased on roughly the timescale over which we observe the cluster--DIB velocity correlation to vanish. 

In this sense the age correlation relations can be read as a qualitative clock. The rate at which young stellar populations ``forget'' the kinematics of their natal medium directly measures the coherence time of collisionless point sources formed from the diffuse ISM velocity field. A quantitative comparison requires a more careful treatment of uncertainties as well as selection effects present in both our maps and the tracers in order to obtain a calibrated absolute rate.

\subsection{Density}
\label{sec:density}

\subsubsection{Comparison with three-dimensional dust maps}
\label{sec:dust_comparison}

While the velocity field is the primary aim of this work, the reconstruction simultaneously yields a fully probabilistic three-dimensional map of the DIB density. This is essentially a 3D dust map, built from a spectroscopic survey with $\sim150$ thousand line-of-sight measurements as opposed to photometric 3D dust maps built from $\sim800$ million line-of-sight measurements. As such, we can ask if the 3D dust map built from a much lower sampling is consistent with photometric 3D dust maps. While DIBs have been shown to correlate well with extinction \citep{Zasowski_2015}, it is possible these maps differ due to the sensitivity of DIBs to variation in dust properties, such as a sensitivity to $R(V)$ at the $\sim10\%$ level shown by \citet{Saydjari_2026_ApJ}. For this reason, and due to uncertainties in both our reconstruction and the photometric dust maps, we do not expect the correlation between them to be exact. 

\begin{figure*}
    \centering
    \includegraphics[width=\linewidth]{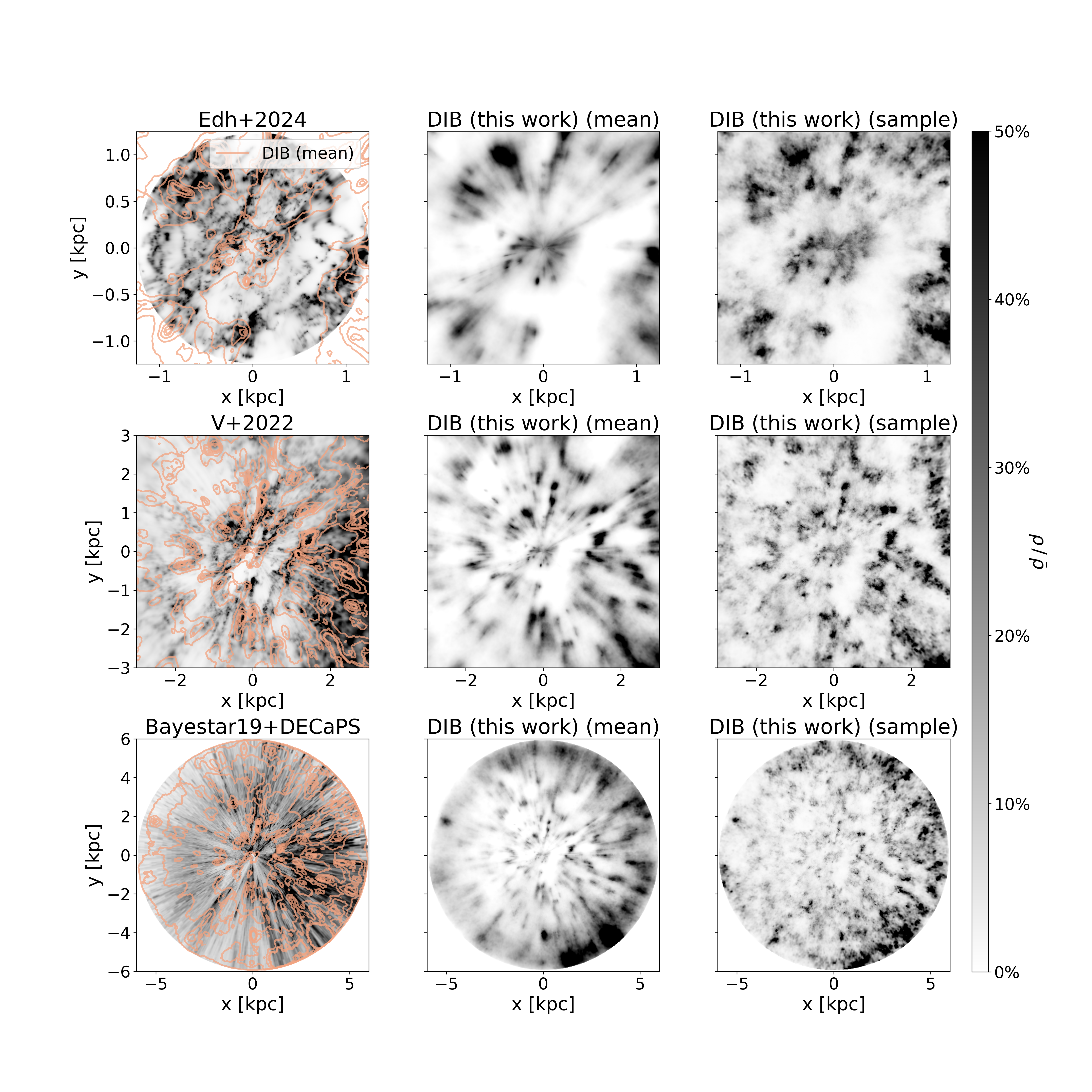}
    \caption{Top-down comparison of projected DIB density and 3D dust. Each row compares the DIBs with dust maps over a different scale: $\pm \rLocal{}$~kpc against \citet{2024A&A...685A..82E} (top), $\pm 3$~kpc against \citet{2022A&A...664A.174V} (middle), and $\pm 6$~kpc against Bayestar19${+}$DECaPS \citep{2019ApJ...887...93G, 2025ApJ...992...39Z} (bottom). All panels show normalized density contrast $\rho/\bar{\rho}$ on a common percentile scale. \emph{First column:} the respective dust map, with contours of the DIB posterior mean overlaid in orange. \emph{Second column:} the DIB posterior mean. \emph{Third column:} a single posterior sample of the DIB density. An animated version of this figure showing all posterior samples is available at \url{https://www.ph-frank.de/resources/dib_release_18_11_26/dust_comparison_samples.html}.}
\label{fig:dust_comparison}
\end{figure*}

We compare against three photometric 3D dust maps: the parsec-resolution map of \citet{2024A&A...685A..82E}, which reaches $\sim\!\rLocal{}$~kpc from the Sun; the \textit{Gaia}-based reconstruction of \citet{2022A&A...664A.174V} at intermediate scales; and the Bayestar19${+}$DECaPS map \citep{2019ApJ...887...93G, 2025ApJ...992...39Z}, which reaches out to the full $6$~kpc (and beyond) of our reconstruction. Figure \ref{fig:dust_comparison} presents this comparison as three rows of increasing scale, each averaged over $z$ through the Galactic plane. In the first column the grayscale shows the dust map and orange contours overlay the posterior-mean DIB density on the same field; the second and third columns show the DIB posterior mean and a single posterior sample on the same normalized contrast scale. The large-scale agreement is clear. The DIB contours trace the principal overdensities of all maps, including solar-neighborhood ISM features like the Radcliffe Wave, the Sagittarius Spur, and the Split \citep{Alves_2020, Kuhn_2021, 2022A&A...664A.174V}. This agreement is reassuring given that the dust and DIB reconstructions rely on independent measurements with different physical origins, dust extinction from grains as opposed to the density of DIB molecular carriers. Only the \textit{Gaia} stellar distance information is common between the maps.

The comparison also makes the complementary strengths of the DIB map apparent. Photometric samples are much larger and can support much higher resolutions locally. However, the \apDIB in APOGEE in the NIR has a highly linear correlation with extinction and can maintain high signal-to-noise through much higher dust columns. As a result, the photometric maps have much better spatial resolution nearby \citep[top row, ][]{2024A&A...685A..82E}, but our NIR DIB map can better probe coherent structure across a larger volume (bottom row) where the dust maps are increasingly dominated by the fingers-of-god effect. 

The contrast between the posterior-mean and posterior-sample DIB panels (second versus third column) illustrates the resolution question directly: the mean contours are smooth, suppressing structure on scales the data cannot constrain, while the posterior sample restores small-scale fluctuations consistent with the inferred power spectrum, giving a visual sense of the resolution and the posterior uncertainty of the reconstruction. Some radial elongation of the DIB contours toward the observer is visible (most prominently at large heliocentric distance), but it is significantly smaller than seen in the 3D dust maps. 

\begin{figure}[b!]
    \centering
    \includegraphics[width=\linewidth]{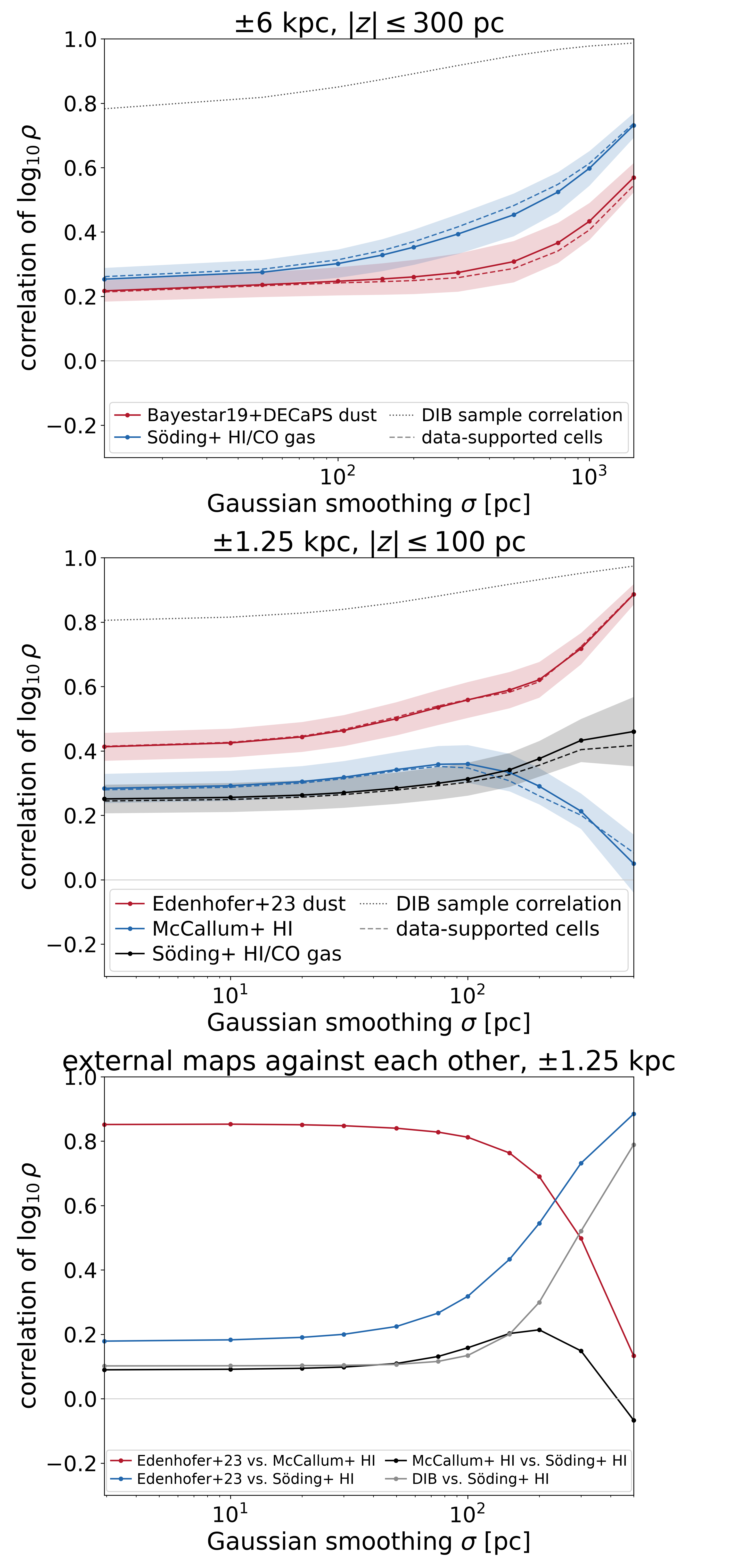}
    \caption{Same as in Figure~\ref{fig:kt_agreement}, but now comparing DIB density with 3D differential dust extinction and gas density maps (see legends). \textit{Top}: the $\pm6$\,kpc maps. \textit{Middle}: the local $\pm1.25$\,kpc maps. \textit{Bottom}: the external local maps compared against each other.}
\label{fig:density_agreement}
\end{figure}

Analogous to the quantitative velocity correlation analysis (Figure \ref{fig:kt_agreement}) we correlate the top down DIB density projections (Figure \ref{fig:density_agreement}) with the external dust as well as gas maps as a function of different spatial smoothing scales. For the large volume map comparisons, the correlation starts at 0.22 (0.25) and rises as a function of smoothing scale to 0.57 (0.73) for the \citet{2025ApJ...992...39Z} and \citet{2025A&A...693A.139S} maps, respectively. For the more local volume, the \citet{2024A&A...685A..82E} and \citet{2025A&A...693A.139S} maps behave similarly to the above discussion, except that the extinction map has a higher correlation than the gas map. For \citet{2024A&A...685A..82E}, the correlation starts at 0.4 and rises to 0.89 at the largest smoothing scale, which is close to the best-possible correlation given the uncertainties on our reconstruction. The \citet{McCallum2026_HI} map, which uses \ion{H}{I} density obtained via morphological matching behaves differently. Its correlation with the DIB density peaks at 0.36 around $100$~pc and falls to 0.05 at $500$~pc. This is not a property of \ion{H}{I} as a tracer, because \citet{2025A&A...693A.139S} also uses \ion{H}{I}.

In fact, the two \ion{H}{I} maps agree with each other at no scale (bottom panel, black, 0.1--0.2, and below zero at $500$~pc). However, each correlates with dust on a different range of scales. The \citet{McCallum2026_HI} atlas follows the \citet{2024A&A...685A..82E} dust closely at small and intermediate scales ($\sigma = 30$~pc to $200$~pc), but carries almost no large-scale correlation. In contrast, the \citet{2025A&A...693A.139S} \ion{H}{I} map correlates well with dust on large scales ($\sim500$~pc), but has little correlation on small scales (at most 0.22 for $\sigma \leq 50$~pc). The DIB density follows each of them simply where they follow the dust. Restricting the domain to the cells in the DIB map constrained by data along the line of sight changes none of these curves by more than the sample spread. 

One caveat is that we do not have a significant number of stars within the Local Bubble in our APOGEE sample. As a result, the DIB density sits 0.3~dex above its value at $200$--$300$~pc within the first $\sim\!100$~pc of the Sun. In contrast, the \citet{2024A&A...685A..82E} dust and the \citet{2025A&A...693A.139S} gas drop by a full dex into the Local Bubble. The problem in our reconstruction is that the innermost volume of the DIB map contains too few sightlines to constrain the field and is prior dominated there. Voxels near the Sun also have a tendency to sweep up residual unresolved density, hence the map is not trustworthy inside the Local Bubble.

\begin{figure*}
    \centering
    \includegraphics[width=\linewidth]{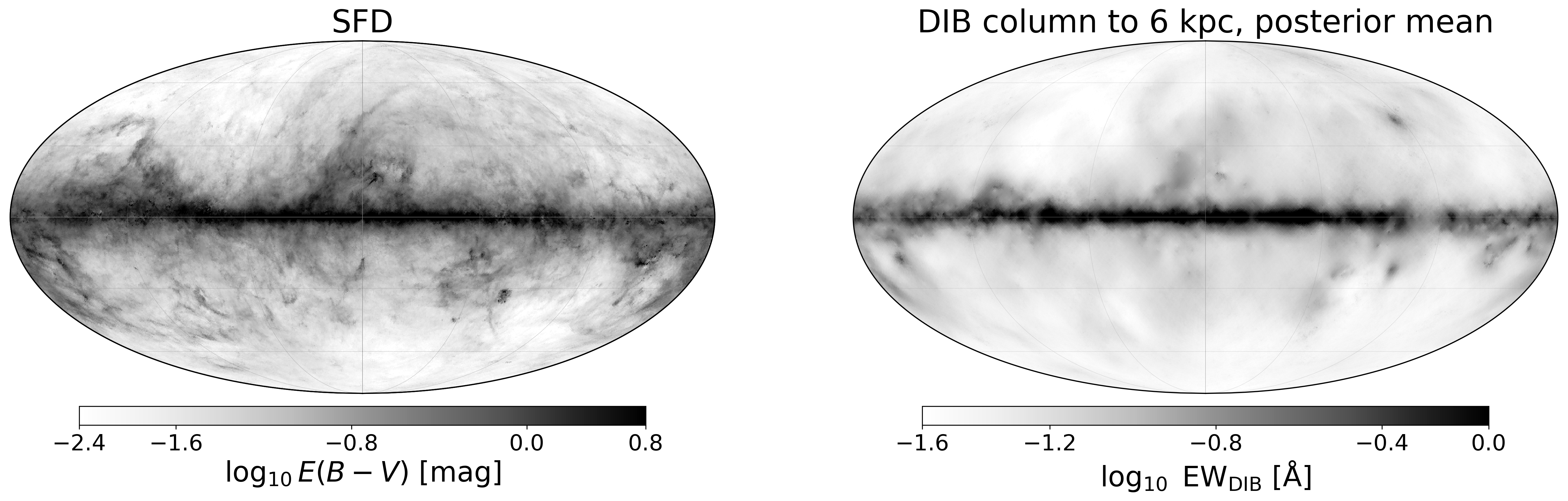}
    \caption{All-sky comparison of the integrated DIB column with dust emission. \emph{Left:} $E(B-V)$ of \citet{1998ApJ...500..525S}. \emph{Right:} our mean DIB equivalent-width integrated to $6$~kpc. Both are shown on a logarithmic scale clipped between the 1st and 99th percentiles of each map.}
\label{fig:sfd}
\end{figure*}

By comparing to a 2D FIR emission dust map \citep[][SFD]{1998ApJ...500..525S}, we can also ask how much of the total column of dust we are sensitive to, either due to our distance limits, sampling density, or sensitivity limits. To compare, we integrate our DIB density map along every line of sight to the edge of the reconstructed volume. Figure \ref{fig:sfd} shows both maps on a HEALPix $N_{\rm side}$-$128$ grid. The DIB column broadly traces SFD across the sky, but sampled at a significantly lower angular resolution due to our limited number of spectroscopic absorption measurements currently available. Hence some small scale features present in \citet{1998ApJ...500..525S} are not visible in our integrated DIB column density map. Furthermore, some parts of the sky are sampled at a much lower density than others (see Figure \ref{fig:allsky_data_plot}), which severely limits the structures that are recoverable in these regions. Nevertheless the Pearson correlation of the two dust maps, computed per posterior sample, is $0.84 \pm 0.02$ over the whole sky and $0.89 \pm 0.02$ over the sky pixels that contain at least one DIB measurement. However, a simple correlation can be pessimistically viewed as a statement that dust nearby is correlated with the total dust density along the line of sight. We therefore also compute the fraction of the SFD column that our map captures. Converting the DIB column to reddening with $W_{\rm DIB}/A_V = 0.102$~\AA\,mag$^{-1}$ \citep{Zasowski_2015} and $R_V = 3.1$, and restricting to sightlines with $E(B-V)_{\rm SFD} > 0.1$, the ratio of the DIB column integrated to $6$~kpc to the SFD reddening has a median of $0.88^{+0.42}_{-0.30}$ for $|b| < 10\degr$ and of $1.29^{+0.54}_{-0.54}$ for $|b| \geq 10\degr$, where the ranges give the 16th–84th percentiles over sightlines. The excess above unity at higher latitudes reflects the unconstrained local volume and the prior-driven density floor, as well as uncertainties in the conversion factors.

\subsubsection{Radial and vertical structure of the disk}
\label{sec:galactic_structure}

\begin{figure*}[t]
    \centering
    \includegraphics[width=\linewidth]{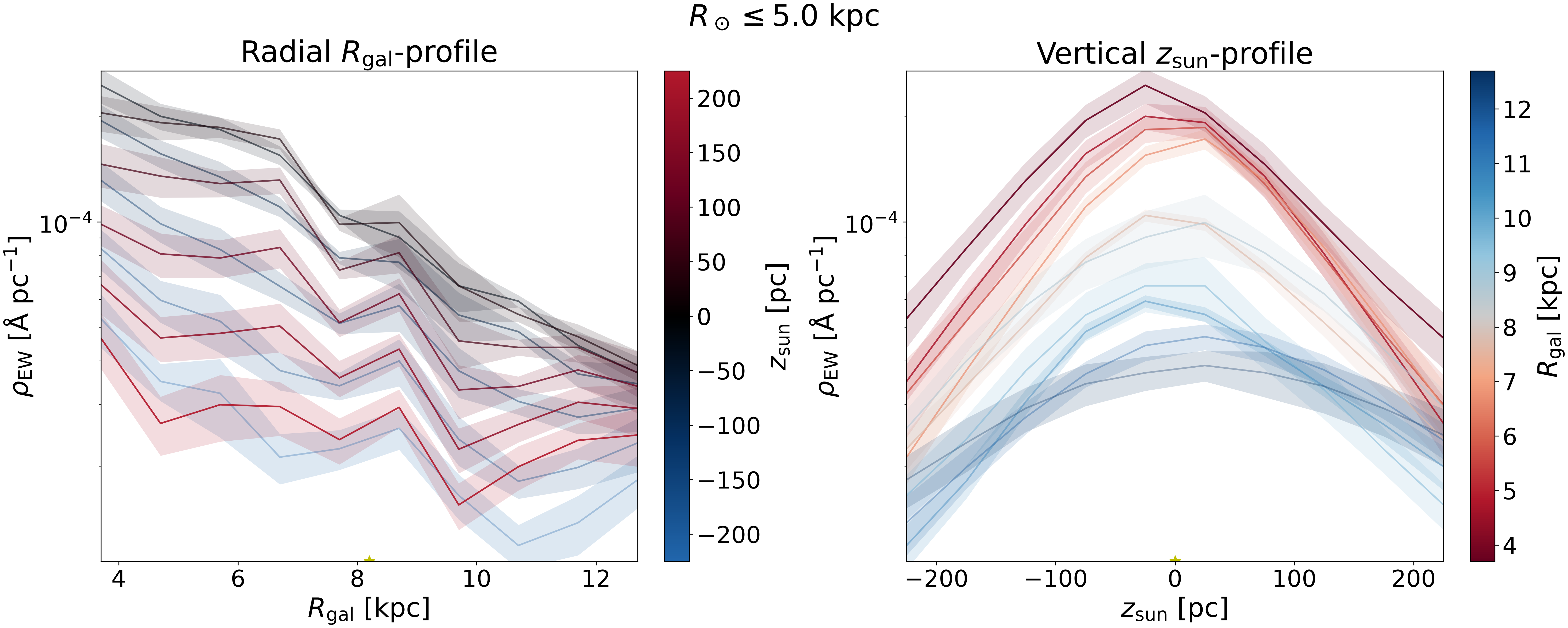}
    \caption{Volume-weighted, angularly averaged DIB density as a function of Galactocentric radius $R_{\rm gal}$ and height relative to the Sun $z_{\mathrm{sun}}$. \emph{Left:} radial profiles at fixed, color-coded $z$. \emph{Right:} vertical profiles at fixed, color-coded $R_{\rm gal}$. Solid lines give the posterior mean over samples and the bands show the posterior spread over samples.}
\label{fig:profiles}
\end{figure*}

Beyond 2D dust density comparisons, the reconstructed DIB density field also can provide constraints on the 3D structure of the Milky Way disk across the 6 kpc radius volume probed by APOGEE. Figure \ref{fig:profiles} shows the volume-weighted, angularly averaged density as a function of Galactocentric radius $R_{\rm gal}$ and height above the Galactic plane $z$. Both fall off approximately exponentially, and we fit the density surface per posterior sample with an exponential disk,
\begin{align}\label{eq:disk}
    \rho_{\rm EW}(R_{\rm gal}, z) &= \rho_0 \, e^{-R_{\rm gal}/R_d} \, e^{-|z|/z_h(R_{\rm gal})} \ , \\
    z_h(R_{\rm gal}) &= z_h(R_0)\, e^{(R_{\rm gal} - R_0)/R_{\rm flare}} \ , \nonumber
\end{align}
where $\rho_0$ is the density normalization, $R_d$ the radial scale length, $z_h(R_0)$ the scale height at the solar radius $R_0$, and $R_{\rm flare}$ the flare scale.

\begin{figure*}[t]
    \centering
    \includegraphics[width=\linewidth]{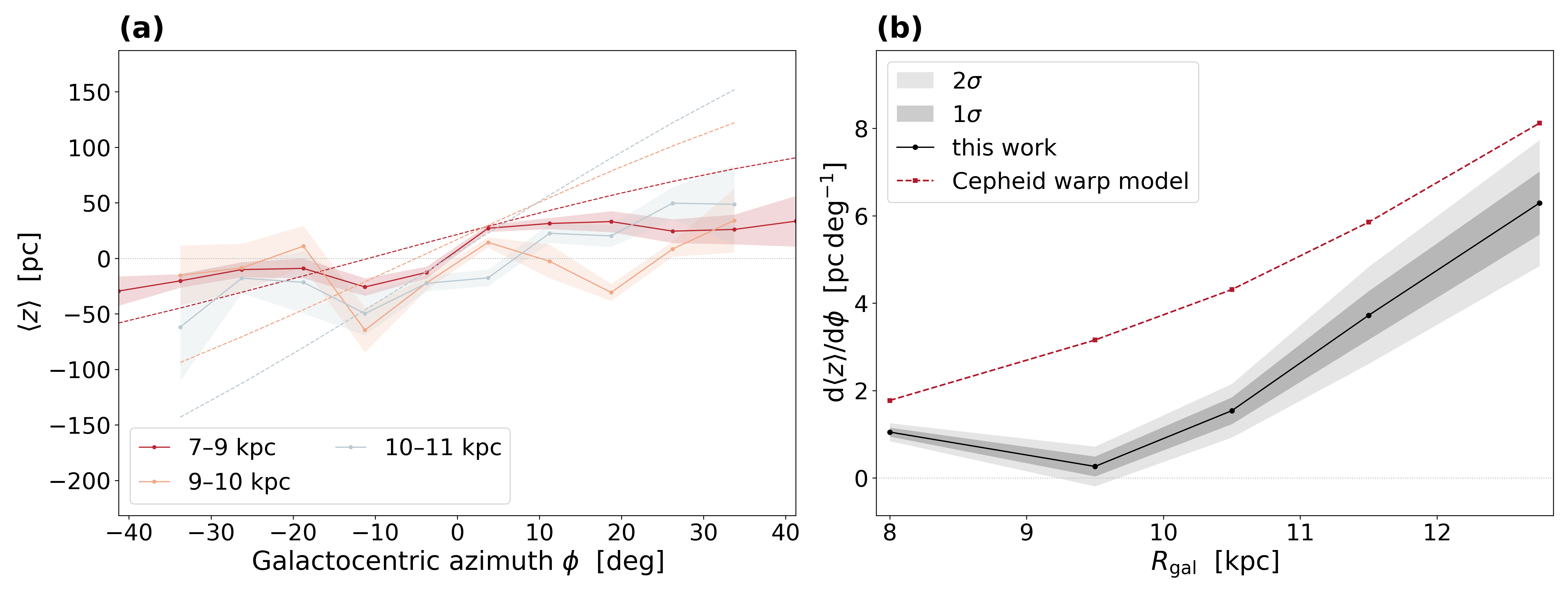}
    \caption{Signatures of the Galactic warp in DIB density. \emph{Left:} Density-weighted mean height $\langle z \rangle$ versus Galactocentric azimuth $\phi$ in rings of Galactocentric radius. Bands are the posterior standard deviation. The dashed lines are the Cepheid warp model of \citet{Chen_2025_warp} evaluated at the same rings. \emph{Right:} The azimuthal gradient $\mathrm{d}\langle z\rangle/\mathrm{d}\phi$ as a function of Galactocentric radius.}
\label{fig:warp}
\end{figure*}

\subsubsection{Dependence on the fit volume}
\label{sec:zh_resolution}

The fit to Equation \eqref{eq:disk} depends on how much of the volume is used. Importantly, the vertical $z$ structure of the disk is not resolved everywhere. The reconstruction has an effective \emph{angular} resolution of $k \approx 1.1\degr$, two and a half times the nominal pixel, set by the angular spacing of the APOGEE sightlines rather than by the pixelation (Appendix \ref{ap:zres}). As a result, the height that this angular resolution limit smears out grows in proportion to distance: $46$~pc at the $2.3$~kpc median distance of stars in our input catalog, but $78$~pc at $4$~kpc and $117$~pc at $6$~kpc. Once this resolution scale exceeds the scale height itself, near $5$~kpc, a fit to the vertical profile measures the angular smoothing rather than the disk. Thus, we fit Equation \eqref{eq:disk} only within 4~kpc of the Sun, which we validate via fits to mock reconstructions. Fit over growing volumes, a mock reconstruction departs from its own, distance-independent truth by 6, 5, 14 and 16\% at $R_{\odot,\max} = 3$, $4$, $5$ and $6$~kpc. The value we find for the observed map drifts by 23\% over the same range (Table \ref{tab:mock_trust} in Appendix \ref{ap:zres}). Inside $4$~kpc, the resolution effect causes a bias of only a few per cent.

Another issue with fitting the disk structure from our posterior is that even limiting to $|R|< 4$~kpc and $|z| \leq 250$~pc, only $\sim$20\% of the voxels have a star behind them along the line of sight. The rest of the voxels are inferred by using the prior as an interpolator between the constrained sightlines. Fitting only the voxels constrained by data, we find $z_h(R_0) = \zhFit{}$~pc while fitting all cells gives $z_h(R_0) = \zhFull{}$~pc. For the other two parameters, the fits are unchanged within their uncertainties regardless of fitting only voxels constrained by data or all pixels,  $R_d = \RdFit{}$ versus $\RdFull{}$~kpc and $R_{\rm flare} = \RflareFit{}$ versus $\RflareFull{}$~kpc, respectively (Appendix \ref{ap:zres}). In what follows, we quote the constrained-voxel fit. Over the $\pm4$~kpc volume, the flare model of Equation \eqref{eq:disk} gives
\begin{align}\label{eq:disk_values}
    R_d &= \RdFit{}~\mathrm{kpc} , \nonumber \\
    z_h(R_0) &= \zhFit{}~\mathrm{pc} , \nonumber \\
    R_{\rm flare} &= \RflareFit{}~\mathrm{kpc} .
\end{align}
We adopt these values as our measurement of the structure of the DIB disk. The quoted uncertainties are the spread over posterior samples. They do not include two systematic effects identified by the mock of Appendix \ref{ap:mock}: (1) the posterior widths are too narrow by a factor of \calLo{}--\calHi{} (Appendix \ref{ap:mock}), and (2) at the same $4$~kpc cut the estimator recovers the scale height of the mock truth to within 5\% across two different mock realizations, which is a $2.5\sigma$ bias in terms of the statistical error of $z_h(R_0)$ alone. We discuss some alternative estimators in Appendix \ref{ap:zres}, which add a free density floor, use only the best-resolved innermost shells, or correct for the resolution explicitly. These give scale heights between $82$ and $100$~pc, which can be viewed as a range that gives the order of magnitude of the systematic uncertainty of $z_h(R_0)$. The scale height agrees with the $108\pm8$~pc that \citet{Zasowski_2015} find for this same DIB.\footnote{\citet{Zasowski_2015} also used APOGEE spectra, from an earlier data release and with a different reduction, different DIB extraction method, and smaller DIB catalog.} The radial scale length is likewise consistent with the $\sim\!5$~kpc found by \citet{Zasowski_2015}, and it is worth noting that forcing a single, radius-independent scale height excluding flaring instead drives $R_d$ to $8.0$~kpc.

Our measurements suggest that the flare we observe in our reconstruction is not an artifact of the distance-dependent smearing. Measuring the scale height against $R_{\rm gal}$ \emph{within} fixed heliocentric shells, where every radial bin shares the same resolution, the disk thickens outward for the shells beyond $3$~kpc, while in the two nearest shells the scale height rises up until $R_{\rm gal} \approx 8.8$~kpc and falls back at $10.7$~kpc within its uncertainties (see Appendix \ref{ap:zres} for details).

\subsubsection{The warp}
\label{sec:warp}
The Galactic warp is often described as a vertical displacement of the midplane parameterized by $z_w(R_{\rm gal}, \phi) = a(R_{\rm gal})\,\sin(\phi - \phi_w(R_{\rm gal}))$. In the Cepheid model of \citet{Chen_2025_warp} the amplitude grows steeply with radius, as $a \propto R_{\rm gal}^{3.08}$, and over the radii we reach in our map the line of nodes they fit twists slowly from $\phi_w = -11\degr$ to $+8\degr$, staying close to the Galactic anti-center direction. Our input catalog footprint reaches the outer disk only through a wedge around the Galactic anti-center, $|\phi| \lesssim 34\degr$ at $R_{\rm gal} = 9$--$11$~kpc and $|\phi| \lesssim 26\degr$ beyond. Over such a narrow wedge the amplitude, phase, and offset of a sinusoid are strongly degenerate. So, instead of trying to constrain a warp model directly, we measure the azimuthal gradient $\mathrm{d}\langle z\rangle/\mathrm{d}\phi$ of the density-weighted mean height of the DIB layer, which constrains $a(R_{\rm gal})$ without depending on $\phi_w(R_{\rm gal})$.

We show $\langle z \rangle$ as a function of azimuth for three different rings in Galactocentric radius (Figure \ref{fig:warp}, left). In the $7$--$9$~kpc ring the centroid rises steadily across the full $\pm 40\degr$ range, with a similar but noisier trend in the other two rings.  We summarize the $\mathrm{d}\langle z\rangle/\mathrm{d}\phi$ slopes fit to each $R_{\rm gal}$ ring in Figure \ref{fig:warp} (right). The measured gradient is positive almost everywhere except around $9.5$~kpc where it is close to zero. It reaches up to 77\% of the Cepheid warp model in the furthest ring, where the warp is largest. We read Figure \ref{fig:warp} as the detection of a coherent vertical distortion of the DIB disk that has $\mathrm{d}\langle z\rangle/\mathrm{d}\phi$ gradients consistent with a warp. It grows with radius as the warp does and approaches the Cepheid prediction in the outermost rings. However, we do not attempt to quote a warp amplitude or constrain a line of nodes due to the limited angular coverage of our current sample. The two aforementioned systematic effects in the reconstructed vertical structure, the angular smearing of our sampling density and the prior-driven density floor (Appendix \ref{ap:zres}), act symmetrically about the plane. They pull the centroid toward zero and could therefore dilute an azimuthal gradient in $\langle z \rangle$, but they cannot produce such a gradient. As shown in Appendix \ref{ap:zres}, including a density floor in our model changes the fit disk parameters only marginally. The fit floor is about $1\%$ of the midplane density at $R_0$, so the density of the floor becomes comparable to the disk density only at $|z| \approx 0.4$~kpc, while the disk fit is dominated by $|z| < 250$~pc. We therefore do not correct for the bias due to the density floor, but note that the floor, $\approx 1.4 \times 10^{-6}\,\nicefrac{\text{\AA}}{\mathrm{pc}}$, sets the level below which our density map should not be interpreted.

\begin{figure*}
    \centering
    \includegraphics[width=\linewidth]{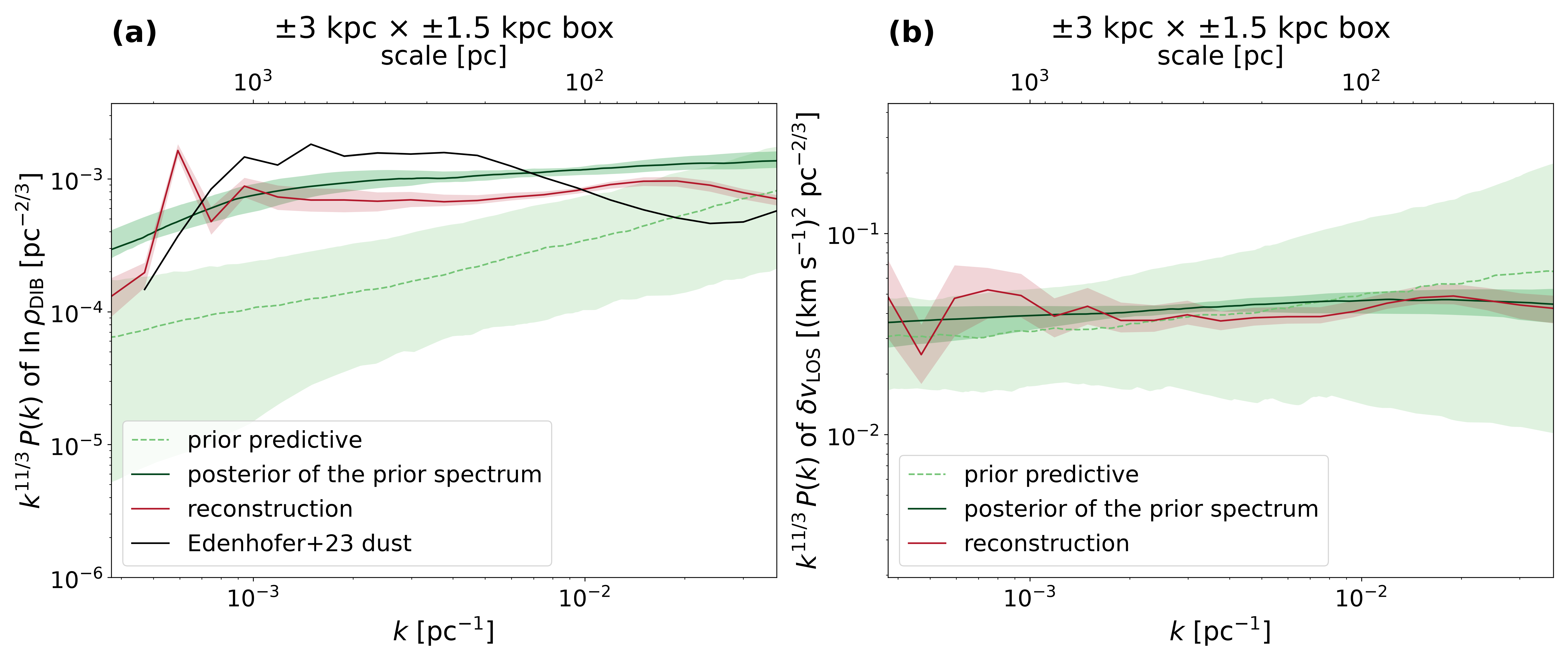}
    \caption{Scaled power spectra, $k^{11/3} P(k)$, of the reconstructed fields compared with their priors. \emph{Left:} 3D power spectrum of $\ln\rho_{\rm DIB}$ (red; median and 16--84\% band over posterior samples), compared with the same statistic for the \citet{2024A&A...685A..82E} 3D dust reconstruction (black). The prior on the power spectrum is shown in light green (dashed) and its posterior in dark green (solid). \emph{Right:} The same for the residual velocity $\delta v_{\rm LOS}$.}
\label{fig:spectra}
\end{figure*}

\subsection{Power spectra and structure functions}
\label{sec:spectra}
In addition to constraining the pixel-by-pixel 3D fields, we also reconstruct the spatial power spectra consistent with the correlation structure of the fields as part of our forward model (\S\ref{sec:cf}, Table \ref{table:params}). These power spectra in the model serve two purposes. First, the spectra and associated structure functions give insight into the physical state of the DIB density and velocity fields, allowing us to study properties of turbulence and confinement where we have resolving power. Second, the inferred prior power spectra act as a data-driven regularization of the map. Specifically, due to the model being statistically homogeneous and isotropic a priori, regions that are data informed at a certain scale are used by the model to constrain the expected power at that scale in other regions of the map where the data is not informative, in order to inter- and extrapolate these regions in a way that is statistically consistent with the rest of the map. This provides a powerful and weakly informative probabilistic interpolator for the full volume across both the density and velocity fields.

In total there are three spectra to compare for each field: the prior predictive distribution of a spectrum, its posterior, and the empirical spectrum of the reconstructed field itself. Figure \ref{fig:spectra} shows all three, computed as three-dimensional spectra of a $\pm 3 \times \pm 3 \times \pm 1.5$~kpc box at $11.8$~pc voxel resolution, which we generated for each posterior sample. We show the spectra compensated by a Kolmogorov slope $k^{11/3}$. Power law slopes are fit between $60$ and $300$~pc, above the voxel scale and well inside the box.

For the density (Figure \ref{fig:spectra}, left),  the data constrain the power spectrum slope to be steeper than the prior spectrum. The prior slope is $-3.1$ while the posterior slope is $-3.5$, and the empirical spectrum of the box is very similar to the posterior of the prior, with a slope of $-3.4$. We can compare this to the power spectrum of the dust density from photometric dust maps, such as \citet{2024A&A...685A..82E}. It agrees with the DIB density spectrum to within a factor of about two on all scales. It lies above it with comparable slope down to $\sim\!200$~pc scales, then steepens and falls below it at smaller scales, in line with the map-level agreement of \S\ref{sec:dust_comparison}.

For the residual velocity (Figure \ref{fig:spectra}, right) the prior predictive and posterior slopes are $-3.4$ and $-3.6$, with the measured box spectrum sitting on the posterior prior slope at every scale. The scalar velocity power spectrum slope predicted for Kolmogorov turbulence in three dimensions is $-11/3 = -3.\bar{6}$, which is consistent with the posterior slope.

\begin{figure*}
    \centering
    \includegraphics[width=\linewidth]{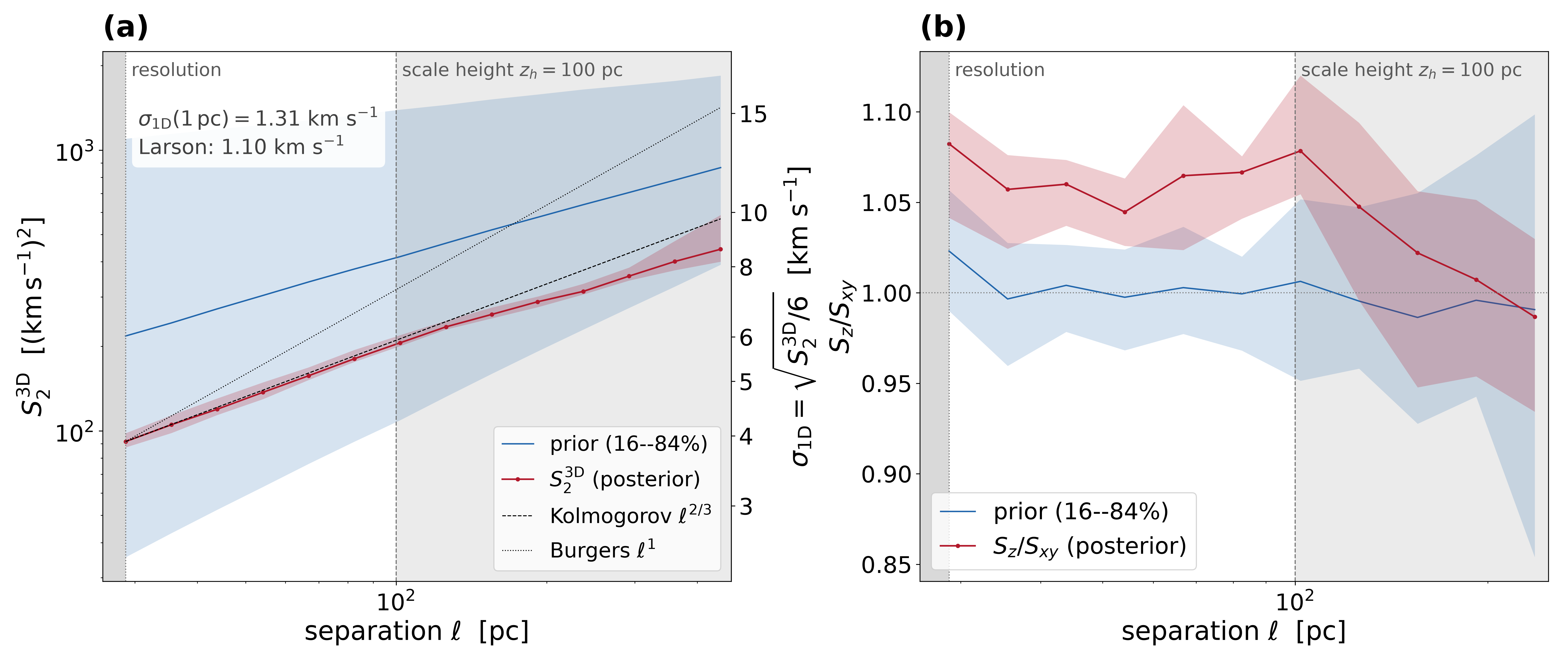}
   \caption{Turbulence statistics of the DIB velocity field in the local volume. \emph{Left:} Second-order structure function $S_2^{\rm 3D}$ (red; median and $16$--$84\%$ posterior band), compared with Kolmogorov and Burgers slopes and with draws from the prior (blue). The right-hand axis gives the equivalent line-of-sight velocity dispersion $\sigma_{\rm 1D}$. \emph{Right:} Ratio of the structure functions measured perpendicular to the disk ($S_z$) and within the plane ($S_{xy}$) for the reconstruction (red) and the prior (blue). In both panels, gray shading marks separations beyond which interpretation is difficult, either because they are below the resolution limit (\resFloor{}~pc) or above the disk scale height ($\sim\!100$~pc).}
\label{fig:turbulence}
\end{figure*}

\subsubsection{Structure functions of the velocity field}
\label{sec:turbulence}

The structure function gives an alternative, real space representation of our power spectra that we use to relate our results to theoretical expectations. We compute the second-order velocity structure function
\begin{equation}\label{eq:s2}
    S_2(\ell) = \left\langle \left| v(\mathbf{x} + \boldsymbol{\ell}) - v(\mathbf{x}) \right|^2 \right\rangle
\end{equation}
in the local volume ($100 < r < 1200$~pc, $|z| < 400$~pc) on the highest ($\sim$10~pc) resolution posterior cubes. We restrict the analysis to cells above the 40th percentile in density, a compromise between having higher SNR for velocities where DIB density is high while maintaining a large enough volume. Because only the line-of-sight component of the velocity is observable, we convert our measurement to a full three-dimensional structure function under the assumption of isotropy, $S_2^{\rm 3D} = 3\langle(\hat{\mathbf{n}}\cdot\delta\mathbf{v})^2\rangle$, for the purposes of comparison (Figure \ref{fig:turbulence}).

To help interpret Figure \ref{fig:turbulence}, we establish the range of separations over which our measurement is trustworthy. The effective resolution of the reconstruction sets the minimum value, which is 9.6~pc for the cube we defined above. To be conservative, we will only trust structures above three resolution elements ($\sim$30~pc). The vertical extent of the disk sets a maximum separation under the isotropy assumption, which we take to be the disk scale height ($\sim100$~pc). The prior's own structure function (Figure \ref{fig:turbulence}, blue, left) is only weakly constrained: its $16$--$84\%$ band spans more than an order of magnitude in amplitude, while the posterior band is about $\pm 6\%$ wide. Within the trustworthy window the posterior lies inside the prior band at about a factor of $2$ below the prior median, and has a slope of $\sTwoSlope{}$ against $0.53$ for the prior. Thus, our velocity structure function measurement is data driven and not prior dominated.

We find the slope of the velocity structure function is consistent with a Kolmogorov turbulent cascade ($2/3=0.\bar{6}$), over the $30$--$100$~pc window. In one of the two mock tests described in Appendix \ref{ap:mock}, we generate a field with a Kolmogorov velocity spectrum as truth. The same estimator used in Figure \ref{fig:turbulence} recovers the slope of the mock test to within 10\%, about $2\sigma$ in the posterior uncertainty. Despite the precipitous drop off in density due to the scale height of the disk, the velocity structure function still has significant amplitude at the scale height. We do, however, observe a flattening in the slope of the velocity structure function above the scale height. We expect some degree of flattening due to $\ell$-independence but interpreting the structure function at these scales requires carefully handling the prior dominated volume above the disk where we have few measurements and low DIB strengths (dust densities).

In addition to the slope, we can test whether the amplitude of the velocity structure function is consistent with expectations. For isotropic, incompressible turbulence the amplitude sets the energy cascade rate $\epsilon$ via $S_2^{\rm 3D} = \frac{11}{3}C_2(\epsilon \ell)^{2/3}$, where $C_2 \approx 2$ is the
Kolmogorov constant for the longitudinal structure function \citep[][\S6.1]{Pope2000}. Fitting over the trustworthy scales ($29$--$100$\,pc) with the slope fixed at $2/3$ (a free fit gives $\sTwoSlope{}$), we find $\epsilon \approx 1.4 \pm 0.6\,(\kms)^{3}\,\mathrm{pc^{-1}} \approx (4.7 \pm 1.9)\times10^{-4}\,\mathrm{cm^{2}\,s^{-3}} \approx (2.4 \pm 1.0)\times10^{-4}\,L_\odot\,M_\odot^{-1}$, where the uncertainty combines the posterior spread, variation between subvolumes, and a 20\% uncertainty on $C_2$. This is comparable to the solar-neighborhood estimates, using the velocity dispersion of gas measurements, of \citet{Hennebelle_2012_AARv}, which are $10^{-3}$ and $10^{-4}\,L_\odot\,M_\odot^{-1}$ for the neutral atomic medium and molecular clouds, respectively. 

Another way of interpreting this amplitude is by extrapolating to predict the velocity dispersion it corresponds to on $1$~pc scales. The line-of-sight velocity dispersion of a region of size $\ell$ follows from the structure function as $\sigma_{\rm 1D}(\ell) = \sqrt{S_2^{\rm 3D}(\ell)/6}$. To do this, we extrapolate our fit to the posterior structure function one and a half decades below our resolution. This gives $\sigma_{\rm 1D}(1~{\rm pc}) \approx 1.3\pm0.07$~\kms, with uncertainties given by the spread over posterior samples. This amplitude can then be compared against the size--line-width relation observed for molecular clouds \citep{1981MNRAS.194..809L, 1987ApJ...319..730S}, the so-called Larson's relation that yields $\sigma_v = 1.10\,(L/{\rm pc})^{0.38}$~\kms. Compared to the 1.1~\kms\ from \citet{1981MNRAS.194..809L} and 0.7~\kms\ from \citet{1987ApJ...319..730S}, our extrapolated value is reasonable, though $\sim20\%$ higher than Larson's relation at parsec scales. For comparison, the prior's structure function extrapolates to $\sigOnePcPrior{}$~\kms. With the current reconstruction, we cannot establish whether the excess in the DIB velocity structure function amplitude relative to the gas is real or if the structure function steepens below our resolution. Future work pushing to higher spatial resolutions with larger DIB catalogs will be required to answer this question.

\subsubsection{Isotropy and modes of the velocity field}
We can investigate the isotropy and mode composition of the velocity field by asking how different the velocity structure function is for in-plane versus out-of-plane directions (Figure \ref{fig:turbulence}, right). To do this, we split pairs by the orientation of their separation vector. $S_{xy}$ uses separations $|s_z|/|s| < 0.3$, i.e.\ within $17\degr$ of the plane. $S_z$ uses separations with $|s_z|/|s| > 0.9$, within $26\degr$ of the disk normal. The right panel of Figure \ref{fig:turbulence} shows their ratio per posterior sample. These cuts balance having a clean separation between in-plane and vertical pairs while maintaining sufficient volume and density for a clean measurement. The result does not depend on the exact values of the cuts defining $S_{xy}$ and $S_z$. Varying the in-plane band between $6$--$27\degr$ and the vertical cone between $18$--$26\degr$ moves $S_z/S_{xy}$ only by $<1\%$. 

The ratio $S_z/S_{xy}$ is expected to be unity for only a specific kind of isotropic turbulence, where the power ratio between compressive (curl-free) and solenoidal (divergence-free) turbulence is 1/2. This is equivalent to treating the velocity field as three independent random scalar fields with a common spectrum. Because we treat the line-of-sight velocity like a scalar field in our model, this is the assumption our prior makes. The same $S_{xy}$ and $S_z$ split applied to draws from the prior returns $S_z/S_{xy} = 1$ to within 2\% for the median at every separation, as we would expect (Figure \ref{fig:turbulence}, right, blue line). To understand why compressive and solenoidal turbulence can depart from $S_z/S_{xy} = 1$, we have to consider the effect of the line-of-sight projection. Because our sightlines lie close to the plane, for vertically separated pairs the line of sight is nearly perpendicular to the separation, so $S_z$ is essentially sampling transverse velocities and $S_z \approx S_{TT}$. Because in-plane separations are uniformly oriented relative to the line of sight, the line-of-sight velocities for in-plane pairs evenly sample transverse and longitudinal velocities, so $S_{xy} \approx (S_{LL} + S_{TT})/2$. For isotropic turbulence with power-law slope $\zeta$ of the second-order velocity structure function, $R = S_{TT}/S_{LL}$ is predicted to be $1/(1+\zeta)$ for a purely compressive field and 1 + $\zeta/2$ for a purely solenoidal field \citep[][\S12.3]{Monin_1975_mit}. Thus, for our measured $\zeta = 2/3$ at small separations, we would expect $S_z/S_{xy} = 3/4 = 0.75$ for a purely compressive field and  $8/7 \approx 1.14$ for a purely solenoidal field. These are limiting values because the finite width of our orientation cuts moves both toward unity, to 0.78 and 1.12, respectively. In future work, these comparisons should also account for the finite range of sightline latitudes used.

At separations below the scale height of the DIB disk, we find $S_z/S_{xy}$ exceeds unity by 4--8\%. Based on the per-sample spread, this excess above unity is significant at $\sim 2\sigma$. Above the scale height, $S_z/S_{xy}$ drops and becomes consistent with one. Taken at face value, our measured $S_z/S_{xy}$ at small scales is closer to the expected value for solenoidal turbulence than compressive turbulence and we can convert this into a compressive-to-solenoidal power ratio of $\sim0.2$. This power ratio is roughly comparable to the mean of $\sim\!0.3$ found within molecular clouds in simulations of supernova-driven turbulence by \citet{2016ApJ...825...30P}. The evolution of $S_z/S_{xy}$ as a function of separation may reflect the flattening of the structure function, be a signature of $S_z$ saturating due to confinement by the disk scale height, or reflect a reversion to our prior. Differentiating these cases, as well as bounding confounding effects from the evolution of the velocity dispersion with $|z|$, should be the subject of future work. An important caveat is that this discussion has worked under the assumption of isotropic turbulence. However, Figure \ref{fig:v_residual_clusters} shows large-scale, non-circular kinematic features that suggest that an isotropic assumption will break down at certain scales.

\subsection{Validation on a mock reconstruction}
\label{sec:mock}
\label{sec:validation}

\begin{figure*}
    \centering
    \includegraphics[width=0.85\linewidth]{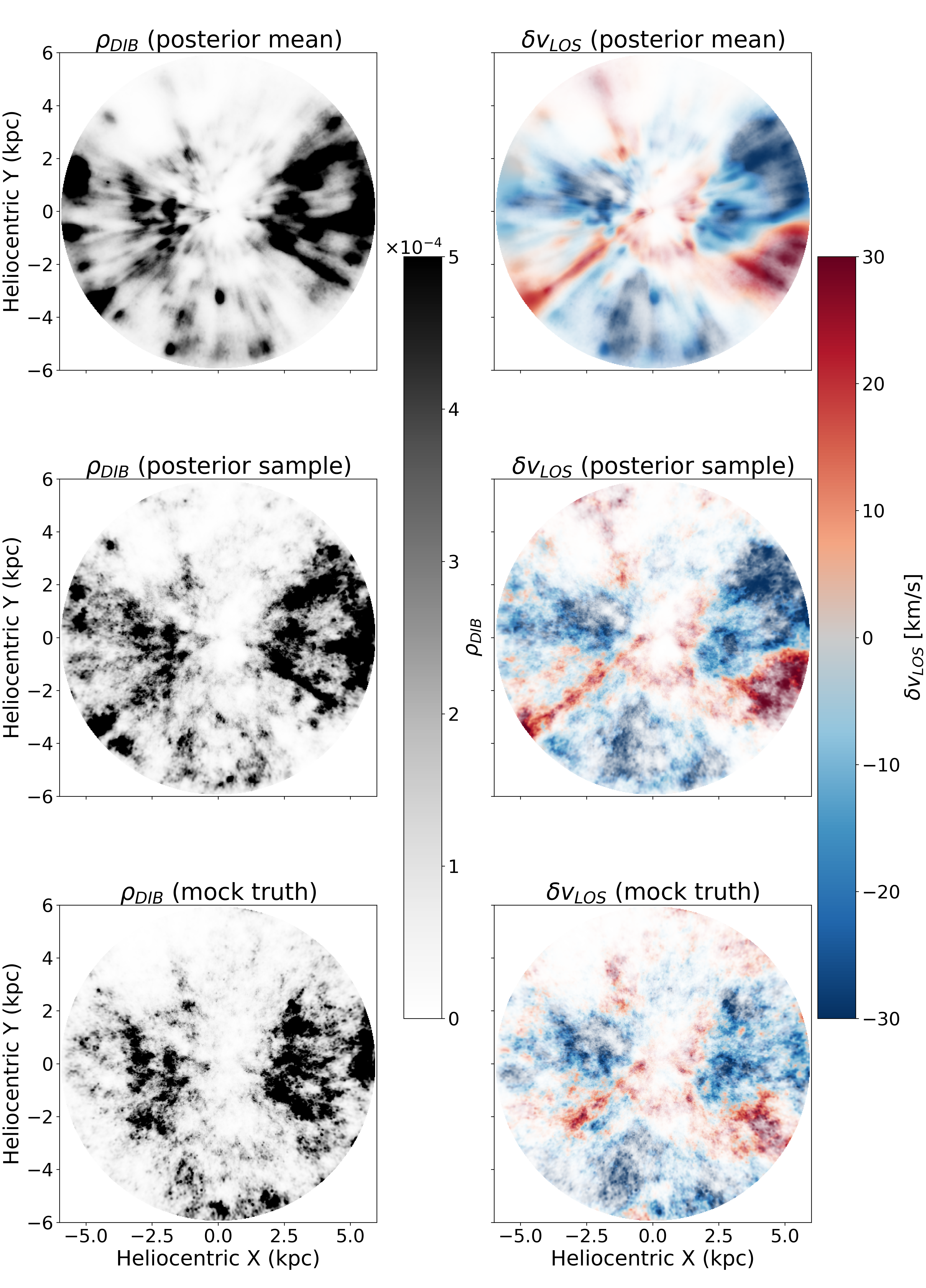}
    \caption{The mock reconstruction against its truth, in the layout of Figure \ref{fig:density_velocity}. The DIB density (left) and density-weighted residual velocity (right) are shown in the $\pm 6$~kpc midplane slab, for the posterior mean over the \NsampMock{} mock samples (top), a single posterior sample (middle) and the truth of the mock catalog (bottom), with the same color caps and density weighting as Figure \ref{fig:density_velocity}. A single sample reproduces the truth's texture for data-constrained voxels. The posterior mean is smoother, carries radial sightline streaks, and saturates in the outer, unconstrained volume where the prior returns to its mean.
}
\label{fig:mock_vs_truth}
\end{figure*}

Everything above rests on our reconstruction being faithful where the data is informative and honest about its uncertainty where it is not. We tested this on a mock catalog whose truth is known. The fields were chosen to be a correlated-field draw under a disk envelope for the density and a stationary correlated field for the velocity. The mock fields were observed with the real star locations and distances, per-visit uncertainties, and detection model. We ran those mock observations through the production pipeline to obtain a reconstruction (Appendix \ref{ap:mock}). Figure \ref{fig:mock_vs_truth} shows the ground truth compared to the reconstruction. The reconstructed fields agree well with the truth where they are constrained by data. A single posterior sample reproduces the texture of the truth in density and velocity. The joint disk fit also recovers the input scale height, $\zhMock{}$~pc compared to the true value of $110$~pc, and recovers the flare. The vertical-resolution effect that limits the disk fit to the volume within $4$~kpc discussed in \S\ref{sec:zh_resolution} is reproduced on the mock with the same angular smearing scale ($\kMock{}$ against $\kReal{}$ on the real data), making this scale mock calibrated rather than inferred only (Appendix \ref{ap:zres}). The mixture likelihood of (non-)detections trusts the right number of detections to within 5 percent (\respMockBelieved{} versus the true \respMockTrue{}), essentially never rejecting a real DIB detection, and the whitened residuals of the measurements classified as detections have the width of the assumed noise (\nwMockEW{} and \nwMockV{} in equivalent width and velocity, respectively, versus the true \nwRef{} for a Student-$t$ with $\nu = 10$).

However, two limitations follow from this test that affect the results and particularly reported uncertainties in a quantitative manner. First, the posterior is too narrow. Over the data-constrained voxels the standardized error compared to the truth has a robust width of \calDens{} for the log density and \calVel{} for the velocity, so the true uncertainties are a factor of \calLo{}--\calHi{}$\times$ the uncertainties quoted here (Appendix \ref{ap:mock}). Second, there is a prior-driven density floor. This exists because a log-normal field cannot reach zero. Thus, we recover a median density of order $10^{-6}\,\nicefrac{\text{\AA}}{\mathrm{pc}}$ in regions that should be essentially empty ($|b| > 60\degr$, $r > 3$~kpc), compared to the $10^{-12}\,\nicefrac{\text{\AA}}{\mathrm{pc}}$ or smaller predicted there by the fit disk model. So the posterior mean sits above the truth in low-density cells, which is visible in Figure \ref{fig:mock_vs_truth} as the saturated outer volume of the mean map. The mock could not test the radial scale length inference, because the prior realization carries too much large-scale structure of its own and significantly deviates from the exponential profile the mock draw is scaled with. More importantly, the mock truth is drawn from the same family of priors that the reconstruction assumes, so this mock reconstruction does not test our ability to constrain structures that are significantly out of distribution with respect to our prior.

\section{Conclusions}
\label{sec:conclusions}
We have generated a new probabilistic 3D spatial reconstruction of the DIB density and radial velocity out to a distance of 6 kpc from the Sun. Our map is based on a new \texttt{apMADGICS} reprocessing of the 15{,}273~\AA\ DIB detected in APOGEE DR17 spectra towards roughly 140,000 stars with \textit{Gaia} distances. Our main contributions and conclusions are as follows: 

\begin{enumerate}
\item We develop a forward model that extends DIB-based kinetic tomography in three ways: a principled mixture likelihood that incorporates low-SNR and non-detection measurements via injection-calibrated detection probabilities; a full, non-perturbative marginalization over heterogeneous \textit{Gaia} distance uncertainties; and a joint, non-parametric prior on both the density and velocity fields whose correlation structure is jointly learned from the data.

\item Furthermore, we introduce a \emph{sparse} variant of Iterative Charted Refinement, which refines the multi-resolution grid only where supported by the survey footprint, and we enable \emph{kernel inference} within the ICR hierarchy, learning the prior power spectra of both fields directly from the observations.
 
\item We compare our 3D DIB density field with independent tracers of the 3D spatial structure of the ISM from 3D dust maps. We find good agreement across all maps, both locally (within $\approx 1$ kpc of the Sun) and on more global scales. 

\item We validate our 3D DIB radial velocity field with independent distance-resolved kinematic tracers from masers and young stellar clusters. Even after subtracting off Galactic rotation, we find good correlation with these tracers, demonstrating that we can recover peculiar motions at the level of $\approx 10$~\kms. We also detect a decorrelation between DIB radial velocity and young stellar radial velocities as a function of stellar age over a timescale of $\approx 100$ Myr, indicating that we can constrain the rate at which young stars kinematically decouple from their birth clouds. 

\item We reproduce the principal features of longitude--velocity diagrams traditionally observed in CO and \ion{H}{I} emission, while adding the distance dimension that emission surveys lack, providing a route to resolving distance ambiguity of longitude-velocity diagram structures in future work.

\item We analyze the structure function of the velocity field and find a slope consistent with an isotropic Kolmogorov turbulent cascade on scales of 30--100\,pc, with preliminary constraints on the compressive-to-solenoidal power ratio. The amplitude is consistent with energy cascade rates expected for molecular clouds and is consistent with Larson's size-linewidth relation when extrapolated to parsec scales.

\item Our 3D reconstruction is well-suited to characterize both the radial and vertical structure of the Galactic disk. Fitting a simple disk model, we measure a radial scale length $R_d = 4.10 \pm 0.15$~kpc, a scale height at the solar radius $z_h(R_0) = 93.3 \pm 1.8$~pc, and a flare growth scale $R_{\rm flare} = 11.6 \pm 0.8$~kpc. We also find azimuthal gradients consistent with a warp in the DIB disk.

\item Future methodological work will sharpen this picture. Modeling the absorption spectra directly, rather than the cataloged DIB equivalent widths and velocities, will carry the full line-profile information into the reconstruction. Moving field priors to GraphGP \citep{Dodge_2026_graphgp}, a recent extension and improvement of ICR, will allow for finer resolution over larger volumes and more flexible Gaussian Process priors. Finally, work on improved posterior sampling will give better calibrated uncertainties than the present variational fit, which the mock shows to be too narrow on a per pixel basis.

\end{enumerate}

Together, these results establish DIB-based kinetic tomography as a powerful probe of the 4D (3D position + 1D radial velocity) structure of the ISM. Our methods will scale naturally to the larger and deeper DIB catalogs anticipated from ongoing and upcoming spectroscopic surveys, including future SDSS-V/APOGEE releases, \textit{Gaia} DR4, and DESI. These catalogs will enable next-generation maps with greater sensitivity and depth, as well as finer spatial and velocity resolution, ultimately leading to unprecedented constraints on ISM dynamics at the scales of individual clouds across the Galactic disk.

\section*{Data availability}
\label{sec:dataproducts}
The data products of this paper are publicly available at \href{\zenodoURL}{doi:\zenodoDOI}. The release contains the reconstruction on its native HEALPix (nested, $N_{\rm side}$ $=128$) times $260$ log-linearly spaced radial shells from $2$~pc to $6$~kpc, as well as interpolated onto three Cartesian cubes of $511 \times 511 \times 255$ voxels centered on the Sun, covering $12 \times 12 \times 6$, $6 \times 6 \times 3$ and $3 \times 3 \times 1.5$~kpc. For each grid the released fields are the DIB density $\rho$ in units of equivalent width per parsec, the line-of-sight velocity $v$ in \kms, and its residuals after subtracting the flat rotation curve of \citet{Reid2019} as well as the internally inferred rotation. The HEALPix product additionally carries the cumulative line-of-sight moment $0$ and $1$ maps. All \Nsamp{} posterior samples are released, as well as companion files with the posterior mean and standard deviation of every field together with one full posterior sample. Finally, we provide binary masks of the sky pixels and cube voxels through which the fitted sightlines pass, i.e.\ the volume the data constrain directly. All files are HDF5 with lossless compression.

\section*{Author contributions}
P.F.~and A.K.S.~conceived of the project and led the primary analysis of this work. P.F.~developed geoVI and the methods in this work, wrote and ran the code, ran the catalog re-calibration, produced the figures, co-ideated testbeds and analysis, and co-wrote the first draft of the manuscript. A.K.S.~produced the DIB catalog input to this work, provided the injection tests, wrote pieces of the code, co-ideated testbeds and analysis, and co-wrote the first draft of the manuscript. C.Z.~co-ideated testbeds and analysis, co-wrote the first draft of the manuscript, and provided general feedback on the results. S.C., D.F., and T.E.~provided partial supervision to various method developments leading towards this work as well as throughout it. S.C.~provided significant feedback on the first draft of the manuscript, as well as general discussions and feedback throughout the project. D.F.~provided a large part of the computational resources used to conduct the analysis. T.E.~supervised the NIFTy related method development and gave comments on the first draft. A.C., A.R-L., J.F-T., D.K., and J.W.~have made substantial contributions to the SDSS collaboration without which this work would not exist, and A.R-L., J.F-T., and J.W.~provided useful comments on the first version of this manuscript.

\begin{acknowledgements}
A.K.S. acknowledges support for this work provided by NASA through the NASA Hubble Fellowship grant HST-HF2-51564.001-A awarded by the Space Telescope Science Institute, which is operated by the Association of Universities for Research in Astronomy, Inc., for NASA, under contract NAS5-26555. C.Z. acknowledges support by the NSF CAREER award \#2442546 (CAREER: Charting the Formation, Transformation, and Destinies of Gas and Young Stars in the Solar Neighborhood). S.E.C. acknowledges support from the U.S. National Science Foundation under NSF CAREER grant No. AST-2441452, and from an Alfred P. Sloan Research Fellowship. J.G.F-T gratefully acknowledges the support provided by ANID Fondecyt Regular No. 1260371.

The authors acknowledge the Interstellar Institute and the Institut Pascal at Université Paris-Saclay for hosting discussions that nourished the development of the ideas behind this work.

A.K.S. acknowledges Sophia S\'{a}nchez-Maes for helpful discussions and much support. This work made use of the Cannon cluster supported by the FAS Division of Science Research Computing Group at Harvard University. This work was supported in part by the Carnegie Theoretical Astrophysics Center (CTAC) through its Theory Scholar program. The authors acknowledge the Interstellar Institute and the Institut Pascal at Université Paris-Saclay for hosting discussions that nourished the development of the ideas behind this work.

We thank Cameren Swiggum for discussions on the age dependence of the coupling between young stars and the interstellar medium, Emily Hunt for insights into the quality cuts of the young cluster sample, Ben Dodge for discussions on optimization, Theo O'Neill for discussions on kinetic tomography and absorption spectroscopy, Juan Soler for providing the HOG velocity map, Lewis McCallum for the \ion{H}{I} velocity map and helpful discussions, Julianne Dalcanton for pointing us to simulation predictions for the clearing time, Mandy Chen for suggesting the energy cascade rate formulation, and Marc-Antoine Miville-Desch\^{e}nes for discussions on interstellar turbulence and for pointing us to Larson's law.

The computations in this paper were run on the FASRC Cannon cluster supported by the FAS Division of Science Research Computing Group at Harvard University. This work used Marlowe \citep{marlowe2025}, Stanford University's GPU-based Computational Instrument, supported by Stanford HAI and Stanford Research Computing.

Funding for the Sloan Digital Sky Survey V has been provided by the Alfred P. Sloan Foundation, the Heising-Simons Foundation, the National Science Foundation, and the Participating Institutions. SDSS acknowledges support and resources from the Center for High-Performance Computing at the University of Utah. SDSS telescopes are located at Apache Point Observatory, funded by the Astrophysical Research Consortium and operated by New Mexico State University, and at Las Campanas Observatory, operated by the Carnegie Institution for Science. The SDSS web site is \url{www.sdss.org}.

SDSS is managed by the Astrophysical Research Consortium for the Participating Institutions of the SDSS Collaboration, including the Carnegie Institution for Science, Chilean National Time Allocation Committee (CNTAC) ratified researchers, Caltech, the Gotham Participation Group, Harvard University, Heidelberg University, The Flatiron Institute, The Johns Hopkins University, L'Ecole polytechnique f\'{e}d\'{e}rale de Lausanne (EPFL), Leibniz-Institut f\"{u}r Astrophysik Potsdam (AIP), Max-Planck-Institut f\"{u}r Astronomie (MPIA Heidelberg), Max-Planck-Institut f\"{u}r Extraterrestrische Physik (MPE), Nanjing University, National Astronomical Observatories of China (NAOC), New Mexico State University, The Ohio State University, Pennsylvania State University, Smithsonian Astrophysical Observatory, Space Telescope Science Institute (STScI), the Stellar Astrophysics Participation Group, Universidad Nacional Aut\'{o}noma de M\'{e}xico, University of Arizona, University of Colorado Boulder, University of Illinois at Urbana-Champaign, University of Toronto, University of Utah, University of Virginia, Yale University, and Yunnan University.

This work made use of the AI assistant Claude (Anthropic) through Claude Code. Under the direction of the authors, it helped write results analysis and plotting code, run validation tests, check the manuscript against the code and its outputs, and draft and edit text. All of its contributions were reviewed by the authors, who take full responsibility for the content of this paper.

\end{acknowledgements}

\bibliographystyle{aa}
\bibliography{bib}

\begin{table*}
\centering
\caption{Latent variables and hyper-parameters of the forward model, with the number of degrees of freedom each carries in the production run and, for the scalar hyper-parameters, their pooled posterior (median and $16$--$84\%$ interval). The two fields are represented identically, so the counts below are per field.\label{table:params}}
{\footnotesize \setlength{\tabcolsep}{3pt} \renewcommand{\arraystretch}{1.05} \newcommand{\tabmeaning}[1]{\parbox[t]{2.0in}{\raggedright #1}}
\begin{tabular}{l l r l l}
\hline\hline
Symbol / name & Meaning & $N_{\rm dof}$ & Prior & Posterior \\
\hline
\multicolumn{5}{l}{\textit{Latent field variables} (standard normal by construction; the model maps them to $\tau$ and $v$)} \\[3pt]
\texttt{dibexcitations} & \tabmeaning{ICR excitations of $\tau = \log\rho$, 7 levels} & $18{,}876{,}016$ & $\mathcal{N}(0,\mathbf{I})$ &  \\
\texttt{vexcitations} & \tabmeaning{ICR excitations of the velocity field} & $18{,}876{,}016$ & $\mathcal{N}(0,\mathbf{I})$ &  \\
\texttt{dibspectrum} & \tabmeaning{non-parametric deviation $\psi_\tau(\log k)$} & $1{,}296$ & $\mathcal{N}(0, C_{\psi_\tau})$ &  \\
\texttt{vspectrum} & \tabmeaning{non-parametric deviation $\psi_v(\log k)$} & $1{,}296$ & $\mathcal{N}(0, C_{\psi_v})$ &  \\
\hline
\multicolumn{5}{l}{\textit{Scalar hyper-parameters}, inferred jointly with the fields} \\[3pt]
$\bar{\tau}$ (\texttt{dib\_offset}) & \tabmeaning{constant prior mean log-density} & $1$ & $\mathcal{N}(-13.0,\, 1.5^2)$ & $-13.63$ $[-13.96, -13.32]$  \\
$\beta_\tau$ & \tabmeaning{power-spectrum slope of $\tau$} & $1$ & $\mathcal{N}(-3.0,\, 0.4^2)$ & $-3.48$ $[-3.59, -3.38]$  \\
 & \tabmeaning{power-spectrum fluctuation amplitude of $\tau$} & $1$ & log-normal, mean $1.4$, std $0.4$ & $1.93$ $[1.84, 2.08]$  \\
 & \tabmeaning{spectrum flexibility of $\tau$} & $1$ & log-normal, mean $0.1$, std $0.05$ & $0.107$ $[0.074, 0.148]$  \\
$v_0$ (\texttt{meanv}) & \tabmeaning{flat rotation curve amplitude} & $1$ & $\mathcal{N}(220,\, 3^2)$~\kms & $219.5$ $[217.2, 222.7]$~\kms  \\
$\beta_v$ & \tabmeaning{power-spectrum slope of $v$} & $1$ & $\mathcal{N}(-3.5,\, 0.4^2)$ & $-3.65$ $[-3.71, -3.60]$  \\
 & \tabmeaning{power-spectrum fluctuation amplitude of $v$} & $1$ & log-normal, mean $20$, std $4$~\kms & $19.6$ $[17.2, 21.5]$~\kms  \\
 & \tabmeaning{spectrum flexibility of $v$} & $1$ & log-normal, mean $0.1$, std $0.075$ & $0.029$ $[0.012, 0.066]$  \\
\hline
\textbf{Total} & \textbf{inferred degrees of freedom} & $\mathbf{37{,}754{,}632}$ & & \\
\hline
\multicolumn{5}{l}{\textit{Fixed structure}} \\[3pt]
$N_{\rm side}$ & \tabmeaning{HEALPix pixelation, refined from $N_{\rm side}$$_0 = 2$} & --- & $128$ &  \\
$N_r$ & \tabmeaning{radial bins, $2$--$6000$~pc} & --- & $260$ &  \\
$r_{\rm lin}$ & \tabmeaning{linear/logarithmic transition radius (last linear bin edge below the configured $250$~pc)} & --- & $\rLin{}$~pc &  \\
$R_0$ & \tabmeaning{assumed solar Galactocentric radius} & --- & $\Rzero{}$~kpc &  \\
\hline
\multicolumn{5}{l}{\textit{Likelihood}, calibrated externally on the injection tests (Appendix \ref{ap:lowsnr}) and fixed during the run} \\[3pt]
$\nu$ & \tabmeaning{Student-$t$ degrees of freedom, detected branch} & --- & $10$ &  \\
$p_v$ & \tabmeaning{flat velocity density, non-detected branch} & --- & $10^{-3}$ &  \\
$a, b$ & \tabmeaning{detection probability, Eq.~\eqref{eq:pdetect_param}} & --- & fit &  \\
$\mu_{\rm n}, \sigma_{\rm n}$ & \tabmeaning{non-detection Gumbel, Eq.~\eqref{eq:gumbel}} & --- & fit &  \\
$b_{\rm n}, M, L$ & \tabmeaning{residual recalibration, Eq.~\eqref{eq:nwr}} & --- & fit &  \\
\hline
\multicolumn{5}{l}{\textit{Inference}} \\[3pt]
 & \tabmeaning{variational family} & --- & geoVI &  \\
 & \tabmeaning{antithetic sample pairs per iteration} & --- & $8$ (mirrored, so $4$ independent) &  \\
 & \tabmeaning{total iterations} & --- & $25$ (geoVI starting at iteration 3) &  \\
 & \tabmeaning{independent runs combined} & --- & $6 \times 8 = \Nsamp{}$ samples &  \\
\hline
\end{tabular}}
\end{table*}

\let\oldclearpage\clearpage
\let\clearpage\relax
\begin{appendix}
\global\let\clearpage\oldclearpage
\section{The low-SNR likelihood: a detection/non-detection mixture}
\label{ap:lowsnr}

This appendix specifies the mixture likelihood introduced in \S\ref{sec:lowsnr}, and the injection-test calibration of its three ingredients: the detection probability, the distribution of non-detections, and the recalibrated distribution of detections.

\subsection{Why a mixture is necessary}

\apMADGICS\ reports, for every visit, the maximum-likelihood DIB parameters and the local curvature of the likelihood surface at that optimum. It finds them by searching for the global optimum among locally optimal fits across a window of wavelength bins around the \apDIB\ rest wavelength. At high signal-to-noise this is a faithful summary of the likelihood and the form of Equation \eqref{ew:pdetect} applies. At low signal-to-noise, however, the search returns the largest noise excursion in the window and hence reported values stop being informative about the true DIB parameters and instead follow the distribution of extrema of the noise. Crucially, the reported local likelihood curvature based error does 
\emph{not} reflect this so simply using the reported error bar would assign potentially high confidence to a measurement that carries no signal.
For tomography, sightlines whose DIB absorption is consistent with zero are exactly the ones that bound the density along the line of sight from below, and discarding them would inflate the reconstructed density in empty regions that give the map its contrast. We therefore keep the low-SNR measurements and model the transition explicitly, as the mixture of Equation \eqref{ew:visit}. The one visit level cut we do apply, ${\rm SNR} > 2$ (\S\ref{sec:lowsnr}), sits just below the turnover of the detection probability and only removes the visits most likely to be pure noise anyway.

We proceed to parametrically model the probability of detecting a true signal peak vs not detecting it, a statistical model for the noise detections, and also re-calibrate the likelihood of the detection branch using a set of injection tests of fiducial DIB signals of varying strength and width, injected into real, absorption-free stellar spectra.

\subsection{Detection probability}

The detection probability is the raised logistic of Equation \eqref{eq:pdetect_param}, a function of the model-dependent signal-to-noise ratio ${\rm SNR} = {\rm SNR}_{\rm proxy}\,|\bar{\rho}_{d^*}/d_\rho|$ with two free parameters, the slope $a$ and the offset $b$. The exponent of $10$ generalizes the logistic to an asymmetric turn-on.

\subsection{The distribution of non-detections}

The non-detection density is Equation \eqref{eq:gumbel}: flat in velocity with $p_v = 10^{-3}$ over the velocity range considered, and a symmetrized Gumbel in $|d_\rho|$, normalized over the signed $d_\rho$ so that each sign carries half of the probability, with the location $\mu_{\rm n}$ and width $\sigma_{\rm n}$ as its two free parameters.

\subsection{Recalibrating the detected branch}

The injection tests also reveal that the reported errors are not perfectly calibrated even where the DIB is confidently detected; this is what the affine recalibration of Equation \eqref{eq:nwr} absorbs, and the Student-$t$ of Equation \eqref{ew:pdetect} is what keeps the few catastrophic outliers surviving the quality cuts from dominating the posterior. The free parameters are the offset $b_{\rm n}$, the mixer $M$ and the lower-triangular scale $L$; $\nu = 10$ is held fixed.

\subsection{Calibration on injection tests}

All of the above is fit simultaneously to the \apMADGICS\ injection tests, in which fiducial DIB signals of known equivalent width and velocity are injected into real, DIB-free APOGEE residual traces spanning the full range of noise levels, and then recovered by the standard pipeline. Writing $H_1$ for the negative log density of the recalibrated detected branch and $H_2$ for that of the non-detection branch, we minimize
\begin{equation}\label{eq:injloss}
    -\sum_{i} \log\left[
        p_i \, e^{-H_1(d_i, \bar{x}_i)} + \left(1 - p_i\right) e^{-H_2(d_i)}
    \right]
\end{equation}
over $\left(a, b, \mu_{\rm n}, \sigma_{\rm n}, b_{\rm n}, M, L\right)$ jointly, with $p_i$ given by Equation \eqref{eq:pdetect_param} evaluated at the injected truth. The sum runs over the 3744 injections passing edge and stellar quality cuts. Optimization used L-BFGS.

\begin{figure*}
    \centering
    \includegraphics[width=\linewidth]{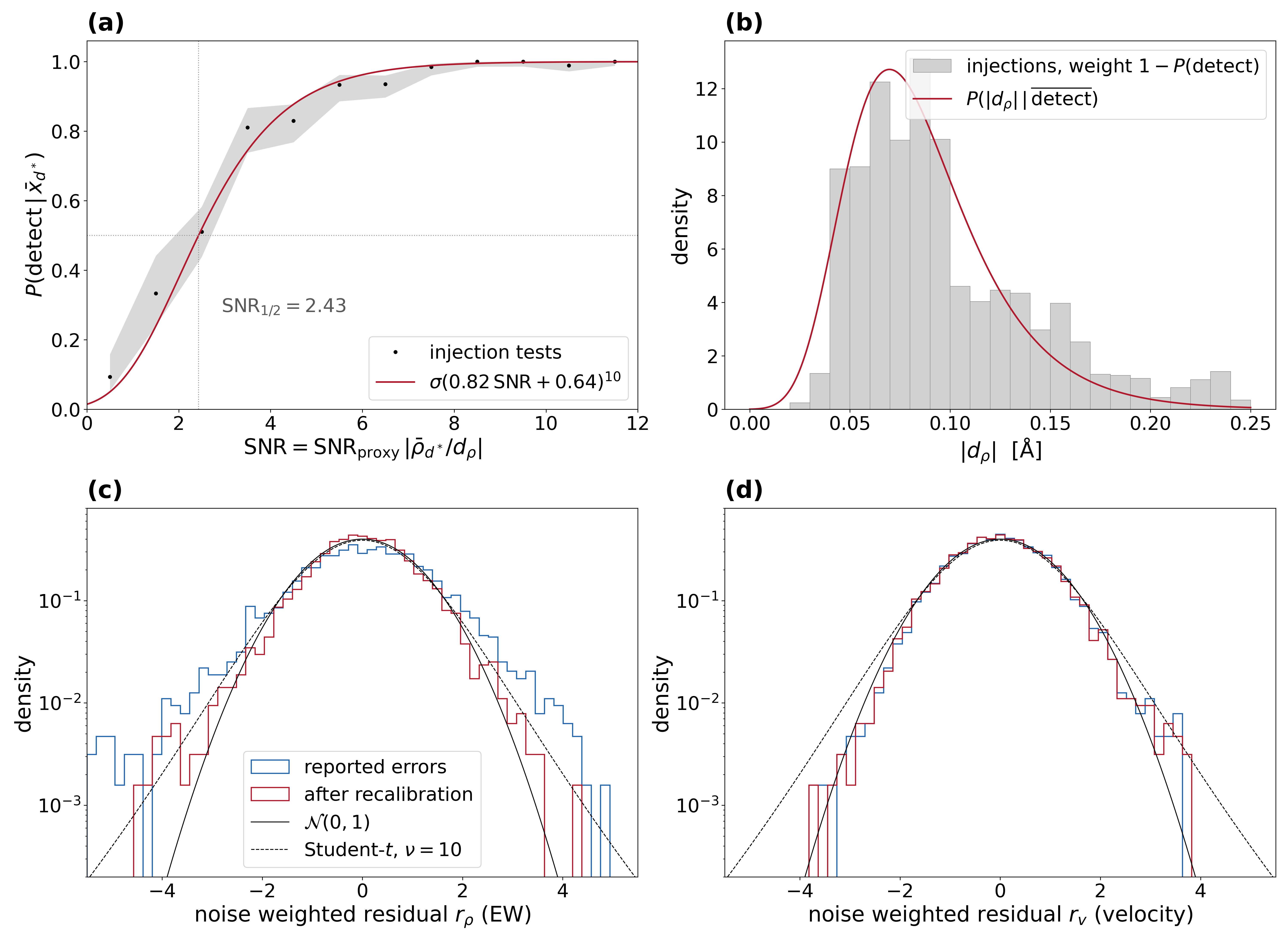}
    \caption{Calibration of the low-SNR mixture likelihood on the \apMADGICS\ injection tests. \emph{Top left:} Detection probability: points give the empirical detection fraction of the injections in bins of the model-dependent SNR of Equation \eqref{eq:pdetect_param}, with $1\sigma$ binomial (Wilson) intervals; the curve is the fitted raised logistic, and the dotted lines mark its half-way point. \emph{Top right:} The fitted symmetrized Gumbel density of non-detections (Equation \eqref{eq:gumbel}), folded onto $|d_\rho|$, against the injections weighted by the model's own non-detection weight $1 - \prob{\mathrm{detect}}$. \emph{Bottom row:} Noise-weighted residuals of the confidently detected injections ($\prob{\mathrm{detect}} > 0.99$) in equivalent width and velocity, using the reported uncertainties (blue) and after the affine recalibration of Equation \eqref{eq:nwr} (red), against a unit Gaussian and against the Student-$t$ with $\nu = 10$ actually used in the likelihood. The recalibration removes the over-dispersion of the equivalent-width residuals; the heavier-than-Gaussian tails that remain are what the Student-$t$ is there to absorb.}
    \label{fig:ap_lowsnr}
\end{figure*}

Figure \ref{fig:ap_lowsnr} shows the result. Panel (a) compares the recovered detection probability against the empirical detection fraction of the injections, binned in SNR, where an injection counts as detected if its recovered velocity is consistent with the injected one. Panel (b) compares the non-detection density against the injections weighted by the model's own non-detection weight $1 - \prob{\mathrm{detect}}$. Panels (c) and (d) show what the recalibration buys: for the confidently detected injections the equivalent-width residual is over-dispersed when the reported errors are taken at face value, with a standard deviation of 1.34 instead of unity, and the recalibration brings it to 1.01 by rescaling the reported precision by 0.77 and removing a signal-proportional offset. The velocity residuals were already close to correctly scaled, at 0.96, and are essentially unchanged. 
That scale was fit with a Gaussian detected branch, whereas in the full tomography likelihood we evaluate the recalibrated residual under the Student-$t$ of Equation \eqref{eq:nwr}. We found this small inconsistency after the production runs were complete and chose to not re-run the full pipeline as we believe the change in the results to be negligible relative to uncertainties. Specifically, refitting the calibration with the Student-$t$ branch leaves the non-detection density unchanged to within 1\%, moves the turnover of the detection probability from ${\rm SNR} = 2.43$ to $2.52$, and moves the scale to \recalScale{}: the equivalent-width errors should have been inflated by 15\% rather than 30\%. We have not re-run the reconstructions, and we do not expect any result in this paper to change significantly, for the following reasons. The mismatch is essentially a single multiplicative factor of 1.12 on every equivalent-width error, uniform across the catalog; the velocity errors and the non-detection density are untouched, and the mixing weight of Equation \eqref{eq:pdetect_param} shifts by less than a tenth of the width of its turn-on. Re-evaluating the delivered posterior with the consistent scale, the mass the fit assigns to the non-detection branch is 0.138 against 0.133, the confidently rejected fraction 9.4\% against \respRejected{}, and the velocity residual is unchanged, so no visit changes its role; the equivalent-width residual width moves from \nwRealEW{} to \nwRealEWscaled{}, i.e.\ the real data is more consistent with the model once the scale is corrected (Appendix \ref{ap:mock}). The mock of Appendix \ref{ap:mock} was generated and fit with the same scale, so it is a self-consistent test of the method that validated it with equivalent-width noise 12\% wider than the instrument's, which errs on the conservative side. We will alleviate this slight inconsistency in future applications.
Another caveat, seen in panel (b), is that the symmetrized Gumbel underestimates the tail of the non-detection distribution beyond $|d_\rho| \approx 0.1$~\AA. This is conservative in the sense that it makes large apparent absorption slightly harder to explain as noise, but it is a genuine imperfection of the chosen Gumbel model.

\subsection{The posterior responsibility}
\label{ap:responsibility}

The detection probability of Equation \eqref{eq:pdetect_param} is the \emph{prior} weight of the detected branch in the mixture of Equation \eqref{ew:visit}, and should not be misread as a statement about which visits the fit accepts. The quantity that expresses what the fit believes about a visit is the \emph{posterior} responsibility of the detected branch,
\begin{equation}\label{eq:responsibility}
    r = \frac{p\,\mathcal{L}_{\rm det}}{p\,\mathcal{L}_{\rm det} + (1-p)\,\mathcal{L}_{\overline{\rm det}}} \ ,
\end{equation}
where $p = \prob{\mathrm{detect}|\bar{x}_{d^*}}$ is the prior weight of Equation \eqref{eq:pdetect_param} and $\mathcal{L}_{\rm det}$, $\mathcal{L}_{\overline{\rm det}}$ are the two branch likelihoods of the observed equivalent width and velocity, so that the prior is weighed by how well each branch explains the measurement. A visit with $\prob{\mathrm{detect}} = 0.9$ whose equivalent width sits five sigma from the model under the detected branch is not believed, whatever the prior says: the non-detection branch explains it better, and the mixture routes its likelihood mass there.

How well this sorts genuine detections from noise can only be tested where the nature of every visit is known, on the mock catalog of Appendix \ref{ap:mock}, which also reports the whitened residuals of the detected branch on the mock and on the real data.

\section{Distance posteriors from \textit{Gaia} quantiles}
\label{ap:cbj_distances}

The per-star likelihood of Equation \eqref{eq:perstar} marginalizes over the line-of-sight distance $d^*$ of each star against a distance posterior $\prob{d^*|\mathrm{CBJ}}$. \citet{2021yCat.1352....0B} publish that posterior not as a density but as three quantiles -- the $16$th, $50$th and $84$th percentiles of the geometric distance -- so we must reconstruct a density from them.

\begin{figure*}
    \centering
    \includegraphics[width=\linewidth]{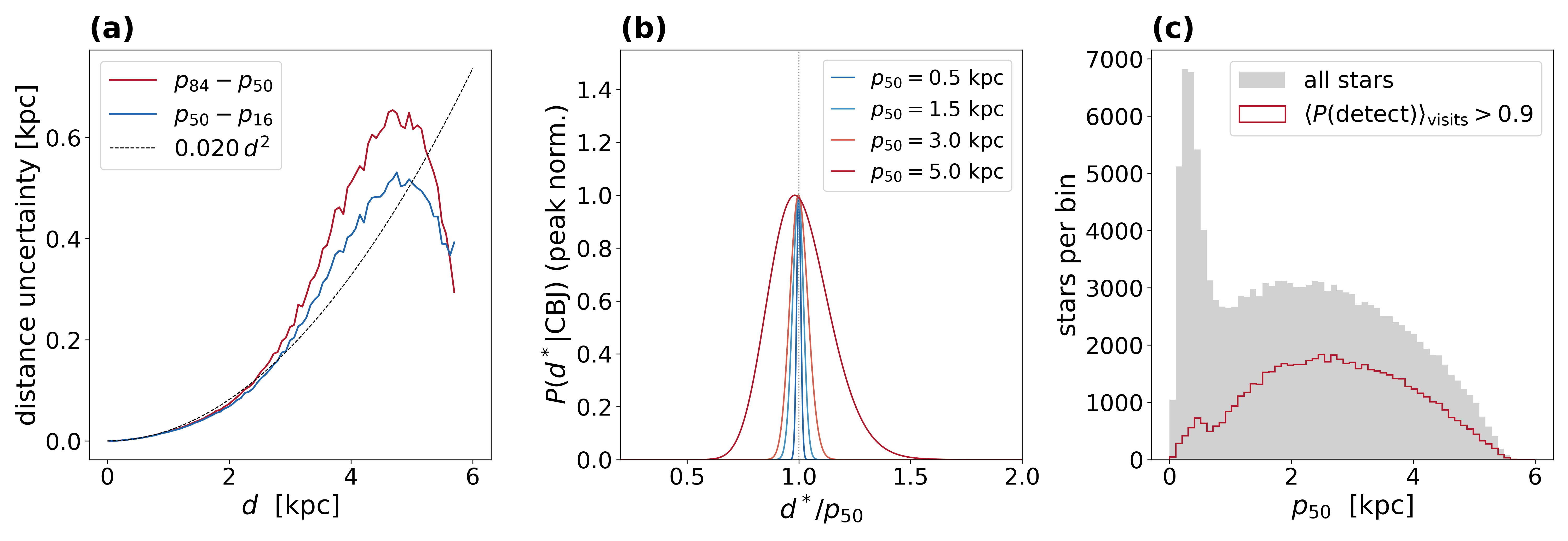}
    \caption{The \citet{2021yCat.1352....0B} distance posterior model. \emph{Left:} Median upper and lower uncertainty of the modeled visits as a function of distance, compared to a quadratic whose amplitude matches the medians between $1$ and $3.5$~kpc; beyond $\sim\!4.7$~kpc the selection $p_{84} < 6$~kpc removes the most uncertain stars. \emph{Middle:} The split-normal distance posteriors $\prob{d^*|\mathrm{CBJ}}$ the likelihood uses, for representative stars at four distances, each scaled by its own median and peak-normalized, showing the growth of both width and asymmetry with distance. \emph{Right:} Distance distribution of all stars admitted to the reconstruction (gray) and of the subset whose detection probability, averaged over the star's visits, exceeds $0.9$ (red).}
    \label{fig:ap_distance}
\end{figure*}

For simplicity we do so with a split normal distribution in $\ln d^*$, the asymmetric log-normal of \S\ref{sec:distances}. Its mode lies at the published median $p_{50}$, and below and above it the density is a half normal with the widths $\sigma_- = \ln p_{50} - \ln p_{16}$ and $\sigma_+ = \ln p_{84} - \ln p_{50}$, the two halves weighted by $2\sigma_\mp/(\sigma_- + \sigma_+)$ so that the density is continuous at the mode and normalized. The cumulative distribution is
\begin{align}
    \prob{d^* < x|\mathrm{CBJ}} &= \frac{2\sigma_-}{\sigma_- + \sigma_+}\,\Phi\!\left(\frac{\ln x - \ln p_{50}}{\sigma_-}\right) & (x \leq p_{50}) , \nonumber \\
    \prob{d^* < x|\mathrm{CBJ}} &= \frac{\sigma_-}{\sigma_- + \sigma_+} + \frac{2\sigma_+}{\sigma_- + \sigma_+} \times \nonumber \\
    &\phantom{=} \left[\Phi\!\left(\frac{\ln x - \ln p_{50}}{\sigma_+}\right) - \frac{1}{2}\right] & (x > p_{50}) .
\end{align}
It reproduces the three published quantiles exactly only when the posterior is symmetric in $\ln d^*$, $\sigma_+ = \sigma_-$; otherwise the probability below $p_{50}$ is $\sigma_-/(\sigma_- + \sigma_+)$ rather than one half. For our stars the posteriors are close to symmetric in $\ln d^*$ (median $\sigma_+/\sigma_- = 1.05$, and $0.80$--$1.40$ for $90\%$ of the stars), so that the model places $0.42$--$0.56$ of the probability below $p_{50}$, $0.13$--$0.18$ below $p_{16}$ and $0.82$--$0.86$ below $p_{84}$ for $90\%$ of the stars, against the nominal $0.5$, $0.16$ and $0.84$. The asymmetry is negligible nearby and grows with distance, where the upper uncertainty exceeds the lower one.

In the likelihood this posterior is not used as a continuous density but as a set of weights on the radial bins of the reconstruction grid (Appendix \ref{ap:grid}): the weight of radial bin $k$ for star $i$ is the probability mass that the star's distance posterior places between the bin edges. The line-of-sight integral of Equation \eqref{eq:perstar} then collapses to a weighted sum over radial bins, evaluated after the per-visit likelihoods of that star have been combined. Note that this order matters, since a star's visits all share one unknown distance and must be marginalized together.

Figure \ref{fig:ap_distance} characterizes what this means for our sample. The Bailer-Jones uncertainties are asymmetric and grow roughly quadratically with distance out to $\sim\!3$~kpc and faster beyond; in the modeled sample the median upper uncertainty peaks at $\sim\!0.65$~kpc near $4.5$--$5$~kpc and falls toward the edge, where the requirement $p_{84} < 6$~kpc removes the stars with the largest uncertainties. Panel (b) shows the reconstructed posteriors for representative stars at increasing distance, scaled by their own median. Panel (c) shows the distance distribution of the modeled stars, split by their visit-averaged detection probability; the low-SNR treatment of Appendix \ref{ap:lowsnr} is what allows the large population of nearby stars with little intervening absorption to be retained rather than cut.

\section{Discretization and the sparse grid}
\label{ap:grid}

The continuous fields of \S\ref{sec:cf} are represented on a spherical grid centered on the Sun, the product of a HEALPix angular pixelation with a hybrid linear--logarithmic radial binning.
Angularly we use HEALPix at $N_{\rm side}$ $= 128$ in the nested scheme, giving $196{,}608$ pixels of $\approx 0.46\degr$ on a side and hence a transverse resolution of $\approx 8$~pc at $1$~kpc and $\approx 48$~pc at $6$~kpc. The ICR hierarchy starts from $N_{\rm side}$ $= 2$ and refines by a factor of two per level.
Radially the grid spans $2$--$6000$~pc in \Nrad{} bins. Out to \rLin{}~pc the binning is linear with a bin width of 4.0~pc, matching the regime in which \textit{Gaia} distances are precise enough to justify it. Beyond that the bin width grows exponentially, reaching 16~pc at $1$~kpc, 48~pc at $3$~kpc and \binMax{}~pc at the $6$~kpc outer edge -- a deliberate match to the growth of the distance uncertainties in Appendix \ref{ap:cbj_distances}, so that the radial resolution tracks the resolution the data can actually support.

The ICR hierarchy has seven levels in total. Angularly it starts at $N_{\rm side}$ $= 2$ and radially at $8$. Refinement at each level is done in a sparse manner, by only refining HEALPix pixels that contain at least one visit. Sparsity is restricted to the angular part of the grid, such that each refinement carries a dense refinement in the radial axis.
The total number of cells actually carried at each level is
\begin{align}\label{eq:icr_levels}
    &\left(48,\, 192,\, 768,\, 3012,\, 10308,\, 28296,\, 55052\right) \ \text{angular} \nonumber \\
    &\qquad \times \ \left(8,\, 12,\, 20,\, 36,\, 68,\, 132,\, 260\right) \ \text{radial} \ ,
\end{align}
so that each field carries $18{,}876{,}016$ latent degrees of freedom, $14{,}313{,}520$ of them at the finest level alone (Table \ref{table:params}).
The effect of sparsity is visible here since the first three levels are dense (a full HEALPix map at $N_{\rm side}$ $\leq 8$), and from level 3 onward the refinement follows the footprint, so that at the finest level only $55{,}052$ of the $196{,}608$ $N_{\rm side}$ $= 128$ pixels are carried, a factor $3.6$ fewer degrees of freedom than a dense grid at the same resolution.

Because the hierarchical Gaussian process generation in ICR is conditional -- each level is generated conditionally on the level above, restricted to a local neighborhood -- a process generated on this sparse grid refined only where there is data is mathematically equivalent to the corresponding dense grid, and the field can be up-sampled onto the full dense grid after convergence without redoing the inference. This is what makes the $12$~kpc volume tractable at parsec-scale central resolution, and it is why the reconstruction can be delivered on a dense cube even though it was never inferred on one.

\section{Validation on a mock catalog}
\label{ap:mock}

\begin{figure*}
    \centering
    \includegraphics[width=\linewidth]{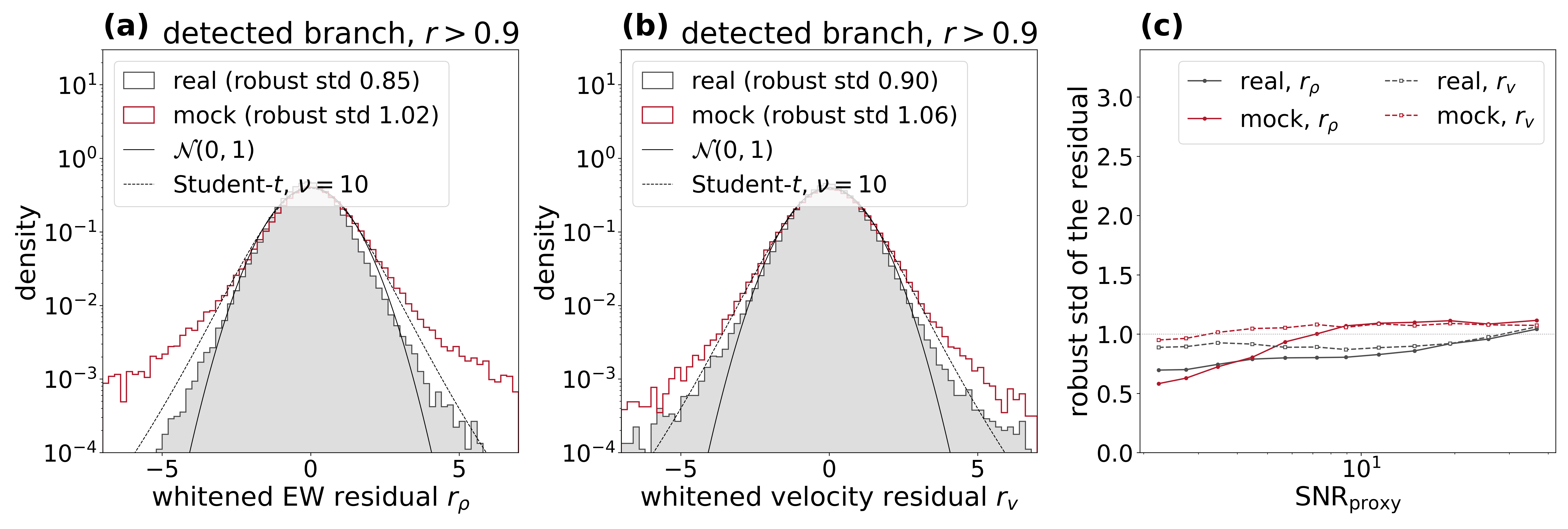}
    \caption{The data model on the real data against the mock. \emph{Left:} Whitened equivalent-width residual of the detected branch for the visits the fit attributes to it with confidence ($r > 0.9$), for the real catalog (gray) and the mock (red), with the $\mathcal{N}(0,1)$ and Student-$t$ ($\nu = 10$) references; the likelihood assumes the latter. \emph{Middle:} The same for the velocity residual. \emph{Right:} Robust width of both residuals as a function of the signal-to-noise proxy. \NsampMock{} mock and \Nsamp{} real posterior samples.}
\label{fig:noise_stats}
\end{figure*}

We validate the full pipeline end to end on a mock catalog whose truth is known (Figure \ref{fig:mock_vs_truth}). The truth density is a draw from the same correlated-field model the fit uses, multiplied by a deterministic disk envelope $e^{-(R_{\rm gal}-R_0)/R_d}\, e^{-|z|/z_h(R_{\rm gal})}$ with a flare, $z_h(R_{\rm gal}) = z_h(R_0)\, e^{(R_{\rm gal}-R_0)/R_{\rm flare}}$, so that the truth carries large-scale structure that the homogeneous, stationary prior does not encode. The input parameters are $R_d = 3.95$~kpc, $z_h(R_0) = 110$~pc and $R_{\rm flare} = 11.7$~kpc and the fluctuation amplitude of the correlated-field is fixed at $1.75$ to match the overall amplitude via the median $|{\rm EW}|$ of the resulting catalog to the real one.
The truth is generated on the full dense $N_{\rm side} = 128$ grid so that nothing about the sparse footprint is baked into it. The observing model is the real one: the stars, their sky positions, per-visit uncertainties and signal-to-noise proxies are taken from the full APOGEE parent sample (SNR $>0$: \NvisAll\ visits toward \NstarAll\ stars, of which \NvisDistCut\ visits toward \NstarDistCut\ stars survive the distance and quality cuts of the fit); each star's true distance is drawn from its own \citet{2021yCat.1352....0B} posterior; the detected branch is drawn from a Student-$t$ through the inverse of the re-calibrated likelihood; and each visit becomes a non-detection with probability $1 - \prob{\mathrm{detect}}$ from the calibrated mixture, with its equivalent width drawn from the calibrated non-detection distribution and its velocity uniformly over $\pm 156$~\kms, the largest $|v|$ in the real catalog; and finally the catalog's SNR $>2$ selection is applied once, on the mock signal-to-noise, exactly as it was applied to the real data. The resulting mock catalog has 205{,}223 measurements toward 126{,}609 stars on 33{,}373 sightlines, 87\% of the visits and 89\% of the stars of the real catalog (\NvisFit\ visits toward \NstarFit\ stars), with 30.0\% of visits non-detections. The reconstruction then uses the production configuration with no change whatsoever, and is stopped, as the production runs were, at a global iteration of 24 or 25. Three such reconstructions with different seeds are pooled into \NsampMock{} posterior samples, the three agree with one another to within $0.45$ of their combined posterior width (see below).
The mock is used to answer three questions that the real data alone cannot.

\emph{Is the vertical-resolution effect real?} The truth has a scale height with no dependence on heliocentric distance, and pushed through the shell estimator of Appendix \ref{ap:zres} it returns one: 99--137~pc in every shell from $0.5$ to $6$~kpc, with a fitted angular smearing of $k = 0.31\degr$, below a pixel. The reconstruction of that same truth returns an apparent scale height rising from 134~pc at $1$--$2$~kpc to $178 \pm 56$~pc at $5$--$6$~kpc, with $k = \kMock{}$ against $\kReal{}$ on the real data. The growth is therefore a property of the survey sampling and the method, and the smearing scale used in \S\ref{sec:galactic_structure} is this way calibrated. The joint decomposition of Equation \eqref{eq:flare_joint} recovers $z_h(R_0) = \zhMock{}$~pc against an input of $110$ and recovers the flare: at $R_{\rm gal} = 12$~kpc it recovers $z_h = \zhMockFlare{}$~pc against the $\zhTruthFlare{}$~pc of the input. The cumulative fit restricted to $r \leq 4$~kpc, the estimator whose numbers we use for the results in \S\ref{sec:galactic_structure}, returns $110.9 \pm 5.6$~pc for a truth that the same estimator puts at 106.1~pc. It hence overstates the scale height by about 5\% at that cut which we regard as an acceptable bias. Every fit here and in \S\ref{sec:galactic_structure} uses only cells with at least one star behind them in their own pixel.

\begin{table}
\centering\small\setlength{\tabcolsep}{4pt}
\caption{Detection-branch responsibility $r$ on the mock, per visit, marginalized over the \NsampMock{} posterior samples and over each star's distance posterior.}
\label{tab:responsibility}
\begin{tabular}{l c c}
\hline\hline
 & true noise draws & true detections \\
\hline
visits & $61{,}565$ & $143{,}658$ \\
mean $r$ & $0.19$ & $0.985$ \\
believed, $r > 0.95$ & $6.3\%$ & $94.2\%$ \\
rejected, $r < 0.05$ & $65.0\%$ & $0.08\%$ \\
in between & $28.7\%$ & $5.7\%$ \\
\hline
mean prior weight $\prob{\mathrm{detect}}$ & $0.43$ & $0.87$ \\
prior weight $> 0.95$ & $10.4\%$ & $61.0\%$ \\
\hline
\end{tabular}
\end{table}

\begin{figure*}
    \centering
    \includegraphics[width=\linewidth]{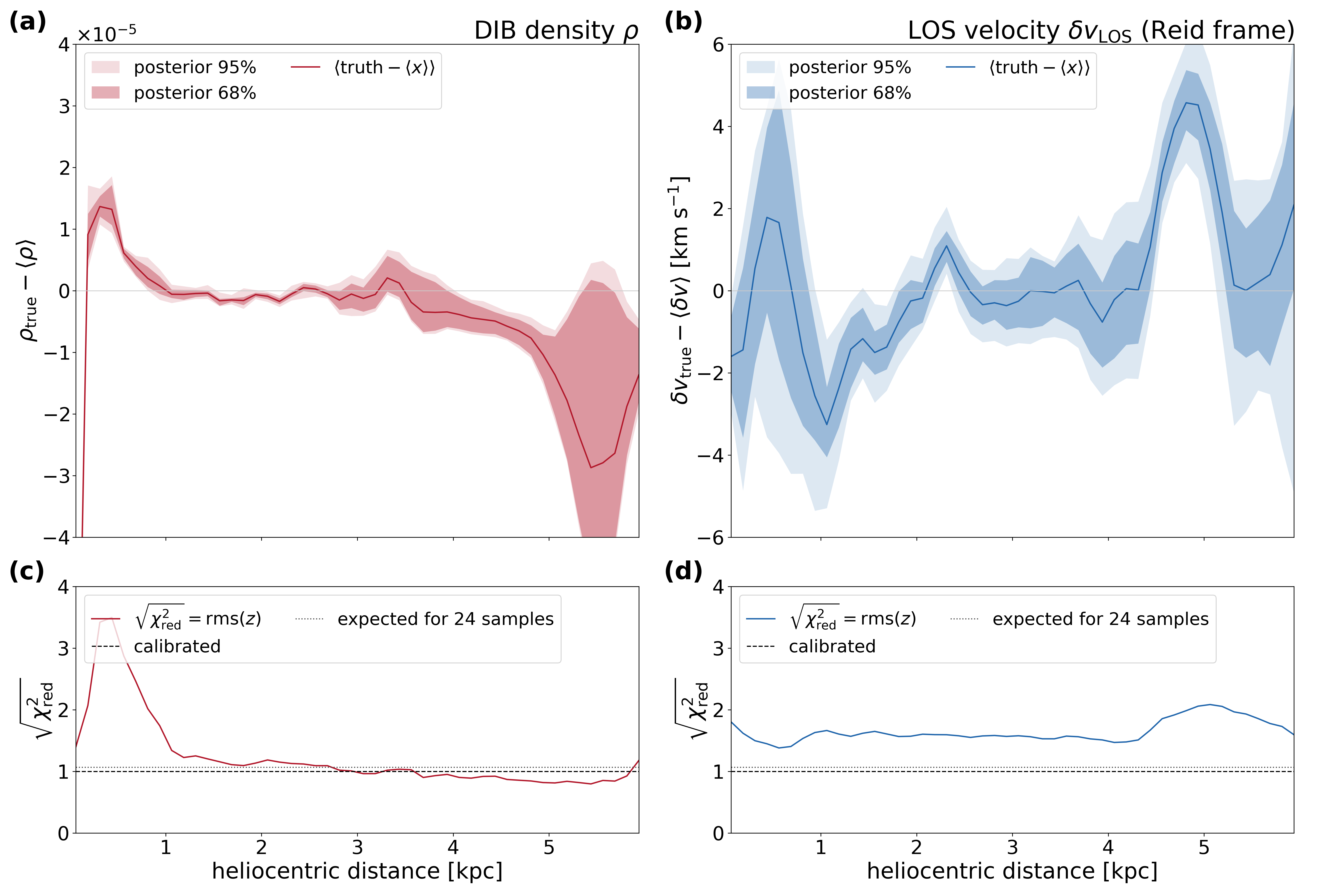}
    \caption{Calibration of the mock posterior against heliocentric distance, over all cells of the reconstructed volume. Left: linear DIB density; right: line-of-sight velocity in the denser half of the cells, with the truth and every sample referred to the same fixed rotation model (Reid frame, $\Theta_0 = 236$~\kms). \emph{Top row:} Expected residual $\langle \mathrm{truth} - \langle x\rangle \rangle$ over the cells of each distance bin (line), with the posterior's $68\%$ and $95\%$ intervals for that bin average (percentiles over the samples of the bin mean of $x_s - \langle x\rangle$, drawn around the expected residual). \emph{Bottom row:} Rms of the standardized error $z = (\mathrm{truth} - \langle x\rangle)/\sigma_x$, i.e.\ $\sqrt{\chi^2_{\rm red}}$; dashed at one, dotted at the value expected for $24$ samples.}
\label{fig:mock_zscore_radius}
\end{figure*}

\emph{How many non-detections does the fit believe?} This test is circular on the real data, because the model equivalent width against which each visit's detection probability is evaluated comes from the reconstruction itself. On the mock the truth is known per visit. Evaluating the mock posterior responsibility $r$ (Equation \eqref{eq:responsibility}) of the detected branch, $r$ sums to 153{,}100 believed detections, or \respMockBelieved{} of the visits against the true \respMockTrue{}. The split is therefore recovered to within 5 percent. For the genuine detections the mean responsibility is $\respDet{}$: 94\% are believed confidently ($r > 0.95$) and 0.08\% rejected ($r < 0.05$), so the fit essentially never discards a real measurement. For the noise draws the mean responsibility is $\respNoise{}$: the fit confidently rejects 65\% of them, confidently believes 6.3\%, and leaves the remaining 29\% in between, down-weighted rather than decided.
Table \ref{tab:responsibility} summarizes the mock responsibilities. Note that the prior weight alone is far less decisive: it averages \priorNoise{} for the noise draws and exceeds $0.95$ for 10.4\% of them, against 61.0\% of the genuine detections. The sorting is therefore done by the likelihood term, not by the prior weight alone.

On the real catalog, where the truth is unknown, the same evaluation gives a mean responsibility of $\respReal{}$: the fit attributes \respNondet{} of the catalog's likelihood mass to the non-detection branch. It confidently believes \respBelieved{} of visits, confidently rejects \respRejected{}, and leaves \respMiddle{} in between.
The believed non-detection fraction of the real catalog, \respNondet{}, is well below the 30\% with which the mock was generated, so the mock is a slightly less informative case.

\emph{Is the noise model right?} The injection tests calibrate the noise model on spectra with a known injected line; the mock tests it on a whole reconstruction, and the comparison with the real data then asks whether the real catalog behaves like data drawn from that model. To this end we look at the whitened residuals of the detected branch, $r_\rho$ and $r_v$, evaluated at the posterior samples, for visits the fit attributes to the detected branch ($r > 0.9$). On the mock, its robust width is \nwMockEW{} in equivalent width and \nwMockV{} in velocity, against the \nwRef{} of the assumed Student-$t$ with $\nu = 10$: the posterior samples reproduce the data to the noise level, with no sign of over-fitting. On the real data the same widths are \nwRealEW{} and \nwRealV{}, and the tails are lighter than assumed -- 0.6\% of visits lie beyond $3\sigma$, against 3.5\% on the mock and $1.3\%$ for the Student-$t$ itself. The real residuals are therefore, if anything, narrower than the model allows: the recalibrated errors overstate the scatter by some 15\% at intermediate signal-to-noise (Figure \ref{fig:noise_stats}). As a function of SNR, from ${\rm SNR} \approx 7$ upward the mock is flat at 1.0--1.1 in both quantities, while the real data sit at 0.80--0.86 in equivalent width and 0.87--0.92 in velocity up to ${\rm SNR} \approx 15$--$20$ and reach 1.04 and 1.06 only in the highest bin. Most of that margin is attributed to the small inconsistency in calibrating the errorbars of the detection branch likelihood against a Gaussian.

\begin{figure*}
    \centering
    \includegraphics[width=\linewidth]{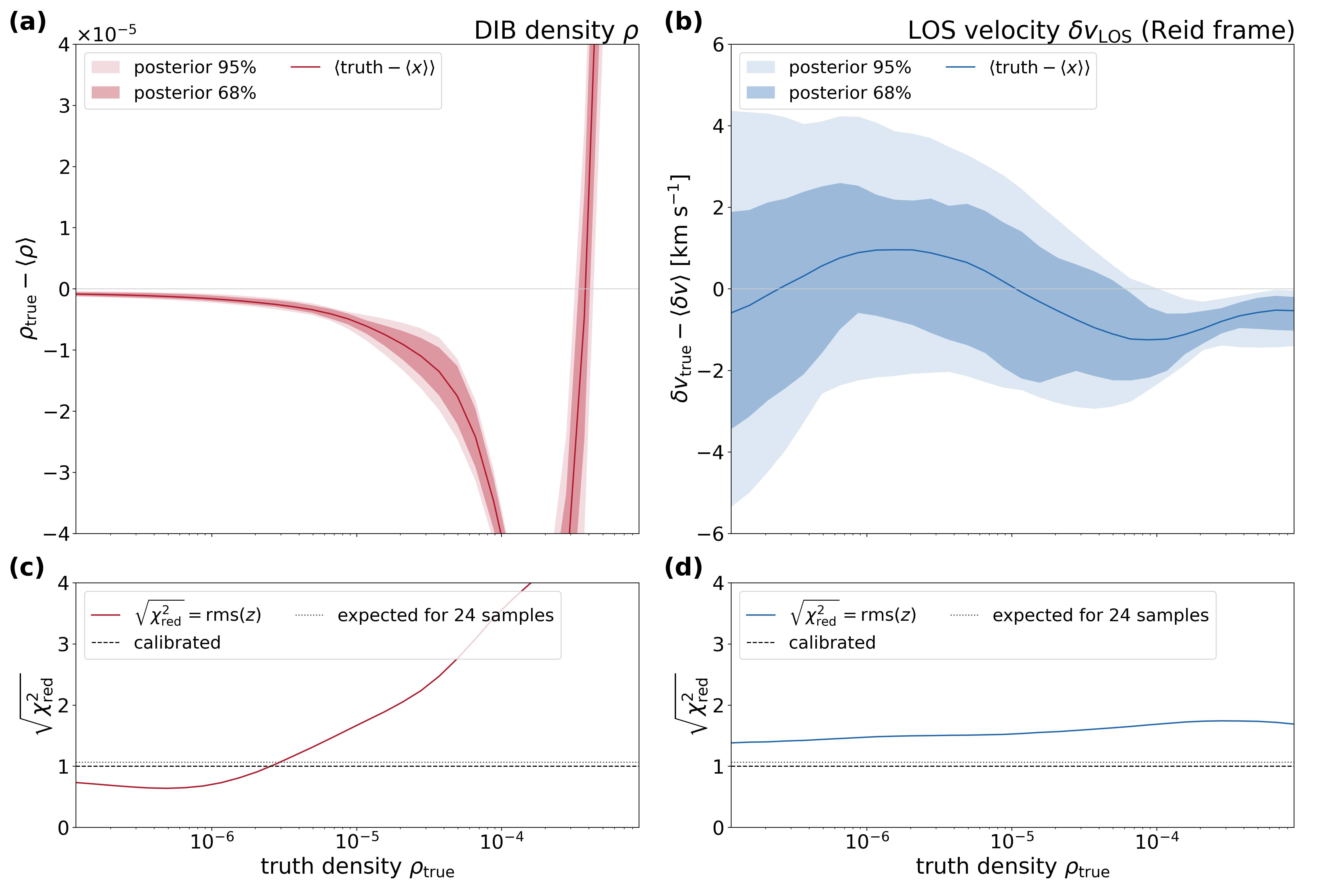}
    \caption{As Figure \ref{fig:mock_zscore_radius}, against the true density of the cell, for all cells within $4$~kpc in bins of equal logarithmic width from $\rho_{\rm true} = 10^{-7}$ to the $99.9$th percentile of the true density (the velocity over all cells of the bin).}
\label{fig:mock_zscore_density}
\end{figure*}

\emph{Is the posterior calibrated?} Variational inference tends to underestimate posterior widths, and only the mock can measure by how much. Figures \ref{fig:mock_zscore_radius} and \ref{fig:mock_zscore_density} compare the truth with the posterior mean and standard deviation for the linear density and for the velocity, cell by cell on the $N_{\rm side} = 128$ grid, against heliocentric distance over the whole reconstructed volume and against the true density within $4$~kpc. The upper panels show the expected residual, the plain mean of $\mathrm{truth} - \langle x\rangle$ over the cells of a bin, with the posterior uncertainty of that same bin average. The lower panels show the rms of the standardized error $z = (\mathrm{truth} - \langle x\rangle)/\sigma_x$, i.e.\ $\sqrt{\chi^2_{\rm red}}$. With $n = \NsampMock$ (mock) samples the reference $\sqrt{\chi^2_{\rm red}}$ is 1.07 and with $n = \Nsamp$ (production) it is 1.04. Overall, over the constrained cells, the posterior is too narrow by a factor of about \calLo{}--\calHi{}. The standardized error $(\mathrm{truth} - \langle x\rangle)/\sigma_x$ has a robust width of \calDens{} for the log density and \calVel{} for the velocity.

The velocity is uniformly under-dispersed. Its expected residual stays within $\pm 2.6$~\kms\ at most distances, reaching $-3.3$~\kms\ near $1$~kpc and $+4.6$~\kms\ near $4.8$~kpc, and stays within $\pm 1.3$~\kms\ at every density. It sits inside the posterior's $95\%$ interval for the bin mean ($\pm 1$--4~\kms) in all but a few bins, most of them between 4.5 and 5.1~kpc, where a slight $+3$--4.6~\kms\ offset appears. The rms of $z$ lies between 1.4 and 1.8 out to 4.5~kpc and between 1.6 and 2.1 beyond, and against density it rises slowly from 1.4 in the emptiest cells, whose posterior is close to the prior, to 1.75 in the densest. The rms error is 10--15~\kms\ within $4$~kpc against a quoted $\sigma$ of 7--11, the same factor of 1.5--1.6.

The density fails in two different ways on the two sides of a calibrated middle. Within the first kiloparsec the truth is denser than the posterior mean and the rms of $z$ is 3.5 there, 2.9 at 0.55~kpc, 2.0 at 0.8~kpc and 1.7 at 1~kpc.
Against density the rms of $z$ is around one at small densities and then climbs steadily.
Up to $\rho_{\rm true} \approx 4\times10^{-4}$ the expected residual is negative, the posterior mean sitting above the truth. Beyond that it turns positive and reaches almost half the true density at the top of the range. This is the smoothing of a resolution-limited field, which cannot follow the densest small-scale structure. Between $1$ and $3$~kpc the density posterior is calibrated, with the rms of $z$ at 1.0--1.3 and the expected residual within zero. Beyond $3$~kpc and below $\rho_{\rm true} \approx 2\times10^{-6}$, the posterior is over-dispersed (rms of $z$ $0.65$--$1.0$) and its mean sits above the truth, by $(1$--$3)\times10^{-5}$ at $5$~kpc and by $(1$--$2)\times10^{-6}$ in the emptiest cells. 

Beyond about $2$~kpc the density error is a bias rather than a width. The posterior mean sits above the disk truth, increasingly with distance. Stated against density contrast, the bias is a floor that fills in underdense volume: within $2$~kpc it touches only cells more than 1.5~dex below the local median ($+0.9$~dex), at $2$--$4$~kpc the cells more than $0.5$~dex below it ($+0.2$ to $+2.3$~dex). Sharp overdense structure is not overestimated anywhere: cells more than 1~dex above the local median are recovered to within $-0.3$ to $-0.1$~dex.
These figures are from the \NsampMock{} pooled samples of three independent geoVI runs, as in production. The three runs agree with one another to within $0.45$ of their combined posterior width, and the pooled $\sigma$ is only 4\% wider than a single run's.

One thing we could not test in the mock is the radial scale length. A single prior realization carries $\pm 0.5$~dex of coherent large-scale structure on top of the imposed envelope, and the truth field itself, fitted over the same windows, returns $R_d$ between 10 and 27~kpc rather than the input $3.95$. Testing $R_d$ needs a realization with the large-scale power inhomogeneously suppressed, which we leave to future work.

\section{Vertical resolution of the reconstruction}
\label{ap:zres}

\begin{figure*}
    \centering
    \includegraphics[width=\linewidth]{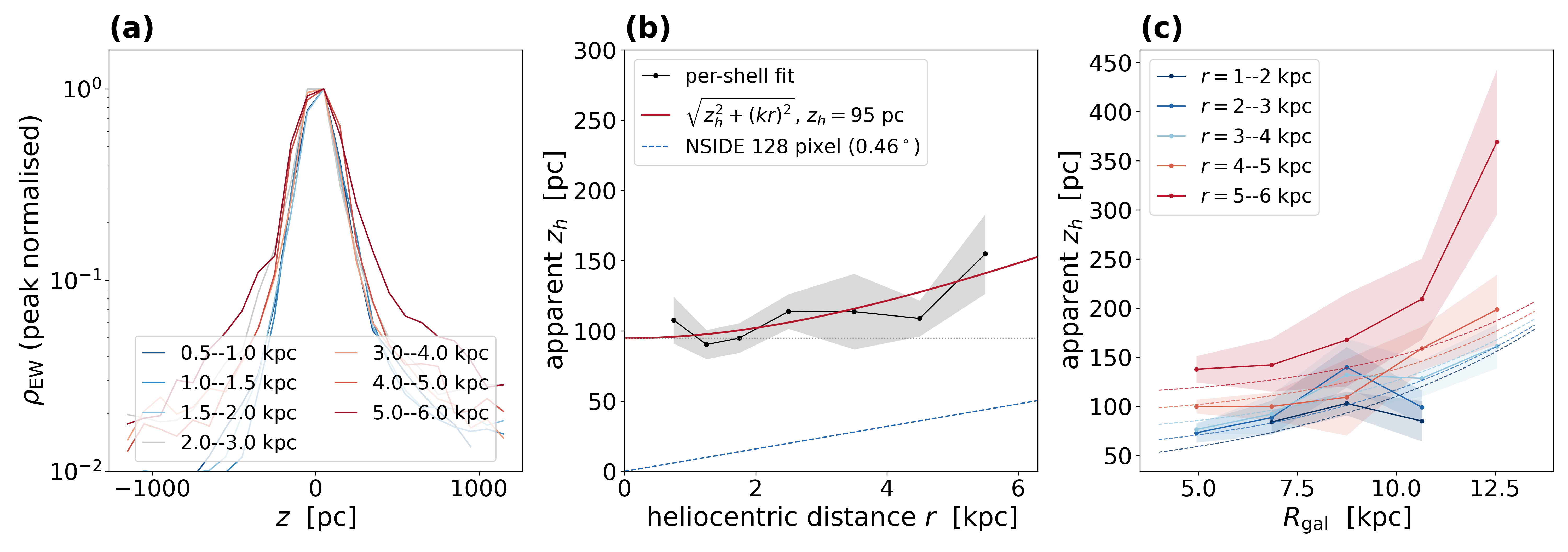}
    \caption{The vertical scale height of the DIB disk is resolution limited beyond a few kpc, but the flare is not. \emph{Left:} Vertical density profiles in shells of heliocentric distance $r$, at fixed Galactocentric radius, peak normalized: the profile broadens systematically with $r$. \emph{Middle:} Apparent scale height per shell, with the fit of Equation \eqref{eq:zres}. It is flat out to $\sim\!2$~kpc and then tracks $k\,r$; the dashed line is the nominal $N_{\rm side} = 128$ pixel scale, and the fitted smearing is about two and a half times it. \emph{Right:} Apparent scale height against Galactocentric radius \emph{within} fixed heliocentric shells, where the resolution is common to every point of a given color. The disk thickens outward in the shells beyond $3$~kpc, while in the two nearest shells the scale height falls back at $R_{\rm gal} = 10.7$~kpc within its uncertainties; together with the joint fit, in which every posterior sample prefers outward thickening, the flare is a property of the Galaxy rather than of the reconstruction; dashed curves are the joint fit of Equation \eqref{eq:flare_joint}.}
\label{fig:zres}
\end{figure*}

The disk fit of \S\ref{sec:galactic_structure} is restricted to cells within $4$~kpc of the Sun. This appendix explains why, since the reasoning also bounds what other questions can be reliably asked of the density map.

The log-normal prior adds a prior-driven density floor, thus far from constraining data, the field returns to a non-zero constant, and a constant added to a vertical profile flattens it. Fitting that floor jointly as a free parameter, over an extended $|z| < 1.5$~kpc range where the outer bins pin it, changes the structure we fit only slightly: the fitted floor is $\sim\!1$--$2\%$ of the midplane density ($1.4 \times 10^{-6}\,\nicefrac{\text{\AA}}{\mathrm{pc}}$ at the $4$~kpc cut, where it leaves $z_h(R_0) = 100.2 \pm 7.5$~pc), and the scale height drift remains at $94.8 \pm 7.6$ to $150.2 \pm 15.2$~pc.

The stronger effect due to limited resolution is seen by fitting the vertical profile in \emph{shells} of heliocentric distance rather than in cumulative cuts, at fixed $R_{\rm gal}$, so that each fit is made at one resolution instead of averaging over many. Figure \ref{fig:zres} shows the result. The apparent scale height is flat at $\sim\!95$~pc out to $\sim\!2$~kpc -- $90 \pm 10$ and $95 \pm 11$~pc in the $1$--$1.5$ and $1.5$--$2$~kpc shells -- and then grows, reaching $155 \pm 28$~pc at $5$--$6$~kpc. 
This linear growth in distance is the signature of a fixed \emph{angular} smoothing. Fitting
\begin{equation}\label{eq:zres}
    z_{h,\rm apparent}^2(r) = z_h^2 + \left(k\,r\right)^2
\end{equation}
gives $k = 1.09\degr \pm 0.34\degr$. Fitting instead over the full two-dimensional $(R_{\rm gal}, r)$ grid, with the flare and the smearing as separate terms,
\begin{equation}\label{eq:flare_joint}
    z_{h,\rm apparent}^2(R_{\rm gal}, r) = \left[z_h(R_0)\, e^{(R_{\rm gal}-R_0)/R_{\rm flare}}\right]^2 + \left(k\,r\right)^2 \ ,
\end{equation}
gives $k = \kReal{}$ and $z_h(R_0) = 82 \pm 10$~pc.
What this costs is set by simple geometry, and it is severe. An angular kernel of $1.12\degr$ subtends
\begin{center}
{\small \setlength{\tabcolsep}{4pt}
\begin{tabular}{lccccc}
\hline\hline
distance [kpc] & $1.5$ & $3$ & $4$ & $5$ & $6$ \\
\hline
$k\,r$ [pc], $k = 1.12\degr$ & $29$ & $59$ & $78$ & $98$ & $117$ \\
elements per $z_h$ & $3.2$ & $1.6$ & $1.2$ & $0.95$ & $0.8$ \\
\hline
\end{tabular}}
\end{center}
against a scale height of $\sim\!93$~pc. The smoothing equals the scale height near $5$~kpc and exceeds it beyond, so the fitted width there is a property of the kernel and not of the disk. The measured shells bear this out: $\sqrt{z_h^2 + (kr)^2}$ predicts $98$, $128$ and $141$~pc for the $1$--$1.5$, $4$--$5$ and $5$--$6$~kpc shells against measured $90$, $109$ and $155$~pc.

\begin{figure*}
    \centering
    \includegraphics[width=\linewidth]{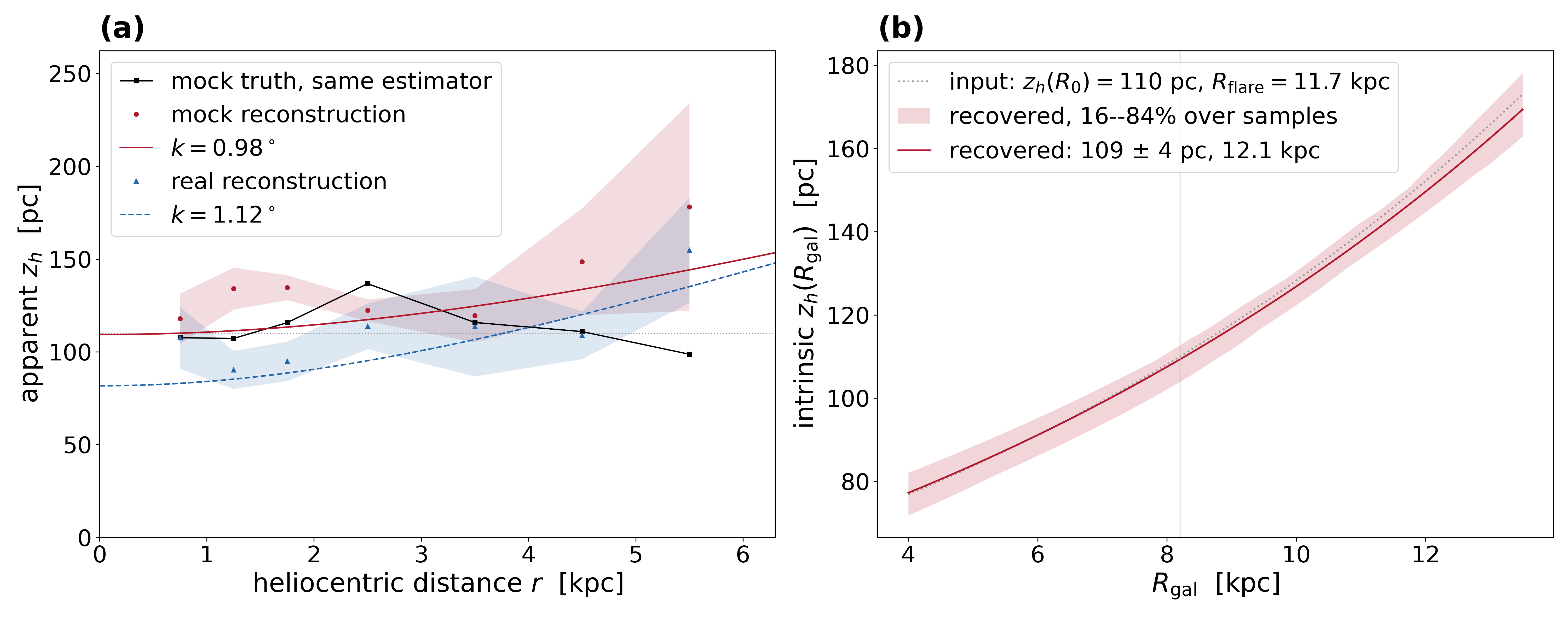}
    \caption{Calibration of the vertical-resolution argument on the mock of Appendix \ref{ap:mock}. \emph{Left:} Apparent scale height per heliocentric shell. Black squares: the mock truth, whose scale height does not depend on distance, pushed through the same estimator. Red circles: the reconstruction of that truth, with the curve $\sqrt{z_h(R_0)^2 + (k\,r)^2}$ from its joint fit of Equation \eqref{eq:flare_joint}. Blue triangles: the real reconstruction, with the same curve from its own joint fit. The two reconstructions share the same angular smearing scale $k$ to within their uncertainties, so the growth is instrumental, and calibrated. The dotted line is the input $110$~pc. \emph{Right:} The intrinsic scale-height profile $z_h(R_{\rm gal})$: the input (dotted) and the reconstruction fitted by Equation \eqref{eq:flare_joint} (red, with the $16$--$84\%$ band of the per-sample fits). The flare is recovered.}
\label{fig:mock_zres}
\end{figure*}

\subsection{Where the smoothing comes from}
The fit $1.1\degr$ is two and a half times the nominal $N_{\rm side} = 128$ pixel size of $0.46\degr$, so it is not the pixelation.
However, the angular sampling of the survey itself is sparse and carries characteristic separations across the pointings. Measuring, for HEALPix cells in the plane ($|b| < 5\degr$), the angular distance to the nearest cell containing at least one target, we do find that the median distance to the nearest occupied cell is 0.46\degr, one pixel, but the distribution has a long tail, with 16\% of plane cells more than 1.8\degr\ from any sightline and 5\% more than 2.9\degr. The fitted $k$ sits between the median and the $84$th percentile of that distribution.

\subsection{Calibration on the mock}
While the survey footprint is very suggestive, a decisive test is the scale at which we can recover the mock parameters. That measures the transfer function $z_{h,\rm recovered}(r)$ directly, with no assumed kernel and no argument about which losses are invertible. Figure \ref{fig:mock_zres} shows the result for the mock of Appendix \ref{ap:mock}, whose truth has $z_h(R_0) = 110$~pc, $R_{\rm flare} = 11.7$~kpc and no dependence on heliocentric distance.

Three things follow from it. First, the shell estimator is unbiased: the truth field itself, binned and fitted identically, returns a flat 99--137~pc across every shell, $k = 0.31\degr$, and a joint fit of $z_h(R_0) = 107.2$~pc, $R_{\rm flare} = 9.0$~kpc. Second, the reconstruction reproduces the real data's behavior: its apparent scale height rises from 134 to 178~pc between $1$ and $6$~kpc, and the joint fit returns $k = \kMock{}$, in agreement with the $\kReal{}$ measured on the real reconstruction. The smearing scale is hence a property of footprint and method. Third, the corrections are right to a few per cent: the joint decomposition returns $z_h(R_0) = \zhMock{}$~pc against $110$, and a flare of $R_{\rm flare} = 12.1$~kpc against $11.7$, i.e.\ $z_h$ at $12$~kpc of 150~pc against 152. The cumulative fit within $4$~kpc -- the number we choose to quote in \S\ref{sec:galactic_structure} -- returns $110.9 \pm 5.6$~pc where the truth through the same estimator gives 106.1, and depends on the fit volume as the real data does.

Table \ref{tab:mock_trust} sets that dependence on the fit volume versus the truth, for different cuts. The mock truth, fit over the same volume, is flat to within 5~pc; the reconstruction departs from it by 6\% at $3$~kpc, 5\% at $4$~kpc, 14\% at $5$~kpc and 16\% at $6$~kpc, and the real data drift by 23\% over the same cuts. This is what fixes the trust radius for Galactic-structure parameters: at $4$~kpc the smearing is a correction of a few per cent, beyond it the shells of Figure \ref{fig:zres} run out of information.

\begin{table}
\centering
\caption{The trust radius for Galactic-structure parameters, from the mock. Each row is the flare-model fit of Equation \eqref{eq:disk} to the data-supported cells within $R_{\odot,\max}$ of the Sun. Column 2: the height smeared by the angular resolution $k = 1.1\degr$ at the edge of that volume. Columns 3--4: the mock truth (input $110$~pc, no dependence on heliocentric distance) pushed through the same estimator, and the reconstruction of that truth from $24$ pooled posterior samples. Column 5: the real reconstruction. The recovered scale height departs from its truth by $6$, $5$, $14$ and $16$ per cent as the volume grows and the real value drifts by $23$ per cent over the same range, so the $4$~kpc cut is where the smearing is still a correction of a few per cent.}
\label{tab:mock_trust}
\begin{tabular}{c c c c c}
\hline\hline
$R_{\odot,\max}$ & $k\,R_{\odot,\max}$ & \multicolumn{2}{c}{mock $z_h(R_0)$ [pc]} & real $z_h(R_0)$ \\
\cline{3-4}
[kpc] & [pc] & truth & recovered & [pc] \\
\hline
3 & 59  & 104.0 & $109.8 \pm 5.5$ & $89.6 \pm 1.8$ \\
4 & 78  & 106.1 & $110.9 \pm 5.6$ & $93.3 \pm 1.8$ \\
5 & 98  & 106.0 & $120.6 \pm 4.9$ & $100.7 \pm 2.0$ \\
6 & 117 & 108.9 & $126.2 \pm 7.0$ & $110.5 \pm 3.0$ \\
\hline
\end{tabular}
\end{table}

\section{The tracer correlation estimator}
\label{ap:correlation}

The correlations of \S\ref{sec:maser_yso} and \S\ref{sec:stars_leave_home} compare the velocities measured for masers and young stellar clusters with the DIB velocity field at the tracer's position. Both the reconstructed velocity field and the tracer velocities are uncertain.

\emph{The DIB velocity field at a tracer position.} A tracer sits at one distance that we do not know exactly, so we average the velocity field along its sightline with the weight $W(r) = p(r) \times \rho_{\rm DIB}(r)$: the tracer's \citet{2021yCat.1352....0B} distance posterior on the radial bins weighted by the DIB density. In the scatter panels of Figures \ref{fig:v_clusters} and \ref{fig:v_residual_clusters} the point is the mean of this average over the posterior samples and the error bars are $16$--$84\%$ quantiles. For the residual comparison both the tracer's velocity and the DIB field are referenced to the same \citet{Reid2019} rotation model on the tracer's sightline.

\emph{Estimation of a correlation coefficient.} For each posterior sample $s$, we correlate the tracers' velocities with the sample's own density-weighted DIB velocities along the sightlines, with an un-centered (not subtracting a mean) coefficient and weights $1/(\sigma_{\rm RV}^2 + 1~\mathrm{km^2\,s^{-2}})$, so that noisy radial velocities are down-weighted. The quoted $\mathrm{CC}$ is the mean over the \Nsamp{} samples, with its uncertainty given by the spread over the posterior samples. For the age-resolved curve of \S\ref{sec:stars_leave_home}, every raw coefficient, for the masers and for each YSO stellar age window, is divided by the coefficient the same estimator would return for a theoretically perfectly coupled tracer subject to the same measurement and statistical DIB velocity map uncertainties. This factor sits at $0.44$ for the masers, $0.55$ for the two youngest cluster windows, $0.50$--$0.51$ at $36$--$52$~Myr, and $0.42$--$0.44$ between $70$ and $125$~Myr.

\emph{Selection cuts.} The cluster sample is restricted to clusters of \citet{Hunt} with at least seven distance-contributing members and at least seven radial-velocity members, a CMD class above $0.5$, a valid age, an $84$th distance percentile below $6$~kpc and $|v_{\rm LSR}| < 60$~\kms, excluding objects classified as globular clusters, rejected candidates, or moving groups. The maser sample is restricted to parallax distances below $5.5$~kpc to place everything confidently in our reconstructed volume. Within each correlation, tracers whose sample-mean or DIB velocity exceeds $30$~\kms\ (residual velocities) or $100$~\kms\ (full velocities) in absolute value are left out of the comparison. This cut removes less than 10\% of the remaining tracers.

\section{External kinetic tomography maps and their reference frames}
\label{ap:external_maps}

The comparison of Figure \ref{fig:kt_comparison} requires every map to be expressed in a common frame, which the published maps are not by design.

\begin{itemize}
    \item \citet{Tchernyshyov_2017} and \citet{Tchernyshyov_2018} publish heliocentric velocities. We remove the full \citet{Reid2019} solar motion and rotation model.
    \item \citet{2025A&A...693A.139S} publish a three-dimensional velocity field, which we project onto the line of sight before differencing. It is heliocentric, and receives the same full subtraction. A distance-independent $\cos l$ dipole of order $8$~\kms\ remains in this field after subtraction, which we attribute to a difference in the assumed frame or to a net radial flow in their reconstruction; their own validation used $(U, V, W)_\odot = (10.6, 10.7, 7.6)$~\kms, $\Theta_0 = 220$~\kms, $R_0 = 8.5$~kpc.
    \item The \ion{H}{I} map of \citet{McCallum2026_HI} is already referred to the local standard of rest, and we therefore do \emph{not} remove the solar peculiar motion from it. We verified this empirically: the map as published shows a dipole of $0.70$~\kms, against $18.83$~\kms\ if the peculiar motion is removed, the latter matching the magnitude of $(U, V)_\odot$.
    \item \citet{Soler_2025} distribute a matched \citet{Reid2019} line-of-sight prediction alongside their HOG map, which we subtract directly. The HOG panels are shown in their native polar $(l, d)$ pixelation.
\end{itemize}

Our own maps are placed in the same frame using $(U, V, W)_\odot = (10.4, 15.1, 8.2)$~\kms, $\Theta_0 = 236$~\kms\ and $R_0 = \Rzero{}$~kpc. The two rows of Figure \ref{fig:kt_comparison} use different slab thicknesses: $|z| \leq 300$~pc for the wide-field row, and $|z| \leq 100$~pc for the local row, the latter chosen because it matches the thickness of the $|b| < 5\degr$ cut of the HOG map at $\rLocal{}$~kpc.

\end{appendix}

\oldclearpage
\end{document}